%% file: main.tex
\documentclass[aps,floatfix,prd,superscriptaddress,reprint,showpacs,nofootinbib,10pt,preprintnumbers,longbibliography]{revtex4-1}
\usepackage[utf8]{inputenc}
\usepackage[pdftex]{graphicx}
\usepackage{float}
\usepackage{amsmath}
\usepackage{amssymb}
\usepackage{mathtools}
\usepackage{siunitx}
\usepackage{amsfonts}
\usepackage{dsfont}
\usepackage{array}
\usepackage{bm}
\usepackage{mathrsfs}
\usepackage{pifont}
\usepackage{multirow}
\usepackage{upgreek}
\usepackage[dvipsnames]{xcolor}
\usepackage{braket}
\usepackage{mathtools}
\usepackage{lmodern}
\usepackage{footnote}
\usepackage{slashed}
\usepackage{dcolumn}
\usepackage{longtable}
\usepackage{bm}
\usepackage{multirow}
\usepackage{verbatim}

\usepackage[pdftex,
  pdftitle={Scattering of two and three physical pions at maximal isospin from lattice QCD},
  pdfauthor={Bartosz Kostrzewa, Liuming Liu, Fernando Romero-Lopez, Martin Ueding, Carsten Urbach},
  bookmarks,
  colorlinks,
  linkcolor=myblue,
  citecolor=mymagenta,
  menucolor=black,
  urlcolor=myblue,
  plainpages=false,
  pdfpagelabels,
  hypertexnames=false]{hyperref}
\usepackage{verbatim}
\usepackage{slashed}
\usepackage{cleveref}
\usepackage{ucs}
\usepackage[caption=false]{subfig}

\newcommand{\Kisolin}{\mathcal{K}^{\text{iso,}1}_{\text{df,3}}}
\newcommand{\Kiso}{\mathcal{K}^{\text{iso,}0}_{\text{df,3}}}

\definecolor{mymagenta}{RGB}{200, 0, 100}
\definecolor{myblue}{RGB}{45, 48, 146}

\graphicspath{{plots/}}

\usepackage{booktabs}
\usepackage{tikz}
\usetikzlibrary{arrows.meta}
\usetikzlibrary{positioning}
\usetikzlibrary{calc}

\newcommand{\Kdf}[0]{ \mathcal{K}_{\text{3,df}} }

\begin{document}
\title{Scattering of two and three physical pions at maximal isospin from lattice QCD}

\author{Matthias Fischer}
\affiliation{Helmholtz-Institut f\"ur Strahlen- und Kernphysik, University of Bonn, 53115 Bonn, Germany}
\affiliation{Bethe Center for Theoretical Physics, University of Bonn, 53115 Bonn, Germany}
\author{Bartosz Kostrzewa}
\affiliation{High Performance Computing and Analytics Lab, University of Bonn, 53115 Bonn, Germany}
\author{Liuming Liu}
\affiliation{Institute of Modern Physics, Chinese Academy of Sciences,
  Lanzhou  730000, China} 
\affiliation{University of Chinese Academy of Sciences, Beijing 100049, China}
\author{Fernando Romero-López}
\affiliation{Instituto de Física Corpuscular, Universitat de València and CSIC, 46980 Paterna, Spain}
\author{Martin Ueding}
\affiliation{Helmholtz-Institut f\"ur Strahlen- und Kernphysik, University of Bonn, 53115 Bonn, Germany}
\affiliation{Bethe Center for Theoretical Physics, University of Bonn, 53115 Bonn, Germany}
\author{Carsten Urbach}
\affiliation{Helmholtz-Institut f\"ur Strahlen- und Kernphysik, University of Bonn, 53115 Bonn, Germany}
\affiliation{Bethe Center for Theoretical Physics, University of Bonn, 53115 Bonn, Germany}
\collaboration{Extended Twisted Mass Collaboration}
\date{\today}
\begin{abstract}
We present the first direct $N_f=2$ lattice QCD computation of two- and three-$\pi^+$ scattering quantities that includes an ensemble at the physical point. We study the quark mass dependence of the two-pion phase shift, and the three-particle interaction parameters. We also compare to phenomenology and chiral perturbation theory (ChPT). In the two-particle sector, we observe good agreement to the phenomenological fits in $s$- and $d$-wave, and obtain $M_\pi a_0 = -0.0481(86)$ at the physical point from a direct computation. In the three-particle sector, we observe reasonable agreement at threshold to the leading order chiral expansion, i.e.\@ a mildly attractive three-particle contact term. In contrast, we observe that the energy-dependent part of the three-particle quasilocal scattering quantity is not well described by leading order ChPT. 
\end{abstract}

\maketitle

\section{Introduction}

Quantum chromodynamics (QCD) describes the interaction of quarks and gluons, while only hadrons (mesons and baryons) are experimentally observable. They are low energy bound states, or resonances of the former fundamental particles.
Understanding the interactions of two or more hadrons is highly relevant for several reasons.
For instance, resonances become visible only when studying the interaction of other hadrons.
And for understanding experimental signatures of particle decays, the interactions of the final states need to be understood.

Lattice QCD, the formulation of QCD on a spacetime lattice, offers the opportunity of first principles, numerical explorations of few-particle scattering amplitudes.
Maybe the most obvious example for the importance of three-particle interactions is the $\omega$-meson, which decays predominantly into three pions with $J^P=1^-$~\cite{PhysRevD.98.030001}.
Another one would be the Roper resonance \cite{Roper}, with both $N \pi$ and $N \pi\pi$ decay channels.
However, since the investigation of three-particle interactions from lattice QCD is in its infancy, three weakly interacting pions with isospin $I=3$ is an interesting and important benchmark system.

\begin{figure}[t!]
  \includegraphics[width=1\linewidth]{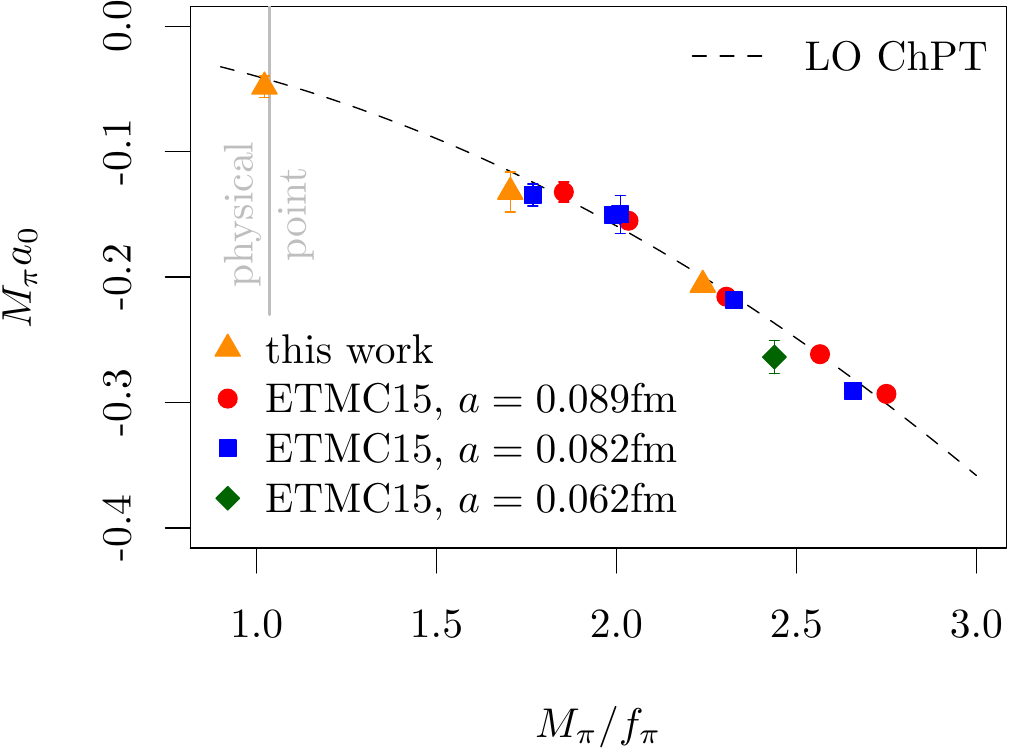}
  \caption{$I=2$ scattering length $M_\pi a_0$ as a function of $M_\pi/f_\pi$
  comparing the $N_f=2+1+1$ ETMC 
  twisted mass results~\cite{Helmes:2015gla} with this work.
  The dashed line represents the leading order ChPT prediction.}
  \label{fig:a0ETMC}	
\end{figure}

The extraction of two-particle scattering amplitudes in Lattice QCD is by now well established for $2 \to 2$ systems, both theoretically~\cite{Luscher:1986pf,Luscher:1990ux,Luscher:1990ck,Kari:1995, Kim:2005gf, He:2005ey, Bernard:2010, Hansen:2012tf, Briceno:2012yi, Briceno:2014oea, Romero-Lopez:2018zyy,Luu:2011ep,Gockeler:2012yj}, and in practice~\cite{Feng:2009ij,Lage:2009zv,Wilson:2015dqa, Briceno:2016mjc, Brett:2018jqw, Andersen:2017una, Guo:2018zss, Andersen:2018mau, Dudek:2014qha, Dudek:2016cru, Woss:2018irj, Woss:2019hse,Helmes:2018nug,Liu:2016cba,Helmes:2017smr,Helmes:2015gla,Werner:2019hxc,Culver:2019qtx,Mai:2019pqr,Doring:2012eu, Fischer:2020fvl, Woss:2020cmp, Bulava:2016mks, Rendon:2020rtw, Alexandrou:2017mpi}   (see Ref.~\cite{Briceno:2017max} for a review). One of the most studied systems is isospin-2 $\pi\pi$ scattering. To illustrate the state-of-the-art, we show in \Cref{fig:a0ETMC} the $\pi\pi$ $I=2$ scattering length $M_\pi a_0$ as a function of $M_\pi/f_\pi$ comparing this work's result to the $N_f=2+1+1$ results of Ref.~\cite{Helmes:2015gla}.
The new $N_f=2$ point at a slightly less than physical value of $M_\pi/f_\pi$ as well as the other two new points are compatible within errors with leading order (LO) ChPT (dashed line).

Over the last few years, theoretical and numerical work investigating three-particle scattering amplitudes from lattice QCD emerged as a hot topic.
The finite-volume formalism exists following three different approaches:
(i) generic relativistic effective field theory (RFT)~\cite{Hansen:2014eka,Hansen:2015zga,Hansen:2015zta,Hansen:2016fzj,Briceno:2017tce,Briceno:2018mlh,Briceno:2018aml,Blanton:2019igq,Romero-Lopez:2019qrt, Hansen:2020zhy, Blanton:2020jnm, Blanton:2020gha}, (ii)
nonrelativistic effective field theory (NREFT)~\cite{Polejaeva:2012ut,Meissner:2014dea,Hammer:2017uqm,Hammer:2017kms,Doring:2018xxx,Pang:2019dfe}, 
and (iii) (relativistic) finite volume unitarity (FVU)~\cite{Mai:2017bge,Mai:2018djl} (see also Refs. \cite{Klos:2018sen,Guo:2017ism,Jackura:2018xnx} and Ref.~\cite{Hansen:2019nir} for a review). 
Lattice data~\cite{Horz:2019rrn,Culver:2019vvu,Mai:2019fba} has been confronted with both the FVU \cite{Mai:2018djl,Culver:2019vvu,Mai:2019fba} and RFT \cite{Blanton:2019vdk} formalisms (see also \cite{Guo:2020kph, Guo:2020spn}).
For a related approach see also Refs.~\cite{Beane:2007qr,Detmold:2008fn,Romero-Lopez:2018rcb,Beane:2020ycc}.

In this article, we present results for scattering quantities of two and three-pion systems with maximal isospin, including for the first time an ensemble at the physical point. 
This work breaks new ground on several fronts: the first direct computation at the physical point of the $I=2$ $s$- and $d$-wave phase shift, and the chiral dependence of the three-$\pi^+$ quasilocal interaction. 

\section{Scattering amplitudes from lattice QCD}
The calculation of scattering amplitudes from lattice simulations proceeds in an indirect
way.
The required physical quantities from the lattice are the finite-volume interacting energies of two and three particles---the finite volume spectrum.
The mapping between the finite volume spectrum and infinite-volume scattering quantities---the so-called quantization condition---is known but highly nontrivial. It is valid up to effects that vanish exponentially with the pion mass, $\sim \exp (-M_\pi L)$.

The two-particle quantization condition (QC2) takes the form of a determinant equation~\cite{Luscher:1986pf,Luscher:1990ux,Luscher:1990ck} (we assume two identical scalars):
\begin{equation}
\det \left[ F_2^{-1}({\bf P}, E^*, L) + \mathcal{K}_2(E^*) \right]= 0\,. \label{eq:qc2}
\end{equation}
Here, $F_2$ and $\mathcal{K}_2$ are both matrices in angular momentum space $\ell,m$. The matrix elements of $F_2$ are kinematical functions (L\"uscher zeta function) that depend on the three-momentum of the system, $\mathbf{P}$ and the center-of-mass (CM) energy, $E^*$.
$(\mathcal{K}_2)_{\ell m, \ell' m'} = \delta_{\ell m, \ell'm'} (\mathcal{K}_2)_\ell$ is simply the infinite-volume scattering K-matrix projected to the corresponding partial wave. 
In order to render the matrices finite-dimensional, a truncation must be applied in $\ell, \ell'$ by assuming that $\mathcal{K}_2$ vanishes for higher partial waves.
Furthermore, the relations between $\mathcal{K}_2$, the phase shift ($\delta_\ell$), and the scattering amplitude ($\mathcal{M}_2$) are trivial.
More details can be found in Ref.~\cite{Briceno:2017max}.

The three-particle quantization condition (QC3) for identical (pseudo)scalars in  the RFT approach reads (G-parity is assumed)~\cite{Hansen:2014eka}:
\begin{equation}
\det \left[ F_3^{-1}(E, {\bf P}, L)  + \Kdf(E^*) \right]= 0\,.  \label{eq:qc3}
\end{equation}
Even though this looks formally identical to Eq.~\ref{eq:qc2}, there are some distinct features.
First, the matrices in Eq.~\ref{eq:qc3} live in a larger $k\,\ell\,m$ space, where $\ell, m$ are the angular momentum indices of the interacting pair, and $k$ labels the three-momentum of the third particle---the spectator.
Next, $F_3$ depends on geometric functions (like $F_2$ itself), but also on $\mathcal{K}_2$.
Thus, two-particle interactions are a \emph{necessary} ingredient for three-particle scattering. Note that an analytical continuation of $\mathcal{K}_2$ below threshold is needed for the QC3.
Finally, $\Kdf$ is a real, singularity-free, quasilocal, intermediate three-particle scattering quantity---which we aim to determine.
As in the case of the QC2, Eq.~\ref{eq:qc3} is infinite-dimensional, and must be truncated.
The truncation in $k$ is due to a cut-off function, whereas for $\ell , m$ one assumes that $\Kdf$ vanishes above some value of $\ell$, see Refs.~\cite{Hansen:2014eka,Hansen:2019nir} for details.
Establishing the connection between $\Kdf$ and the physical scattering amplitude, $\mathcal{M}_3$ requires a set of integral equations, derived in Ref.~\cite{Hansen:2015zga} and solved in Ref.~\cite{Briceno:2018mlh}.
In this work, we focus only on the extraction of $\Kdf$. 

In a finite volume, partial waves mix and, thus, $F_2$ and $F_3$ are nondiagonal in $\ell,m$. The correct labels are then irreducible representations (irreps) of the discrete symmetry group, which we label as $\Gamma$.
The subduction of angular momenta into irreps is known~\cite[Table~2]{Dudek:2012gj}.
Therefore, one block-diagonalizes the quantization conditions into irreps, see Refs.~\cite{Gockeler:2012yj,Doring:2018xxx, Blanton:2019igq, Blanton:2019vdk}.

\begin{table}[th]
  \centering
  \begin{tabular*}{.49\textwidth}{@{\extracolsep{\fill}}cccccc}
    \toprule\hline
      Ensemble & $L^3 \times T$ & $M_\pi / \si{\mega\electronvolt}$ & $a M_\pi$ & $M_\pi/f_\pi$ &  \#\,confs. \\ 
    \midrule\hline
      cA2.60.32 & $32^3 \times 64$ & $340$ &  $0.1578(1)$\phantom 0 & $2.235(6)$ & 337 \\
      cA2.30.48 & $48^3 \times 96$ & $242$ &  $0.11199(4)$ &  $1.705(1)$ &  1403 \\ 
      cA2.09.48 & $48^3 \times 96$ & $134$ & $0.06205(4)$ & $1.022(1)$ & 1604 \\ 
    \bottomrule\hline
  \end{tabular*}
  \caption{%
    $N_f=2$ Ensembles used in this work. The lattice spacing is $a = 0.0914(15)\,\si{\femto\meter}$,
and $c_\text{SW} = 1.57551$. For the decay constant we use the normalization $f_\pi = \sqrt{2} F_\pi$.
  $M_\pi / f_\pi$ has been corrected for finite-size effects according to Refs.~\cite{Gasser:1986vb,Gasser:1987ah,Gasser:1987zq}.
  }
  \label{tab:etmc-ensembles}
\end{table}

\section{Lattice computation}

This work uses $N_\mathrm{f} = 2$ flavour
lattice QCD ensembles generated by the Extended Twisted Mass collaboration (ETMC)
\cite{Abdel-Rehim:2015pwa}, including one ensemble at the physical 
pion mass---see \Cref{tab:etmc-ensembles}.
For the ensemble generation the Iwasaki gauge
action~\cite{Iwasaki:1985we} was used together with Wilson clover
twisted mass fermions at maximal twist~\cite{Frezzotti:2000nk}. The latter 
guarantees scaling towards the continuum with only $O(a^2)$ artefacts
in the lattice spacing $a$~\cite{Frezzotti:2003ni}.
The presence of the clover term
(with coefficient $c_\mathrm{sw}$) has been shown to further reduce
the $O(a^2)$ artefacts, in particular isospin-breaking effects of
the twisted-mass formulation, which have been empirically found to be very small
for masses and decay constants~\cite{Abdel-Rehim:2015pwa}.
For the two-pion scattering length with $I=2$, discretisation artefacts are only of order $O(a m_q)^2$, with $m_q$ the up/down quark
mass~\cite{Buchoff:2008hh}.
Another possible source of $O(a^2)$ effects that should be mentioned is the $\pi^0$ contamination in the correlation functions due to the breaking of parity in twisted mass.
However, it is also important to realise that at maximal isospin there is no mixing with other flavour states due to broken isospin symmetry.
Parametrically, $O(a^2)$ artefacts are $\sim 2.5\%$ and $O(a m_q)^2 \leq 0.4$\% for this lattice spacing, and thus well below our statistical uncertainty.

The two- and three-$\pi^+$ energy spectrum is measured from Euclidean correlation functions of operators with the corresponding quantum numbers. By means of the single pion operators ($\pi^+ = - \bar{\mathrm u} \gamma_5 \mathrm d$), we construct two-particle operators as
\begin{equation}
\mathcal{O}_{\pi\pi}(p_1, p_2) = \sum_{x,y} e^{ip_1 x + i p_2 y}\, \pi^+(x)\, \pi^+(y)\,,
\end{equation}
where $p_i$ labels the momentum of each single pion, and similarly for three pions
\begin{equation}
  \begin{split}
    \mathcal{O}_{\pi\pi\pi}(p_1, p_2, p_3) = \sum_{x,y, z}& e^{ip_1 x + i p_2 y + ip_3 z}\\
    & \times \pi^+(x)\, \pi^+(y)\, \pi^+(z)\,.
  \end{split}
\end{equation}
Correlation functions are computed using the stochastic Laplacian-Heaviside smearing \cite{Peardon:2009gh,Morningstar:2011ka} with algorithmic parameters as in Ref.~\cite{Dimopoulos:2018xkm}.
In addition, operators that transform under a specific irrep of a discrete symmetry group are constructed following Ref.~\cite{Werner:2019hxc}.
In the two-pion case we use the irreps $A_1^{(+)}, E^{(+)}, B_1$ and $B_2$, in the three pion channel $A_1^{(-)}, E^{(-)}, A_2, B_1$ and $B_2$, for all $\mathbf{P}^2 \leq 4$ with $\mathbf{P}$ the centre-of-mass momentum.
We refer to \Cref{tab:summarylevels} in the appendix for an overview.
We extract the spectrum in each irrep independently using the generalized eigenvalue method (GEVM)~\cite{Michael:1982gb,Luscher:1990ck,Blossier:2009kd} and also the GEVM/PGEVM method~\cite{Fischer:2020bgv}, see the appendix for more details.

A technical issue of lattice calculations with (anti)periodic boundary
conditions in the time direction is the presence of so-called thermal
states, i.e.\@ effects from states that propagate backwards in time
across the boundary.
They vanish with $M_\pi T \to\infty$, but at finite values of $T$, these effects are significant and need to be treated accordingly.
In fact, thermal pollutions are one of the major systematic uncertainties in our calculation.
We deal with them as follows: using the operators discussed above we build correlator matrices which are input to the GEVM/PGEVM which in turn have so-called principal correlators as output.
From the latter energy levels and corresponding error estimates are extracted from bootstrapped, fully correlated fits to the data with fit ranges chosen by eye.
We use five different \emph{treatments} to arrive from a correlator matrix at an energy level.
Details of those five treatments are explained in \Cref{sec:thermal}.

As also explained in \Cref{sec:thermal}, the different energy levels per principal correlator  (up to five) are then combined using a correlated weighted average.
However, to account for the spread between the different methods we use a procedure discussed in Ref.~\cite{Werner:2019hxc} to widen the resampling distribution: for energy level $E$ we compute the scaling factor 
\begin{equation}
  \label{eq:s}
  w = \sqrt{\frac{(\delta E)^2 + \sum_Y(\Delta E_Y)^2}{(\delta E)^2}}\,,
\end{equation}
where $\delta E$ is the statistical uncertainty of the weighted average and $\Delta E_Y$ is the difference between method $Y$ and the weighted average.
By scaling the resampling distribution of the weighted average with $w$, we obtain a distribution that reflects both the statistical and the systematic uncertainties, while still being usable in the bootstrap analysis chain.
The energy levels are publicly available~\cite{Nf2-3pi-I3-scattering-data}.

\begin{figure*}[ht!]
    \centering
        \subfloat[$s$-wave\label{fig:swavephys}	]{\includegraphics[width=0.48\textwidth]{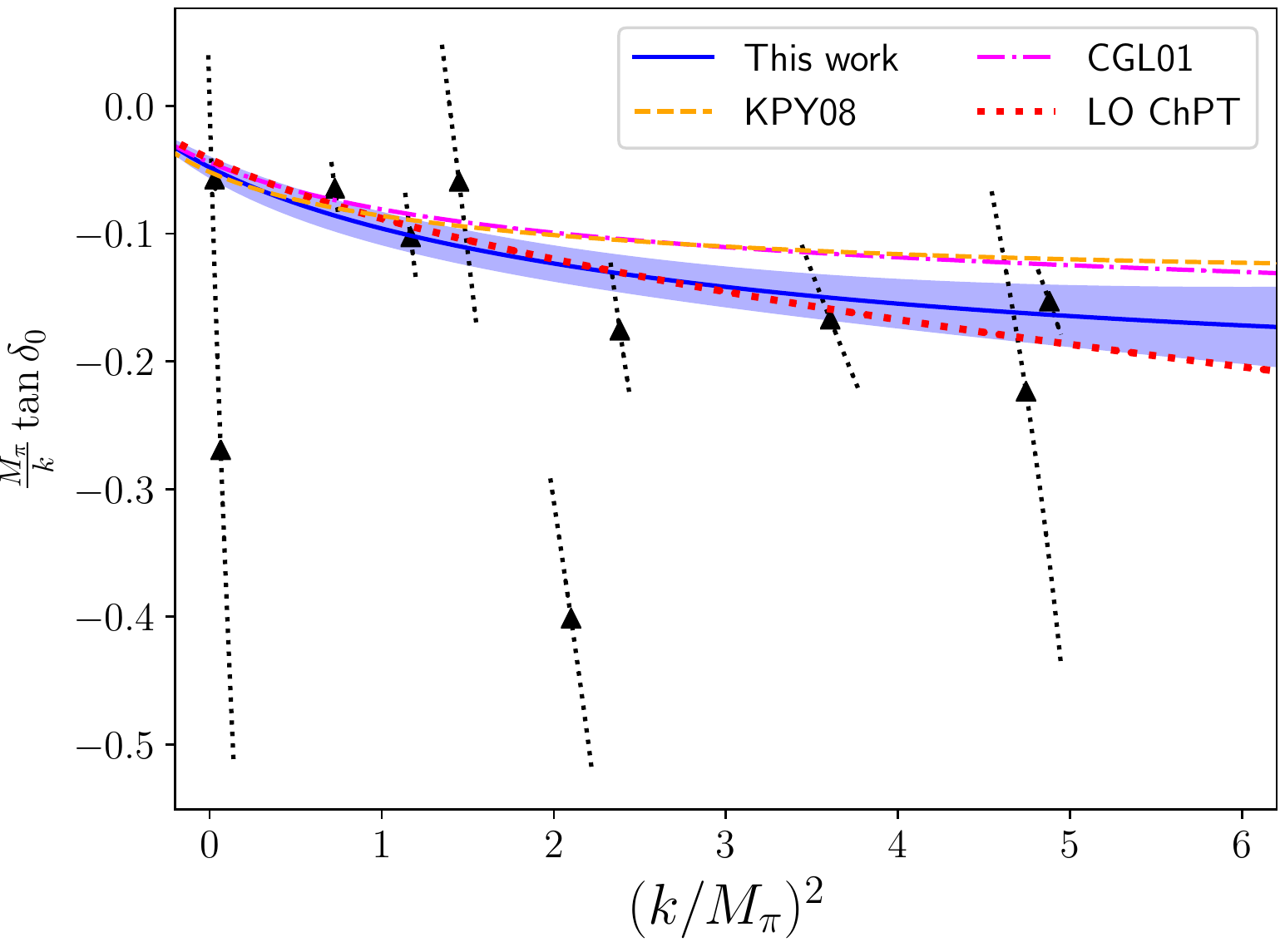}}
        \qquad
        \subfloat[{$d$-wave}\label{fig:dwavephys}]{\includegraphics[width=0.48\textwidth]{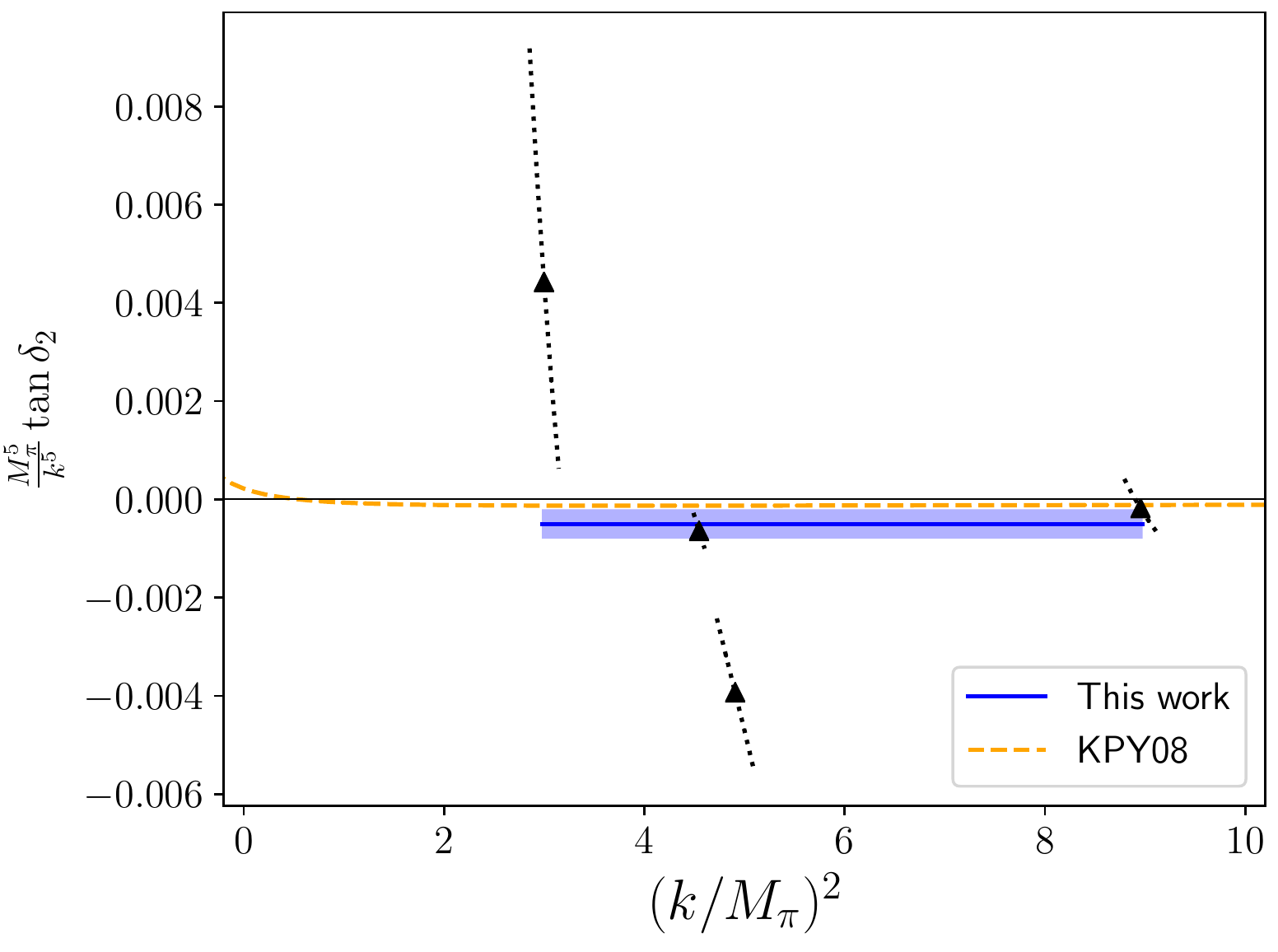}}
   \caption{  $s$- and $d$-wave phase shift at the physical point (ensemble cA2.09.48) compared to the fits to experimental data (KPY08) in Ref.~\cite{Kaminski:2006qe} and (CGL01) in \cite{Colangelo:2001df}. For $s$-wave we use a model that incorporates the Adler-zero, whereas for $d$-wave we fit to a constant in the region for which we have data.}\label{fig:2pions}
\end{figure*}

The finite-volume scattering formalism is applicable under the assumption that exponential finite volume effects are negligible. On the physical point ensemble, we have $M_\pi L\approx 3$, which implies $e^{-M_\pi L} \sim 5\%$ and might be considered to be at the edge of feasibility.
However, based on a ChPT analysis, finite-volume effects are also proportional to $[M_\pi/(4\pi F_\pi)]^2$, which at the physical point reduces finite-volume effects sizably. Moreover, as argued in Ref.~\cite{Romero-Lopez:2018rcb}, if the volume-dependent mass is used to analyze the multi-particle energy levels, the leading finite-size effects cancel.
For the other two ensembles we have $M_\pi L > 5$, which is safe concerning finite volume effects.

\section{Results} 
In the case of two pions, by keeping only $s$-wave interactions in $A_1$ irreps, the projected QC2 becomes a one-to-one correspondence of an energy level to a phase shift point  \cite{Luscher:1990ck, Kari:1995}. For the analysis, we need an appropriate phase shift parametrization. We use a model that incorporates the expected Adler zero \cite{adlerref,Blanton:2019vdk}:
\begin{equation}
\frac{k}{M_\pi} \cot \delta_0  = \frac{\sqrt{s}M_\pi}{(s - 2 z^2)} \left( B_0 + B_1 \frac{k^2}{M_\pi^2} + \ldots \right) \,, \label{eq:Adler}
\end{equation}
with $s$ the center-of-mass energy squared and $k^2 = s/4-M_\pi^2$.
We will fix the position of the Adler zero to its leading order Chiral Perturbation Theory (LO ChPT) value: $z^2 = M_\pi^2$. Even though higher order corrections are to be expected, its value has been seen to be compatible with LO ChPT when left free \cite{Yndurain:2002ud, Pelaez:2004vs, Kaminski:2006qe}. Note that in Eq.~\ref{eq:Adler} with fixed Adler zero, we have $M_\pi a_0 = 1/B_0$. 

\begin{table}[th]
  \centering
  \begin{tabular*}{.49\textwidth}{@{\extracolsep{\fill}}crrrr}
      \hline
      & $1/B_0=M_\pi a_0$ & $B_1$ & $B_2$  & $\chi^2/$dof \\ \hline 
      cA2.60.32      &  -0.2090(54)   & -2.3(3)       &  ---      &    19.06/(16-2)         \\ 
      cA2.60.32       &   -0.2110(57)& -3.1(6)&  0.4(2)              &     15.96/(16-3)         \\
      cA2.30.48   & {-0.132(16)} &  -1.4(5) & --- &  27.15/(16-2)  \\  
      cA2.09.48  & -0.0477(90) &  -1.4(1.2) & --- & 11.08/(10-2)\\
      \hline
  \end{tabular*}
  \caption{$s$-wave fit results for the various ensembles using Eq. \ref{eq:Adler} with fixed $z^2_2=M_\pi^2$. Here we use only the two-pion levels in the $A^+_{1}$ and $A_1$ irreps.}
\label{tab:2piadler}
\end{table}

We perform a correlated two-parameter fit to the energy levels. The results for the three ensembles are shown in \Cref{tab:2piadler}. In all cases, the magnitude of the $B_i$ coefficients decreases with increasing order, indicating that the expansion converges quickly enough even at the heaviest pion mass. Still, for the heaviest ensemble (cA2.60.32), we also attempt a fit with a quadratic term in $k^2$, $B_2$ and observe a small, barely significant value for $B_2$ and no substantial change in $B_0$ and $B_1$. Based on ChPT, better convergence is expected for lighter pions.

The $s$-wave phase shift is visualised for the physical point ensemble in the left panel of \Cref{fig:2pions}.
In this plot we also compare to other results in the literature.
For the other two ensembles the corresponding plots can be found in the left panels of \Cref{fig:2pionsA30} and \Cref{fig:2pionsA60}, respectively, in the appendix.

One interesting point to discuss is the suitability of the $\delta_0$ parametrization. It has been customary to use a standard effective range expansion parametrization (ERE) for isospin-2 $\pi\pi$ scattering:
\begin{align}
\begin{split}
\frac{k}{M_\pi} \cot \delta_0 =  \frac{1}{M_\pi a_0} &+ \frac{1}{2} M_\pi r \left( \frac{k}{M_\pi} \right)^2 \\&+ M_\pi^3 P  \left( \frac{k}{M_\pi} \right)^4. \label{eq:ERE}
\end{split}
\end{align}
However, the presence of the Adler zero limits the radius of convergence to $k^2 \sim 0.5 M_\pi^2$.
For this reason, explicitly incorporating the Adler zero must improve the radius of convergence, and has been shown to provide a better description of the data~\cite{Blanton:2019vdk}.
Here, we compare again the two fit models.
The ERE results are shown in \Cref{tab:2piERE}.
As can be seen, the values of $\chi^2$ in the case of the ERE fits are always larger than their Adler-zero counterparts given in \Cref{tab:2piadler}.
This further supports the  usage of the Adler-zero parametrization for $I=2$ $\pi\pi$ scattering.

\begin{table}[h!]
  \centering
  \begin{tabular*}{.48\textwidth}{@{\extracolsep{\fill}}rllll}
    \hline
    Ensemble & $M_\pi a_0$ & $M_\pi r$ & $M_\pi^3 P$ & $\chi^2/$dof \tabularnewline  \hline
    cA2.60.32      &  -0.2198(55) &  1.1(2)           &   ---      &     28.12/(16-2)   \tabularnewline
    cA2.60.32      &   -0.2177(56)&  2.1(5)& -0.16(8) & 24.26/(16-3)     \tabularnewline
    cA2.30.48      &    -0.186(15)&  1.5(4)&--- &  31.98/(16-2) \tabularnewline
    cA2.09.48         &  -0.064(11) & 3.9(1.1)&--- & 14.00/(10-2) \tabularnewline \hline
  \end{tabular*}
\caption{Two-particle fits to the standard effective range expansion (ERE) model in Eq. \ref{eq:ERE}.}
\label{tab:2piERE}
\end{table}

Similarly, the $d$-wave phase shift can be obtained from most of the
nontrivial irreps when neglecting $\ell>2$ waves
\cite{Luu:2011ep,Gockeler:2012yj}. Since we have few data points, we
attempt the following fit (see \Cref{tab:dwave}):
\begin{equation}
\frac{k^5}{M_\pi^5} \cot \delta_2 = \frac{1}{M_\pi^5 a_2} \,. \label{eq:dwave}
\end{equation}
The best fit curve for the physical point ensemble is show in the right panel of \Cref{fig:2pions} and compared to Ref.~\cite{Kaminski:2006qe}.
Again, for the other two ensembles the corresponding plots can be found in the appendix in the right panels of \Cref{fig:2pionsA30} and \Cref{fig:2pionsA60}, respectively.

In the three pion case we need to parametrize $\Kdf$.
For this, we expand $\Kdf$ about threshold up to linear terms of relativistic invariants \cite{Blanton:2019igq}:
\begin{equation}
\Kdf = \Kiso + \Kisolin \Delta \,, \quad \Delta = \frac{(E^*)^2 - 9M_\pi^2 }{9 M_\pi^2}\,, \label{eq:kdf}
\end{equation}
where $\Kiso$ and $\Kisolin$ are the numerical constants to be determined. This parametrization has no momentum dependence, and thus receives the name ``isotropic''. It is the three-particle equivalent of keeping only $s$-wave interactions. At the next order in the expansion, $O(\Delta^2)$, three new parameters arise, for which also the $d$-wave must be included \cite{Blanton:2019igq}. This is beyond the scope of the present analysis.

Following the strategy outlined in Ref.~\cite{Blanton:2019vdk}, we perform a simultaneous $s$-wave only fit to two-$\pi^+$ $A_1$ levels, and all three-$\pi^+$ levels. For this, we use the $\delta_0$ model in Eq.~\ref{eq:Adler} and the $\Kdf$ parametrization in Eq.~\ref{eq:kdf}---four parameters in total, see \Cref{tab:3piadler}. As can be seen the best fit values for $B_0$ and $B_1$ agree well between the two-particle and the global fit, with even smaller errors in the case of the latter.
For convenience, we provide the full covariance matrices of the fits in \Cref{tab:3piadler} in the appendix, see \Cref{eq:cor1,eq:cor2,eq:cor3}.

We have also performed fits including only the constant term $\Kiso$, the results of which can be found in the appendix.
We observe that for the ensembles with larger than physical pion mass value the inclusion of the linear term seems necessary.

\begin{table}[th]
  \centering
  \begin{tabular*}{.48\textwidth}{@{\extracolsep{\fill}}crrr}
    \hline
    & $M^5_\pi a_2$ & $\chi^2/$dof & CM energy range  \\ \hline
    cA2.60.32      &  -0.0037(08)&     15.03/(12-1) &  $[3.2 M_\pi, 4.4 M_\pi]$        \\ 
    cA2.30.48      &  -0.0072(11)&   23.78/(10-1)  &  $[2.8 M_\pi, 4.2 M_\pi]$   \\ 
    cA2.09.48      &  -0.0005(03)&     7.33/(4-1)    &   $[4.0 M_\pi, 6.3 M_\pi]$     \\
    \hline
  \end{tabular*}
  \caption{$d$-wave two-pion fits to Eq. \ref{eq:dwave}. Here we use only non-$A_1$ two-pion levels. The last column shows the energy range for which data is used. }
  \label{tab:dwave}
\end{table}

In \Cref{fig:spectrum-prediction} in the appendix we provide as an example for the physical point ensemble the measured energy spectrum in the two- and three particle sectors separately.
In that figure we also compare to the noninteracting energy levels.
Moreover, we give the energy levels predicted by our fits, see \Cref{tab:2piadler,tab:dwave,tab:3piadler}

\section{Discussion}
Starting with $\delta_0$, we show in \Cref{fig:swavephys} all phase shift data points, and include the best fit curve from the two- and three-$\pi^+$ global fit. As can be seen, the difference to LO ChPT is small, and due to $B_1 \neq 0$. In addition, our results agree within $<2\sigma$ with Refs.~\cite{Kaminski:2006qe,Colangelo:2001df}. We obtain $M_\pi a_0 = -0.0481(86)$ (see  \Cref{tab:3piadler} and recall $1/B_0 = M_\pi a_0$), which also agrees well with all phenomenological determinations \cite{Yndurain:2002ud, Pelaez:2004vs, Kaminski:2006qe,Colangelo:2001df,Caprini:2011ky,Albaladejo:2012te}, and other lattice results obtained indirectly by extrapolating to the physical point using ChPT \cite{Yamazaki:2004qb,Beane:2005rj, Beane:2007xs,Feng:2009ij,Beane:2011sc,Yagi:2011jn,Fu:2013ffa, Sasaki:2013vxa, Helmes:2015gla,Mai:2019pqr}, see \Cref{fig:a0ETMC}. 

In \Cref{fig:a0ETMC} we also compare to results from $N_f=2+1+1$ calculations from Ref.~\cite{Helmes:2015gla} and with LO ChPT.
Within the uncertainties we do not observe a significant difference between $N_f=2$ and $N_f=2+1+1$ results.
Moreover, as was found in all previous investigations of two pions at maximal isospin, LO ChPT describes the mass dependence extraordinarily well.
At the physical point, LO ChPT predicts $M_\pi a_0 \simeq -0.04438$, which agrees within error bars with the value we report here, see above.
Unfortunately, our determination here suffers from relatively large
statistical uncertainties and, thus, cannot compete with
determinations based on chiral extrapolations. A summary of various
determinations from the literature is compiled in \Cref{tab:a0summary}.

\begin{table}[th]
  \centering
  \begin{tabular*}{.48\textwidth}{@{\extracolsep{\fill}}lll}
    \hline
    & $N_f$ & $M_\pi a_0$  \\
    \hline
    LO ChPT    &       & $-0.04438$ \\
    CGL01 (2001)     &       & $-0.0444(10)$ \\
    CCL11 (2011)   &       & $-0.0445(14)$ \\
    CP-PACS (2004) & 2     & $-0.0431(29)(-)$ \\
    NPLQCD (2006)  & 2+1   & $-0.0426(6)(3)$ \\
    NPLQCD (2008)  & 2+1   & $-0.04330(42)_\mathrm{comb}$\\
    ETM (2010)     & 2     & $-0.04385(28)(38)$\\
    ETM (2015)     & 2+1+1 & $-0.0442(2)(^{+4}_{-0})$ \\
    Yagi (2011)    & 2     & $-0.04410(69)(18)$ \\
    Fu (2013)      & 2+1   & $-0.04430(25)(40)$ \\
    PACS-CS (2014) & 2+1   & $-0.04263(22)(41)$ \\
    GWU (2019)     & 2 & $-0.0433(2)$ \\
    This work      & 2     & $-0.0481(86)(-)$   \\ 
    \hline
  \end{tabular*}
  \caption{Summary of some lattice and phenomenological determinations
    of the isospin-2 $\pi\pi$ scattering length at the physical
    point. Note that the lattice determination of ETM~(2015) is the only
    one with chiral and continuum extrapolations.
    We list LO ChPT, ChPT and Roy
    equations~\cite{Colangelo:2001df} denoted as CGL01,
    CCL11~\cite{Caprini:2011ky},
    CP-PACS~\cite{Yamazaki:2004qb}, NPLQCD~(2006)~\cite{Beane:2005rj}, 
    NPLQCD~(2008)~\cite{Beane:2007xs}, ETM~(2013)~\cite{Feng:2009ij}, 
    ETM~(2015)~\cite{Helmes:2015gla}, Yagi et al.~\cite{Yagi:2011jn},
    Fu~\cite{Fu:2013ffa} and PACS-CS~\cite{Sasaki:2013vxa}, and GWU~\cite{Mai:2019pqr}}
  \label{tab:a0summary}
\end{table}

Regarding the $d$-wave phase shift, we have mild statistical evidence that it is repulsive at the physical point in the considered energy region. We observe agreement within $\gtrsim 1\sigma$ with Ref.~\cite{Kaminski:2006qe}, as shown in \Cref{fig:dwavephys}. An interesting feature of the phenomenological fits to $\delta_2$ is that there is a sign change near threshold, which yields an attractive phase shift at threshold \cite{Yndurain:2002ud, Pelaez:2004vs, Kaminski:2006qe,Bijnens:1997vq}. We cannot confirm or deny such behaviour, as the explored energy region is too far above threshold. For larger pion mass values, we obtain a similar behaviour. The $d$-wave phase shift is more repulsive for the two larger pion mass values---see \Cref{tab:dwave} and the appendix.

\begin{table*}[t]
  \centering
  \begin{tabular*}{.75\textwidth}{@{\extracolsep{\fill}}crrrrrr}
    \hline
    & $M_\pi a_0$ & $B_1$ & $B_2$ & $M_\pi^2\Kiso$ & $M_\pi^2\Kisolin$ & $\chi^2/$dof \\  \hline 
    cA2.60.32    &   -0.2061(49) & -1.9(2) & --- & 4500(1500)& -6200(1800) & 58.89/(43-4)       \\  
      cA2.60.32    &   -0.2070(52) & -2.2(5) & 0.1(2)&4300(1500)& -6000(1800) & 58.50/(43-5)     \\ 
    cA2.30.48     &   -0.156(15) &  -1.9(4) & ---& 1800(3800)& -4300(3800) &  46.18/(33-4)    \\ 
    cA2.09.48     &   -0.0481(86) & -1.3(1.1)& --- & 0(800)& -200(500) &
    19.06/(19-4)    \\
    \hline
  \end{tabular*}
\caption{Two- and three-pion fits using the Adler-zero form ($z^2=M_\pi^2$, fixed).  Since we only include $s$-wave interactions, we use two-pion levels in the $A_1$ irrep, and all irreps for three-pions. Recall that $1/B_0 = M_\pi a_0$.}
\label{tab:3piadler}
\end{table*}

\begin{figure*}[t]
    \centering
        \subfloat[$\Kiso$ \label{fig:K0}]{\includegraphics[width=0.48\textwidth]{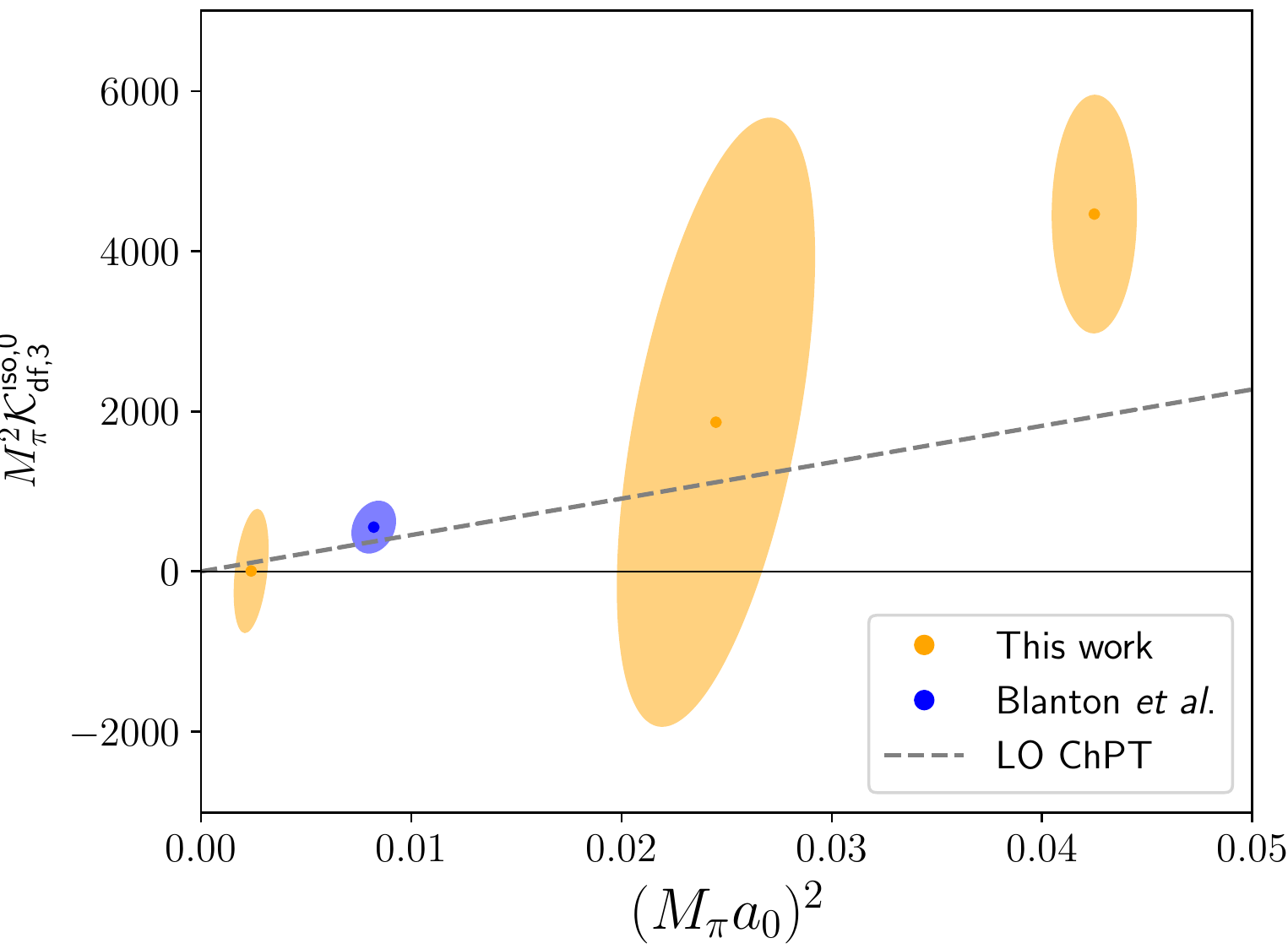}}
        \qquad
        \subfloat[$\Kisolin$\label{fig:K1}]{\includegraphics[width=0.48\textwidth]{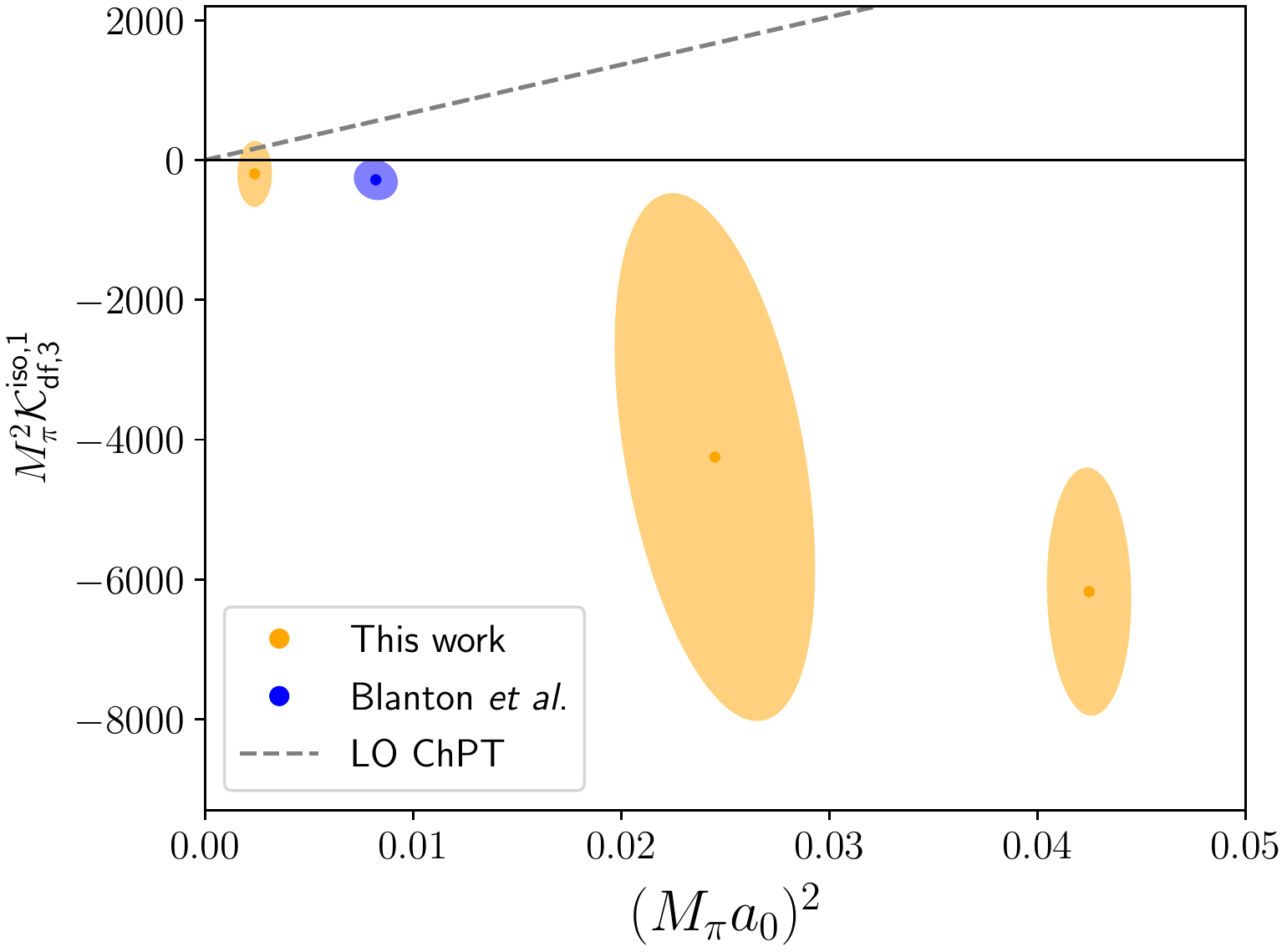}}
   \caption{ Constant(left) and linear(right) terms of $\Kdf$ as a function of the $s$-wave scattering length. We also include the results of Ref. \cite{Blanton:2019vdk}.  }\label{fig:3pions}
\end{figure*}

We show our results in the three-particle sector in \Cref{fig:3pions}. As can be seen in \Cref{fig:K0}, there is significant evidence that $\Kdf$ at threshold ($\Kiso$) is positive (attractive). Even though we find reasonable agreement with the LO ChPT~\cite{Blanton:2019vdk} prediction, the data suggests that NLO effects can be significant, and it may be worth to extend the ChPT result to one loop in future work. For $\Kisolin$,  the situation is somewhat different. All evidence points to a negative value, very far from the ChPT results. While one could conclude that a NLO ChPT description is required, there is a subtlety in the LO ChPT prediction: it assumes that the connection between $\Kdf$ and $\mathcal{M}_3$---which involves integral equations---is trivial in LO ChPT~\cite{Blanton:2019vdk}
\begin{equation}
\Kdf = \mathcal{M}_{3,\text{df}} \left[ 1 +  O(M_\pi^2/F_\pi^2) \right]\,,
\end{equation} 
where $\mathcal{M}_{3,\text{df}}$ is the divergence-free three-to-three amplitude \cite{Hansen:2015zga}. As argued in Ref.~\cite{Blanton:2019vdk}, this induces large errors in $\Kisolin$ (up to 50\% for 200 MeV pions). The situation is expected to be more dramatic for heavier pions, like our two results at 242 and 340 MeV, for which the largest difference is seen. In order to address this rigorously, the integral equation must be systematically solved, which is beyond the scope of this work.

\section{Conclusion}
We have presented the first $N_f=2$ lattice calculation of two- and three-$\pi^+$ scattering at the physical point.
In the two pion channel we observe very good agreement with other lattice calculations and ChPT or ChPT combined with Roy-Steiner equations for the $s$-wave phase shift.
In particular, for the whole range of pion mass values we have available here we do not observe a significant deviation from LO ChPT or a significant difference to $N_f=2+1+1$ lattice results.
For the $d$-wave our uncertainties are relatively large.
However, thanks to the physical point ensemble we can directly compare to phenomenology and observe reasonable agreement.
For the $d$-wave phase shift smaller scattering momenta would be desirable in order to be able to shed light on a possible sign change at small $k^2$-values.

For the three pion case, we observe reasonable agreement with other lattice calculations, phenomenology, and ChPT.
By including two ensembles at heavier pion masses, we have gained insight on the chiral dependence of three-$\pi^+$ scattering quantities for the first time. 
We use an isotropic parametrisation of $\mathcal{K}_{3,\mathrm{df}}$, the real, singularity free, quasilocal, intermediate three particle scattering quantity.
Here we find good agreement to LO ChPT for the constant term in $\mathcal{K}_{3,\mathrm{df}}$ in an expansion about threshold, but an opposite sign compared to LO ChPT for the next-to-leading term.
We have discussed possible explanations for this. 
On the other hand, qualitative agreement is found for both terms with the other available lattice calculation of these quantities.

This letter represents a step towards exploring and understanding the hadronic spectrum of QCD, and shows that three-particle quantities can be extracted with current techniques.
In the very near future we expect more lattice calculations of three-body observables with increasing accuracy and describing systems with growing complexity---e.g.~three-particle resonances such as the $\omega$.

\begin{acknowledgments}
   We thank all members of ETMC for the most enjoyable collaboration. We also thank P. Hernández, A. Rusetsky, and S. Sharpe for comments on the manuscript.
  The authors gratefully acknowledge the Gauss Centre for Supercomputing
  e.V.\@ (www.gauss-centre.eu) for funding this project by providing
  computing time on the GCS Supercomputer JUQUEEN~\cite{juqueen} and the
  John von Neumann Institute for Computing (NIC) for computing time
  provided on the supercomputers JURECA~\cite{jureca} and JUWELS~\cite{juwels} at J\"ulich
  Supercomputing Centre (JSC). Parts of the the results were created
  within the EA program of JUWELS Booster and we are thankful for
  the help and support of the
  JUWELS Booster Project Team (JSC, Atos, ParTec, NVIDIA).
  This project was funded in part by the DFG as a project in the
  Sino-German CRC110.
  FRL acknowledges the support provided by the European projects H2020-MSCA-ITN-2015/674896-ELUSIVES, H2020-MSCA-RISE-2015/690575-InvisiblesPlus, the Spanish project FPA2017-85985-P, and the Generalitat Valenciana grant PROMETEO/2019/083. The work of FRL also received funding from the EU Horizon 2020 research and innovation program 
under the Marie Sk{\l}odowska-Curie grant agreement No. 713673 and ``La Caixa'' Foundation (ID 100010434, LCF/BQ/IN17/11620044).
  The open source software packages tmLQCD~\cite{Jansen:2009xp,Abdel-Rehim:2013wba,Deuzeman:2013xaa}, 
  Lemon~\cite{Deuzeman:2011wz}, 
  QUDA~\cite{Clark:2009wm,Babich:2011np,Clark:2016rdz},
  R~\cite{R:2019}, hadron~\cite{hadron:2020} and paramvalf~\cite{paramvalf}
  have been used.

\end{acknowledgments}

\nocite{Wolff:2003sm}

\bibliography{bibliography}

\clearpage

\begin{appendix}
\input{appendix}

\end{appendix}

\end{document}

%% file: appendix.tex
\widetext

\section{Extraction of the energy levels}
\label{sec:energies}

\begin{figure}[tbh]
  \begin{minipage}{0.85\linewidth}
    \input{quark-diagrams}
    \label{fig:diagrams}
  \end{minipage}
\end{figure}

In this section, we provide more details regarding the extraction of
energy levels from the correlation functions of one, two and three charged
pions. All the required quark contraction diagrams are shown in
\Cref{fig:diagrams}. For the observables in question we have
determined the integrated autocorrelation times using the method put
forward in \cite{Wolff:2003sm} and found that we can treat our
measurements as decorrelated. The statistical analysis is performed via bootstrap.

\subsection{Thermal Pollutions}
\label{sec:thermal}

Given the individual pion momenta $p_i, i=1,2,3$, we adopt the
following convention to express the total momentum
$\mathbf{P}=\sum_i\mathbf{p_i}$ and the relative momenta $\mathbf{q}_j, j=1,2$ 
\begin{equation}
\mathbf{p_1} = \mathbf{P} -\mathbf{q_1}
- \mathbf{q_2}\,,\quad \mathbf{p_2} = \mathbf{q_1}\,,\quad \mathbf{p_3} = \mathbf{q_2}\,.
\end{equation}
The spectral composition of a three-pion correlation function (with periodic
boundary conditions) reads
\begin{equation}
    \sum_m \sum_n \bra n O_\Gamma({\mathbf P}, {\mathbf q_1}, {\mathbf q_2}) \ket m\,
    \bra m O_\Gamma^\dagger({\mathbf P}, {\mathbf q_3}, {\mathbf q_4}) \ket n
     \mathrm e^{- E_n \cdot (T-t)} \mathrm e^{-E_m t} \,.
\end{equation}
The double sum is over all states $m, n$ with the correct quantum
numbers. The desired signal arises when $m$ is the vacuum, and $n$
the three-pion state, or \emph{vice versa}.
Usually one would expect that all other contributing to the spectral
decomposition are exponentially suppressed compared to this ground
state. Here this is not the case, because there are nonzero
contributions to the spectral decomposition for finite $T$ for
instance when $m$ is an 
intermediate two-pion state and $n$ is a one-pion state. Such so-called
thermal pollution states have a time dependence proportional to
$\exp(-\Delta E\, t)$, with $\Delta E = E_{2\pi} - E_{\pi}$, which can
dominate the correlation function for large enough $t$ when
$\Delta E < E_{3\pi}$. There is
an additional backward propagating part as well which goes as \(\exp(-\Delta E
\cdot (T - t))\). Together they either form a cosh (sum) or a sinh
(difference). For three pions we only have time-even operators and
therefore everything will have a cosh-shape. The amplitude of the cosh
will be proportional to $\exp(-(E_{2\pi} + E_{\pi}) T)$, which
vanishes for $T\to\infty$.

The size of the pollution
will depend on the individual momenta of the involved pions through the energy $E_{2\pi}$ and $E_\pi$. The most
significant pollution will be the one leading 
to the smallest $\Delta E$, which usually corresponds to the
smallest involved momenta.

The thermal pollutions depend also on the frame and irrep.
Let us illustrate this for a specific example: assume that $n$ is a
one-pion state $\ket{\mathbf p_1}$ and $m$ a two-pion state
$\ket{{\mathbf p_2}, {\mathbf p_3}}$ with free energies given by 
the dispersion relation. In this specific case only summands where
$\bra{\mathbf p_1} O_\Gamma \ket{{\mathbf p_2}, {\mathbf p_3}} \neq 0$
contribute, i.e.\@ the three-pion operator $O_\Gamma$ must couple to
the momenta $\mathbf p_1$, $\mathbf p_2$ and $\mathbf p_3$. 

The individual particle momenta that couple to a multi-particle
operator can be inferred from group theory. Consider the frame
${\mathbf P}^2=0$, then the three-pion operator will be in some irrep
$\Gamma^-$, and the single pion always in the $A_1^-$. Therefore, the
two-pion system needs to be in the opposite parity irrep $\Gamma^+$
such that $A_1^- \otimes \Gamma^+ = \Gamma^-$.  Note that only the
irreps for ${\mathbf P}^2 = 0$ have a parity index, that is, in moving
frames parity is not a good quantum number. In this situation, the
momenta of the two-pion system can only take the values that actually
couple to the irrep of the three-particle operator. 

\begin{figure}
    \centering
    \subfloat[$\mathbf P^2 = 0$ in $A_1^-$ \label{fig:thermal-states-A1u}]{
        \includegraphics[width=0.48\linewidth, page=1]{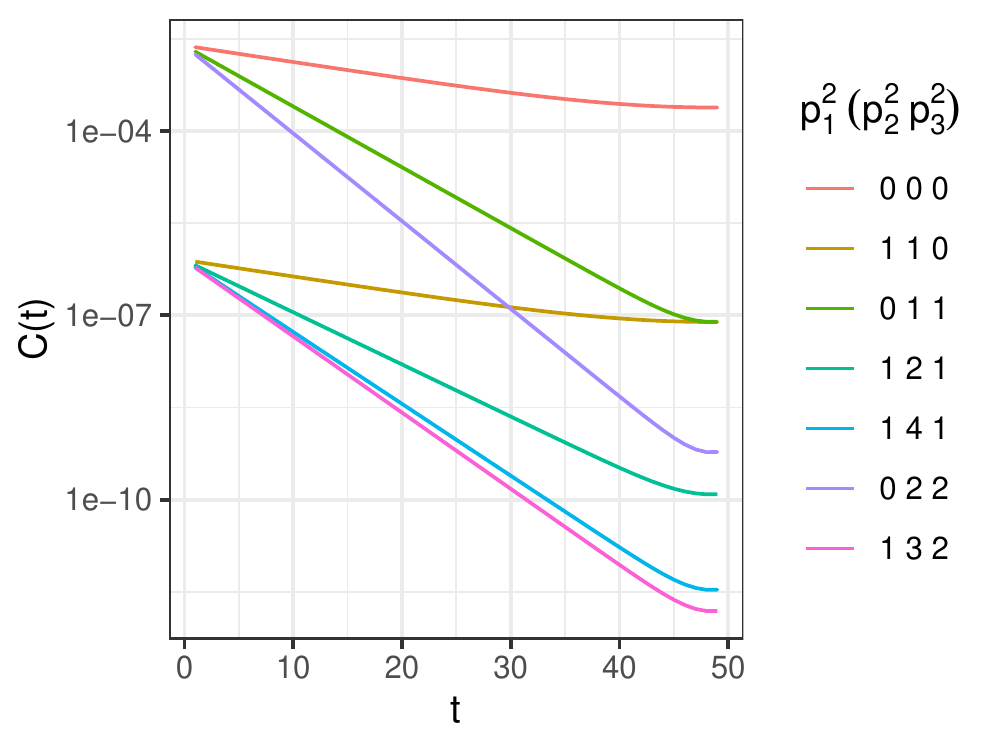}
    }
    \hfill
    \subfloat[$\mathbf P^2 = 2$ in $B_1$ \label{fig:thermal-states}]{
        \includegraphics[width=0.48\linewidth, page=2]{thermal_states_selection_small_paper}
    }
    \caption{%
        Possible thermal contributions to the three-pion correlator matrix in
        the cA2.09.48 ensemble. Each line corresponds to a particular
        combination of the individual particle momentum magnitudes of the
        one-pion ($\mathbf{p}_1$) and the two-pion system
    ($\mathbf{p}_2$ and $\mathbf{p}_3$). 
    }
    \label{fig:thermal-states-both}
\end{figure}

The allowed contributions are generated from all permutations of the three-pion
individual momenta. Using the measured pion rest mass $M_\pi$ and the
free particle dispersion relation (assuming weak interactions between
the two pions) we can thus estimate the relevant energies $E_\pi(\mathbf{p}_1)$ 
and $E_{2\pi}(\mathbf{p}_2, \mathbf{p}_3)$. Using these together with the
$T$-values we can now estimate for every ensemble, irrep and total
momentum which thermal contribution is -- up to unknown matrix
elements -- largest. Since we are able to remove only a single thermal
state, this is the only way to single out the relevant parameters for
the possible subtraction of these polluting states.
\Cref{fig:thermal-states-both} shows the contributing thermal states for
two example cases, left the $A_1^-$ irrep in the ${\mathbf P}^2=0$
frame, right the $B_1$ irrep in the ${\mathbf P}^2=2$ frame. The
different correlators shown correspond to different combinations of
single and two pion momenta. For these cases the
largest contribution is coming from $(\mathbf{p}_1^2 = 0,  \mathbf{p}_2^2 = 0,
\mathbf{p}_3^2 = 0)$ and $(\mathbf{p}_1^2 = 1,  \mathbf{p}_2^2 = 2, \mathbf{p}_3^2 =
1)$, respectively. The other possible contributions are suppressed by
two orders of magnitude or even exponentially.

To be precise, in order to find the dominating contribution for each
irrep, ensemble and frame, we take the largest thermal contribution at
$t = 10$, from which we can estimate $\Delta E$.
To illustrate this procedure further, we will look at irrep $\Gamma =
B_1$ with ${\mathbf P}^2 = 2$. The three-particle momenta that couple
to the operator below our 
threshold are listed in \Cref{tab:momenta-3pi}. As the three particles are
indistinguishable, we can partition them at will into a one-pion and two-pion
state. The two-particle momenta must again be a valid two-particle system,
otherwise they cannot be an intermediate thermal state.
\Cref{tab:momenta-2pi} lists the two-particle contributions in the $B_1$
irrep.

\begin{table}[h!]
  \centering
  \begin{tabular}{@{}rllll@{}}
      \hline
    \({\mathbf P}^2\) & Irrep & \({\mathbf p_1}\) & \({\mathbf p_2}\) &
    \({\mathbf p_3}\)\tabularnewline
    \hline
    2 & $B_1$ & \((0, 1, -1)\) & \((0, 0, 1)\) & \((1, 0, 0)\)\tabularnewline
    2 & $B_1$ & \((1, 1, 0)\) & \((1, 0, 0)\) & \((-1, 0, 0)\)\tabularnewline
    2 & $B_1$ & \((-1, 1, 0)\) & \((1, 0, 0)\) & \((1, 0, 0)\)\tabularnewline
    \hline
  \end{tabular}
  \caption{%
    Possible three-pion individual momenta in the $\Gamma = B_1$
    irrep with total momentum ${\mathbf P}^2 = 2$.
  }
  \label{tab:momenta-3pi}
\end{table}

\begin{table}[h!]
  \centering
  \begin{tabular}{@{}rlll@{}}
    \hline
    \({\mathbf P}^2\) & Irrep & \({\mathbf p_2}\) & \({\mathbf p_3}\)\tabularnewline
    \hline
    1 &  $B_1$ & \((-1, 0, 1)\) & \((1, 0, 0)\)\tabularnewline
    2 &  $B_1$ & \((1, 1, -1)\) & \((0, 0, 1)\)\tabularnewline
    4 &  $B_1$ & \((0, -1, 1)\) & \((0, 1, 1)\)\tabularnewline
    \hline
  \end{tabular}
  \caption{%
    Possible two-pion individual momenta in the $B_1$ irrep
    for different values of $\mathbf{P}^2$ of the two-pion subsystem.
  }
  \label{tab:momenta-2pi}
\end{table}

Thus, again for the example of the $B_1$ irrep, we have to go through
the following possibilities:
\begin{itemize}
    \item
        We take $(0, 1, -1)$ for the one pion and $(0, 0, 1)$ and $(1, 0, 0)$
        for the other two. The two-pion system has $\mathbf{P}^2 = 2$, but the
        lowest contribution \emph{in that irrep} has larger momenta. So this
        does not contribute.

    \item
        The single pion has ${\mathbf p_1} = (1, 1, 0)$ and the two-pion system
        gets ${\mathbf p_2} = (1, 0, 0)$ and ${\mathbf p_3} = (-1, 0, 0)$. The
        two-pion system therefore has total momentum ${\mathbf P}^2 = 0$, but
        there is no contribution to $B_1$ in that moving frame. Therefore this
        example does not contribute to the thermal states.

    \item
        A contribution is obtained using $(1, 0, 0)$ for the one pion
        momentum, and $(-1, 1, 0)$ and $(1, 0, 0)$ for the two-pion
        system. In the latter, we have 
        ${\mathbf P}^2 = 1$, which corresponds to the first entry in
        \Cref{tab:momenta-2pi} (albeit after an inconsequential global
        rotation). This contributes as a thermal state, incidentally it is the
        largest one as shown in \Cref{fig:thermal-states}.
\end{itemize}
Of course, there are many more possibilities to check for. Using this
method we determine the leading thermal state for every correlator 
matrix and can use this as input for thermal state treatments, detailed below.

\subsection{General technicalities}

Multi-particle correlators in general are contaminated with excited
states at early times, and 
with thermal pollution at late time slices. Fitting too early will overestimate
the energy, while fitting too late may underestimate it. In order to obtain a
robust energy estimate, we use combinations of different methods to attenuate
these issues.

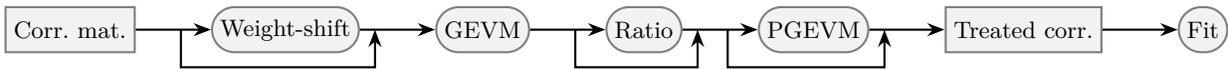
\begin{figure*}
    \centering
    \begin{tikzpicture}[
            >=Stealth,
            thick,
            sdata/.style={
                rectangle,
                minimum size=6mm,
                draw=black!50,
                fill=black!05,
            },
            smethod/.style={
                rectangle,
                minimum size=6mm,
                draw=black!50,
                rounded corners=3mm,
                fill=black!05,
            },
            skip loop/.style={to path={-- ++(0,-0.5) -| (\tikztotarget)}},
        ]
        \node (corr) [sdata] {Corr.\@ mat.};
        \node (wsw) [smethod, right=of corr] {Weight-shift};
        \node (gevm) [smethod, right=of wsw] {GEVM};
        \node (ratio) [smethod, right=of gevm] {Ratio};
        \node (pgevm) [smethod, right=of ratio] {PGEVM};
        \node (pc) [sdata, right=of pgevm] {Treated corr.};
        \node (fit) [smethod, right=of pc] {Fit};

        \draw[->] (corr) -- (wsw);
        \draw[->] (wsw) -- (gevm);
        \draw[->] (gevm) -- (ratio);
        \draw[->] (ratio) -- (pgevm);
        \draw[->] (pgevm) -- (pc);
        \draw[->] (pc) -- (fit);

        \path ($ (wsw.west) - (4mm,0) $) edge[->, skip loop] ($ (wsw.east) + (2mm,0) $);
        \path ($ (ratio.west) - (4mm,0) $) edge[->, skip loop] ($ (ratio.east) + (2mm,0) $);
        \path ($ (pgevm.west) - (4mm,0) $) edge[->, skip loop] ($ (pgevm.east) + (2mm,0) $);
    \end{tikzpicture}
    \caption{%
        Treatment of correlator matrices before fitting.
    }
    \label{fig:flowchart}
\end{figure*}

The order of application of these methods is illustrated with a flow chart in
\Cref{fig:flowchart}. The detour arrows indicate optional parts of the chain.
We will explain the different methods in order. First the correlator
matrices can optionally be treated with 
weight-shift-reweight~\cite{Dudek:2012gj} to suppress thermal states
at the cost of larger statistical uncertainty. Then we independently
use the original and treated correlator matrix and apply the GEVM,
which yields the principal correlators. These principal correlators
can be used to build ratios~\cite{Feng:2009ij,Helmes:2017smr} or left
as-is. All variants can optionally be fed into the Prony Generalized
Eigenvalue Method (PGEVM)~\cite{Fischer:2020bgv} with $t_0 = 2$ fixed
to suppress excited states (The PGEVM with $\delta_0$ 
fixed, see Ref.~\cite{Fischer:2020bgv} for details, turned out to not
be reliable).

The resulting treated correlators are evaluated by looking at the so called
effective mass. The simplest definition of it is the “log effective mass”
\begin{equation}
 m_\text{eff}(t) = - \log\,{\frac{C(t)}{C(t+1)}} \,,
\end{equation}  
which assumes a signal proportional to $\exp(-Et)$ only. There are
generalizations that take back-propagation, shifting or weighting into account.
Depending on the treatment of the correlator we choose the appropriate
effective mass.

We don't use all of the possible treatments in our analysis, but
only the following five: no treatment (i.e. all optional parts are left
out), only PGVM, only ratio, only weight and shift and finally the
combination of weight and shift with PGEVM. In more detail this means:
\begin{description}
\item[No treatment]
  When no thermal states contribute (like in $E$ irreps in the
  two pion channel), a simple
  cosh-like model is fitted:
  \begin{equation}
    C(t) = A_0 \left[\exp(-E_0 t) + \exp(-E_0 \cdot (T - t))\right] \,.
  \end{equation}
  If thermal states are present in the given irrep, a two-state model
  \begin{equation}
    \begin{split}
      C(t) = &A_0 \left[\exp(-E_0 t) + \exp(-E_0 \cdot (T - t))\right] 
      + A_1 \left[\exp(-E_1 \, t) + \exp(-E_1 \cdot (T -
        t))\right]\,,\\
    \end{split}
  \end{equation}
  with constrained second energy $E_1$  will be fitted to the data
  (for how $E_1$ and its error is determined, see \Cref{sec:thermal})
  The constraint is implemented by augmenting the $\chi^2$
  function to be minimized by a term
  \begin{equation}
    \chi^2_\mathrm{add} = \left(- \frac{E_1 - \bar E_1}{\delta E_1}\right)^2 \,,
  \end{equation}
  where $E_1$ is the fit parameter, $\bar E_1$ is the determined central
  value for the thermal energy and $\delta E_1$ the statistical
  uncertainty on $E_1$.
  
\item[PGEVM]
  This method works well when there are no significant thermal state contributions. We
  fit a simple exponential model at early times.
  
\item[Ratio]
  We take the ratio of the principal correlator obtained from
  the GEVP (no weight-and-shift applied)
  and form ratios with the one-pion correlation function:
  \begin{align}
    \label{eq:ratio-2}
    R_2(t) &= \frac{C_{2\pi}(t) - C_{2\pi}(t + 1)}{C_\pi(t)^2 - C_\pi(t + 1)^2}\,,
    \\
    \label{eq:ratio-3}
    R_3(t) &= \frac{C_{3\pi}(t) / C_\pi(t) - C_{3\pi}(t + 1) / C_\pi(t
  + 1)}{C_\pi(t)^2 - C_\pi(t + 1)^2} \,.
  \end{align}
  The ratio $R_3$ is chosen as a double ratio such that in the numerator,
  thermal state contributions $\propto \exp(-\Delta E\, t)$ are
  removed, since $\Delta E\approx E_\pi$. The resulting
  $\sinh$-like correlator needs to be divided by another $\sinh$-like
  expression, that's why we take the difference also in the
  denominator. Among different ratio 
  expressions we have tested, this one works best in the sense that
  the plateau is longest.
  An exponential model is fitted to the ratios where the signal behaves
  like $R_2(t) \sim \exp(-(E_{2\pi} - 2 E_\pi) t)$ and $R_3(t) \sim
  \exp(-(E_{3\pi} - 3 E_\pi) t)$.
  Note that for the ratios we do not include backwards propagating
  parts and thus do not extend fit ranges too far towards $T/2$.
  
\item[Weight-shift]
  The correlator matrix has the leading thermal state
  removed~\cite{Dudek:2012gj} and, 
  therefore, the principal correlators can be fitted with a cosh-like
  model which incorporates the weight-shift-reweight procedure.
  
\item[Weight-shift and PGEVM]
  In general the additional suppression of excited states by the application
  of the PGEVM works well after weight-shift has been applied beforehand.
  The resulting correlator is fitted with an exponential model. Fit ranges
  can be chosen early enough such that the neglect of backwards
  propagating parts is not significant.
\end{description}

\begin{figure}
  \centering
  \includegraphics[width=1\linewidth]{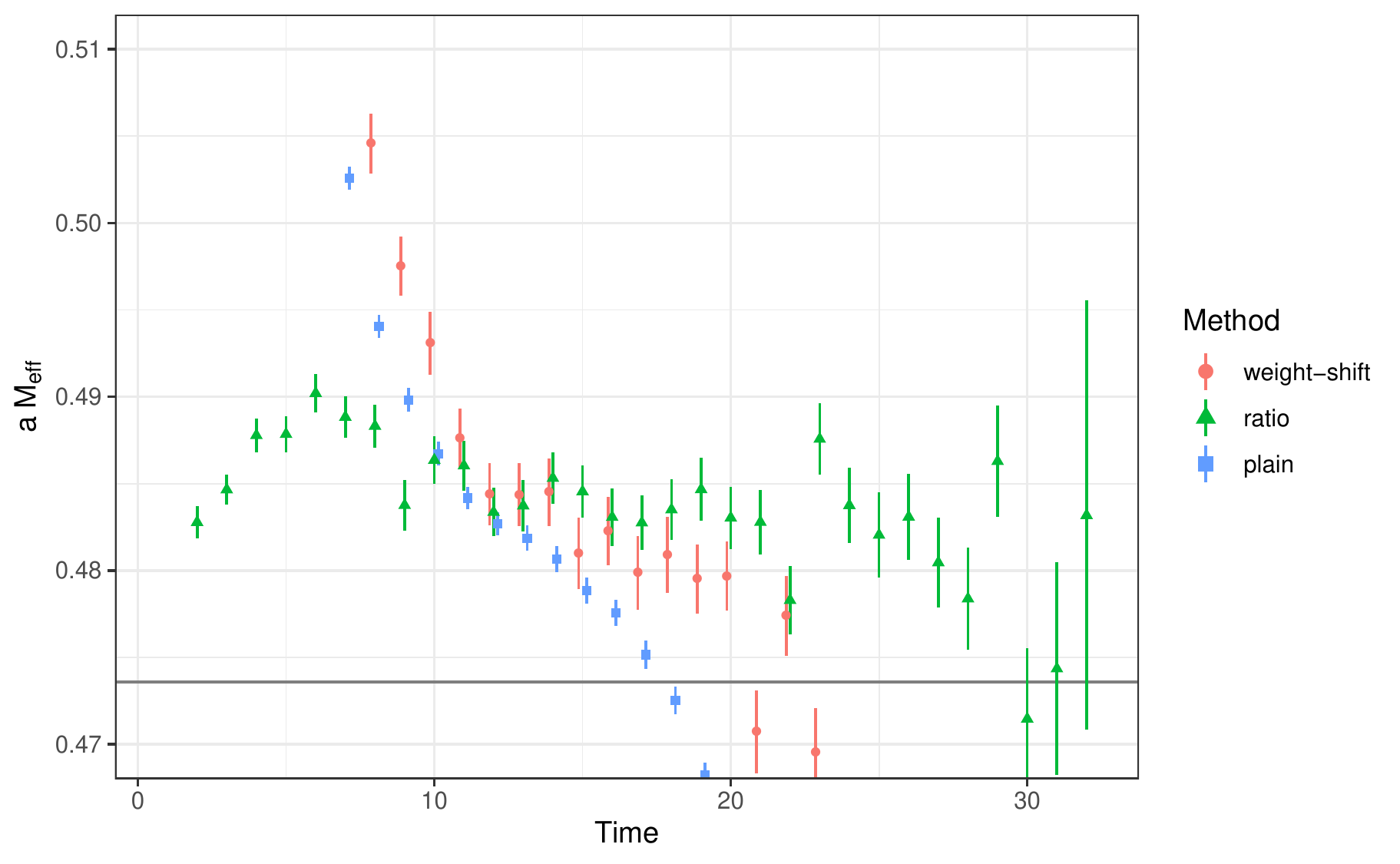}
  \caption{%
    Effective mass for the three-pion ground state ($A_1^-$, $\vec P^2=0$)  on the cA2.60.32
    ensemble. Shown are in blue the plain correlator without any thermal
    state treatment, in red the correlator treated with
    weight-shift-reweight and in green the ratio $R_3$ shifted upwards by
    $3 M_\pi$. The solid line marks the noninteracting
    energy. }
  \label{fig:effmass-ratio-3pi}
\end{figure}

\Cref{fig:effmass-ratio-3pi} shows a comparison between no
treatment, weight-shift and the ratio $R_3$ for a case with
heavy thermal pollution. One can see how the effective mass of
the plain correlator does not
show any plateau due to the high degree of thermal pollution.
The effective mass of the weighted correlator, however, exhibits
a plateau between $t_1 = 12$ and $t_2 = 14$, but still shows a
drop beyond. However, this three time slice plateau can only be identified
when compared to the effective mass given by the ratio $R_3$.
This likely stems from the second leading thermal state as visible in
\Cref{fig:thermal-states-A1u}. The ratio however has a long
plateau that is compatible with the weight-shift method a posteriori. In
general we see that with the ratio method it is possible to fit energy
levels with strong thermal pollution when other methods fail to produce
a plateau. The statistical uncertainty from the energy determination
with the ratio is also lower than with other methods in most cases.

In some cases the thermal states are so pronounced that no plateau can be
identified, even after applying the PGEVM. In these cases the method is
not used for that particular level. These cases work much better with either
the multi-state model, weight-shift-reweight or the combination of
weight-shift-reweight and the PGEVM. The ratio method seems to be the
most robust one, it shows plateaus even when other methods fail to produce one.
Also, the statistical uncertainty seems to be lower compared to the other
methods in general.

For every principal correlator we attempt to extract the energy with
all the five methods detailed above. If a plateau can be identified,
we use the extracted energy level. All 
such determinations per principal correlator are combined with a
correlated weighted average. In order to incorporate the systematic
spread between the central values, we also compute a systematic error
scaling factor as introduced in Ref.~\cite{Werner:2019hxc}: for energy
level $E$ we compute the scaling factor $w$~\Cref{eq:s}, as mentioned
in the main text.

\begin{figure}
    \centering
    \includegraphics[width=0.8\linewidth]{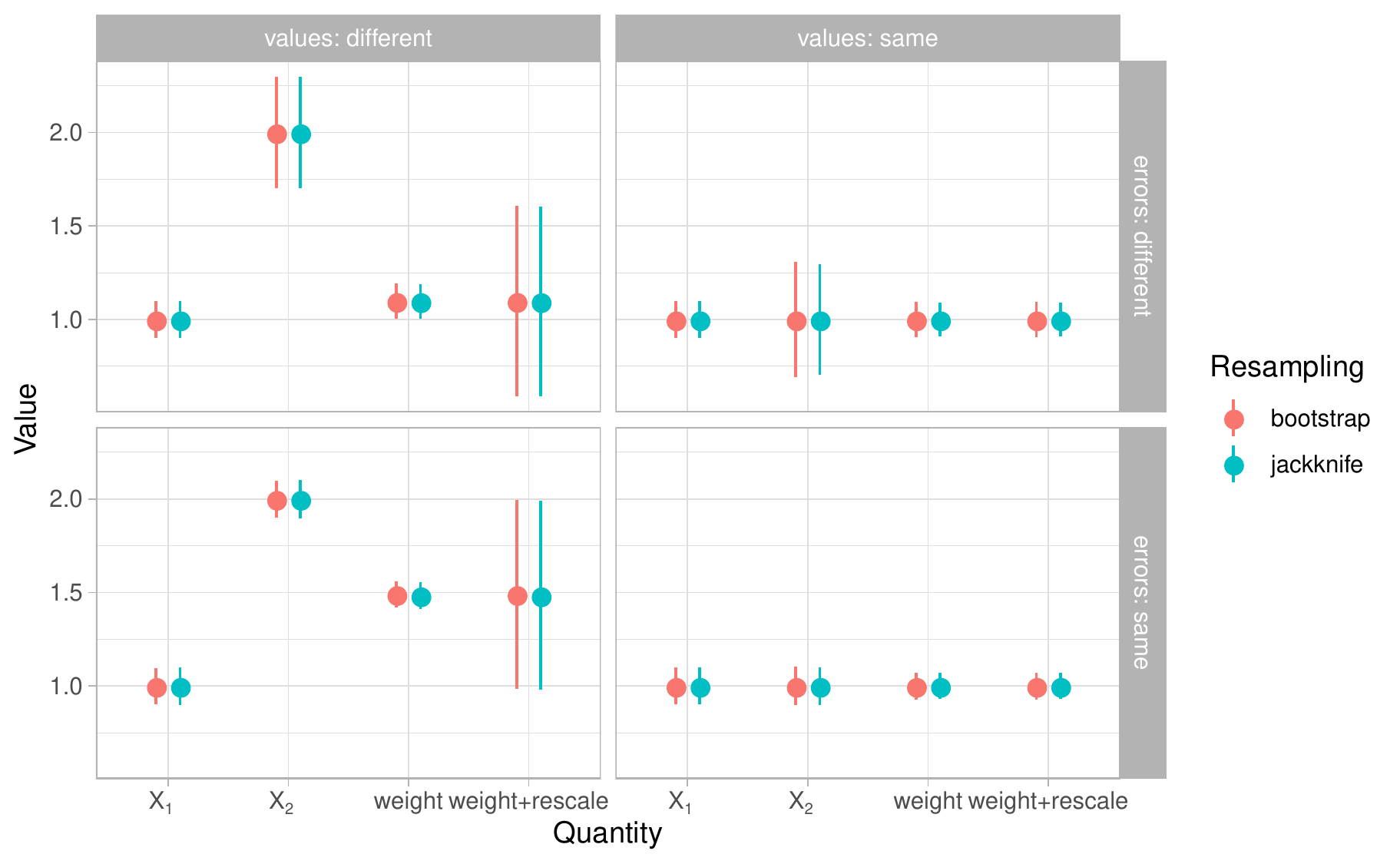}
    \caption{%
        Combination of artificial fit results from two different methods to a weighted
        average and finally the rescaled distribution. Columns show different
        and same central values, rows show different and same statistical
        errors in the two measurements.
    }
    \label{fig:dist-combine}
\end{figure}

To illustrate this method to incorporate the systematic error into the
resampling distribution, we use two artificially generated data
points with central values $X_1$ and $X_2$ and corresponding standard
errors generated in four ways, where either
the central values and/or errors are chosen to be the same or different.
All combinations thus give four cases, which are shown in the quadrants
of \Cref{fig:dist-combine} (upper left: different mean, different
errors; upper right: same mean, different errors; lower left: different
mean, same errors; lower right: all the same). The central values with
standard errors for $X_1$ and $X_2$ are shown as the first two pairs
of points in each quadrant. The third pair shows the weighted average
of the two estimates and the fourth pair the result after the
rescaling. One can nicely see how the weighted average gravitates
toward the data point with the smaller uncertainty
(hence higher weight) and how the rescaling incorporates the spread
between the central values. The method works well for both bootstrap
and jackknife resampling. 

In order to choose appropriate fit ranges for the different methods, we proceed
iteratively, selecting fit ranges by eye guided by the $p$-value of
the fit.
Energy levels are included in the further analysis only if a plateau of at least five time slices length could be identified for the $T=96$ lattices and of at least four time slices for the $T=64$ lattice.
Some energy levels show significant tension
between the different fitting methods after this first iteration.
In these cases, we re-evaluate the plateaus to arrive at our final
choices.

\section{Fitting the spectrum}
\label{sec:fittingspec}

Here, we aim to extend the discussion of the fitting procedure of the spectrum to the quantization condition. The summary of the frames, irreps and energies used in this work is shown in \Cref{tab:summarylevels}.

\input{irrep_ranges.tex}

\subsection{General technicalities}

In both, the two and three-particle sector, we define the $\chi^2$ as:
\begin{equation}
\chi^2 =\sum_{ij} (E^{\text{data}}_i  - E_i^{\text{predicted}}) \left( C\right)^{-1}_{ij}  (E^{\text{data}}_i  - E_j^{\text{predicted}}),
\end{equation} 
where $C$ is the covariance matrix of the energy levels, estimated from the bootstrap samples. Best  fit parameters are obtained using the Levenberg-Marquardt algorithm.

The range of validity of the quantization conditions is limited by the first inelastic threshold. This is  $E^*=4M_\pi$ ($5M_\pi$) for the two-particle (three-particle) quantization condition. We generally include levels up to that threshold, however, for the physical point ensemble (cA2.09.48), we have included levels higher up in energy. Since the $2\pi \to 4\pi$, and $3\pi \to 5\pi$ couplings are very small, we expect this to be a valid approximation. In fact, phenomenological studies set the first relevant inelasticity to be the $\rho\pi\pi$ channel ($E^* \sim 8 M_\pi$ for physical kinematics) \cite{Yndurain:2002ud, Pelaez:2004vs, Kaminski:2006qe}.

\begin{figure*}[h!]
    \centering
        \subfloat[{$s$-wave}\label{fig:swave30}]{\includegraphics[width=0.48\textwidth]{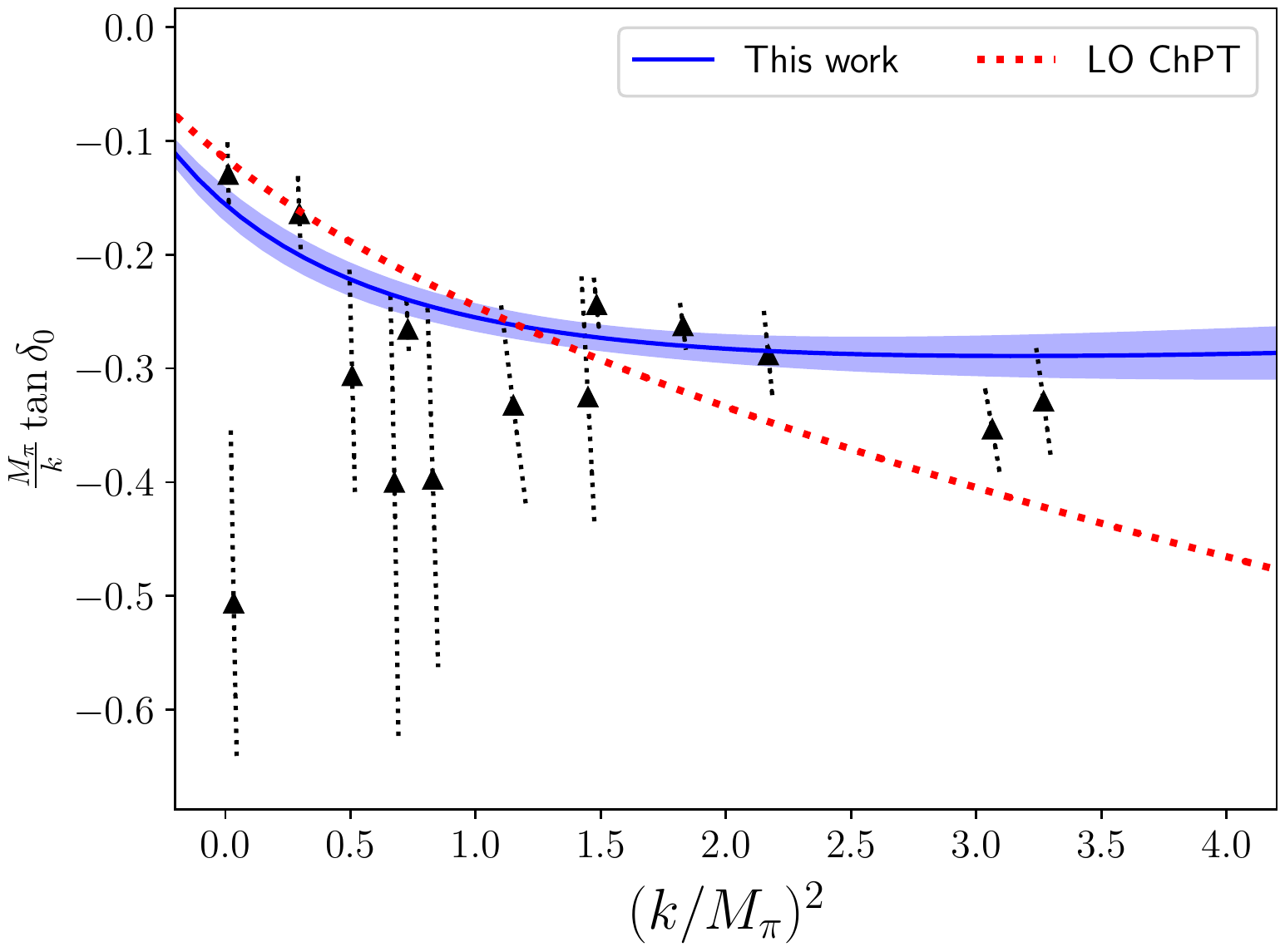}}
        \subfloat[{$d$-wave}\label{fig:dwave30}]{\includegraphics[width=0.48\textwidth]{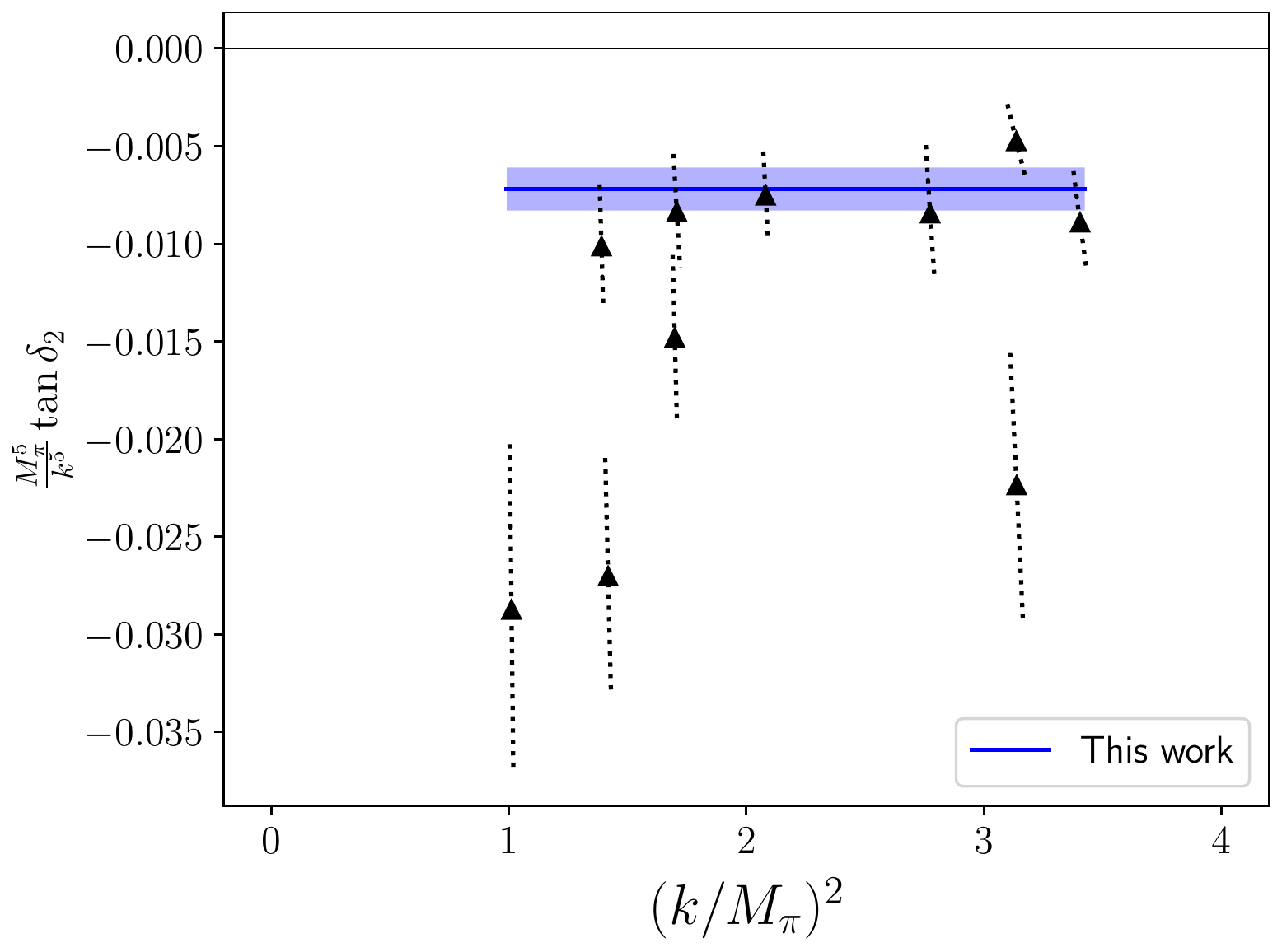}}
 \caption{  $s$- and $d$-wave phase shift for the ensemble cA2.30.48. For $s$-wave we use a model that incorporates the Adler-zero, whereas for $d$-wave we fit to a constant in the region for which we have data.  Two points have been omitted in the plot due to the very large errorbars.}\label{fig:2pionsA30}
\end{figure*}

\begin{figure*}[h!]
    \centering
        \subfloat[{$s$-wave}\label{fig:swave60}]{\includegraphics[width=0.48\textwidth]{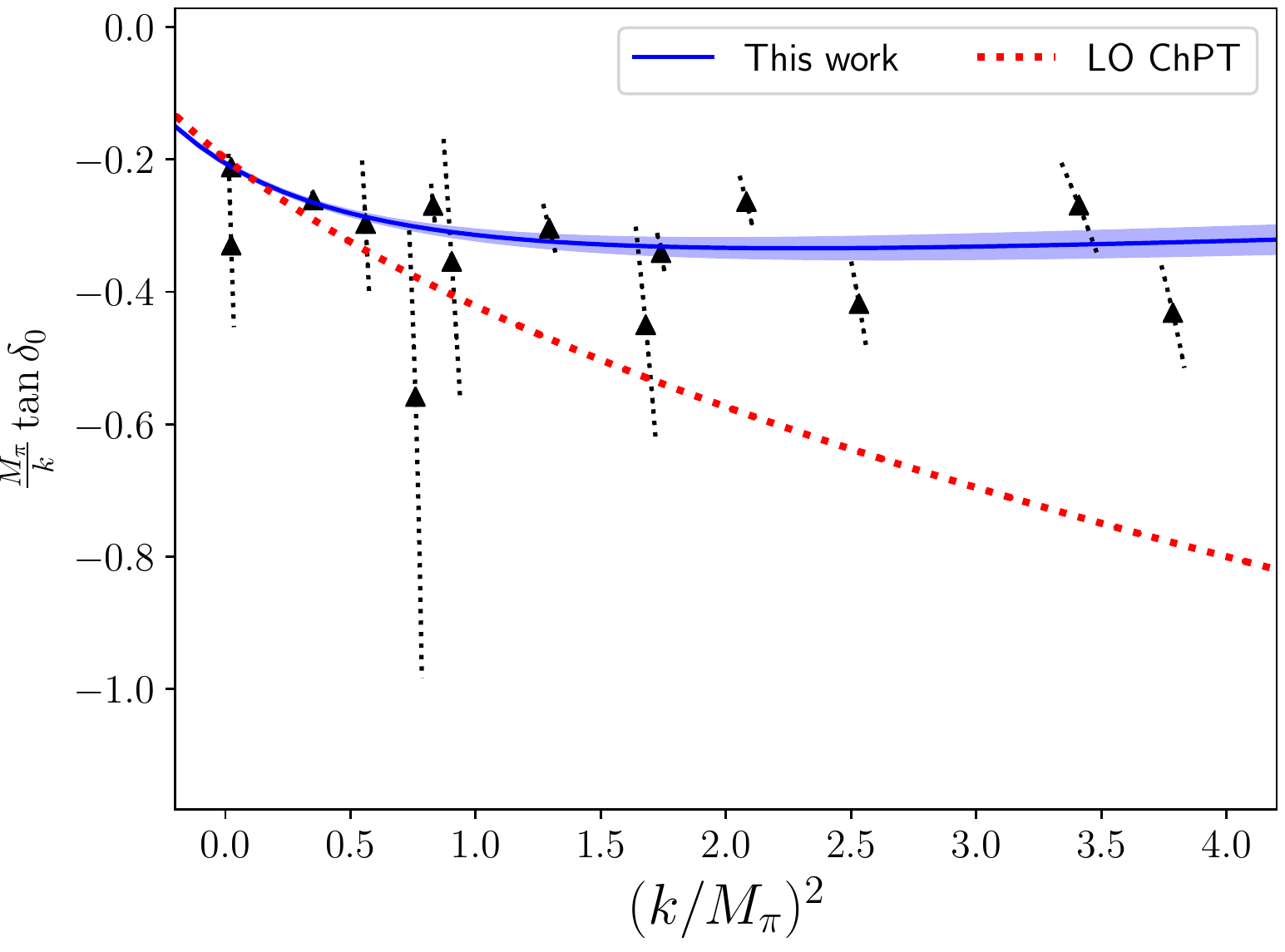}}
        \qquad
        \subfloat[{$d$-wave}\label{fig:dwave60}]{\includegraphics[width=0.48\textwidth]{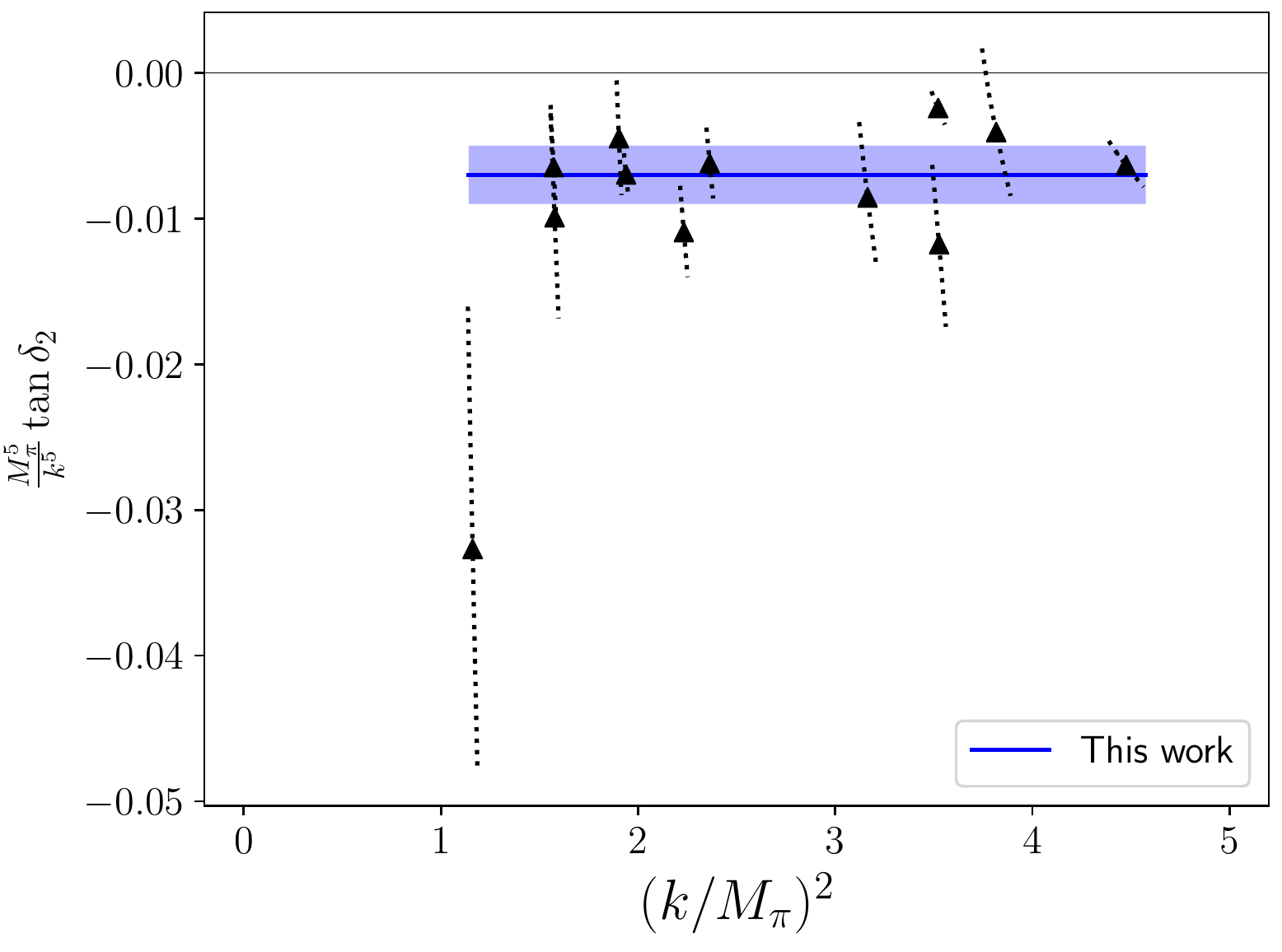}}
 \caption{  $s$- and $d$-wave phase shift for the ensemble cA2.60.32. For $s$-wave we use a model that incorporates the Adler-zero, whereas for $d$-wave we fit to a constant in the region for which we have data.  Two points have been omitted in the plot due to the very large errorbars.}\label{fig:2pionsA60}
\end{figure*}

As mentioned in the main text, we show here additional two-pion phase shift plots:
\Cref{fig:2pionsA30} for cA2.30.48, and \Cref{fig:2pionsA60} for cA2.60.32.
In the case of the $s$-wave phase shift, we also compare to LO ChPT.
As can be seen, the ChPT prediction describes less accurately the data at heavier pion masses--- compare to \Cref{fig:2pions}.

\clearpage
\subsection{Additional discussion on three-pion fits}

First, we perform a global fit to two- and three-particle levels that includes only a constant term in $\Kdf$. This is shown in \Cref{tab:3piadlerK0}. As can be seen, the quality of the fit is significantly worse for the heavier ensembles than in the linear fits of \Cref{tab:3piadler} in the main text. For the ensemble at the physical point (cA2.09.48), the value of $\chi^2$ is basically the same, but in both cases $\Kdf$ is compatible with zero. We thus conclude that the linear model of $\Kdf$ in Eq. 6 in the main text is more appropriate for this system.
\begin{table*}[h!]
  \centering
  \begin{tabular*}{.75\textwidth}{@{\extracolsep{\fill}}crrrrr}
    \hline
    & $1/B_0$ & $B_1$ & $M_\pi^2\Kiso$ & $M_\pi^2\Kisolin$ & $\chi^2/$dof \\  \hline 
    cA2.60.32    &   -0.2050(49) & -1.7(2) & 900(1000)&-- & 71.08/(43-3)       \\ 
    cA2.30.48     &   -0.149(14) &  -1.7(4) &  -2000(1400)& --- &  47.59/(33-3)    \\ 
    cA2.09.48     &   -0.0482(86) & -1.3(1.1) & -200(600)& -- &    19.24/(19-3)    \\
    \hline
  \end{tabular*}
\caption{Two- and three-pion fits using the Adler-zero form ($z^2=M_\pi^2$, fixed). Here we assume that $\Kdf$ is given by a constant:  $\Kdf = \Kiso$. }
\label{tab:3piadlerK0}
\end{table*}

Next, the full covariance matrices of the fits in \Cref{tab:3piadler} in the main text are provided. We use the form $C = D R D $,  with $D$ being a diagonal matrix with the standard errors of the parameters. We ordered the entries as:  $\left( 1/B_0, B_1, M_\pi^2 \Kiso, M_\pi^2 \Kisolin \right)$. 
\begin{align}
  \label{eq:cor1}
  \begin{split}
    \text{cA2.09.48: } & \ D = \text{diag }(0.0086,1.1,800,500),  \\ &
    \ R = \left(
    \begin{array}{cccc}
      1. & 0.73 & -0.37 & -0.02 \\
      0.73 & 1. & -0.25 & 0.11\\
      -0.37& -0.25 & 1. & -0.71 \\
      -0.02 & 0.11 & -0.71 & 1. \\
    \end{array}
    \right),
  \end{split} \\
  \label{eq:cor2}
  \begin{split}
    \text{cA2.30.48: } & \ D = \text{diag }(0.015, 0.4, 3800,3800) , \\ &
    \ R = \left(
    \begin{array}{cccc}
      1.0 & 0.80 & -0.55 & 0.41 \\
      0.80 & 1.0 & -0.40 & 0.35\\
      -0.55 & -0.40 & 1.0 & -0.93 \\
      0.41 & 0.35 & -0.93 & 1.0 \\
    \end{array}
    \right) ,
  \end{split} \\
  \label{eq:cor3}
  \begin{split}
    \text{cA2.60.32: } & \ D = \text{diag }( 0.0049, 0.2,1500,1800 ) ,  \\ &
     \ R = \left(
    \begin{array}{cccc}
      1.0 & 0.36 & -0.02 & 0.05 \\
      0.36 & 1.0 & 0.10 & 0.22 \\
      -0.02 & 0.10 & 1.0 & -0.78 \\
      0.05 & 0.22 & -0.78 & 1.0 \\
    \end{array}
    \right),
  \end{split}
\end{align}
We observe a large correlation within the two and three-particle sectors separately ---  the pairs $1/B_0, B_1$, and $M_\pi^2 \Kiso, M_\pi^2 \Kisolin$ are highly correlated. In contrast, the correlation between the two- and three-particle sectors is milder. 

\subsection{Two- and three-pion spectrum}

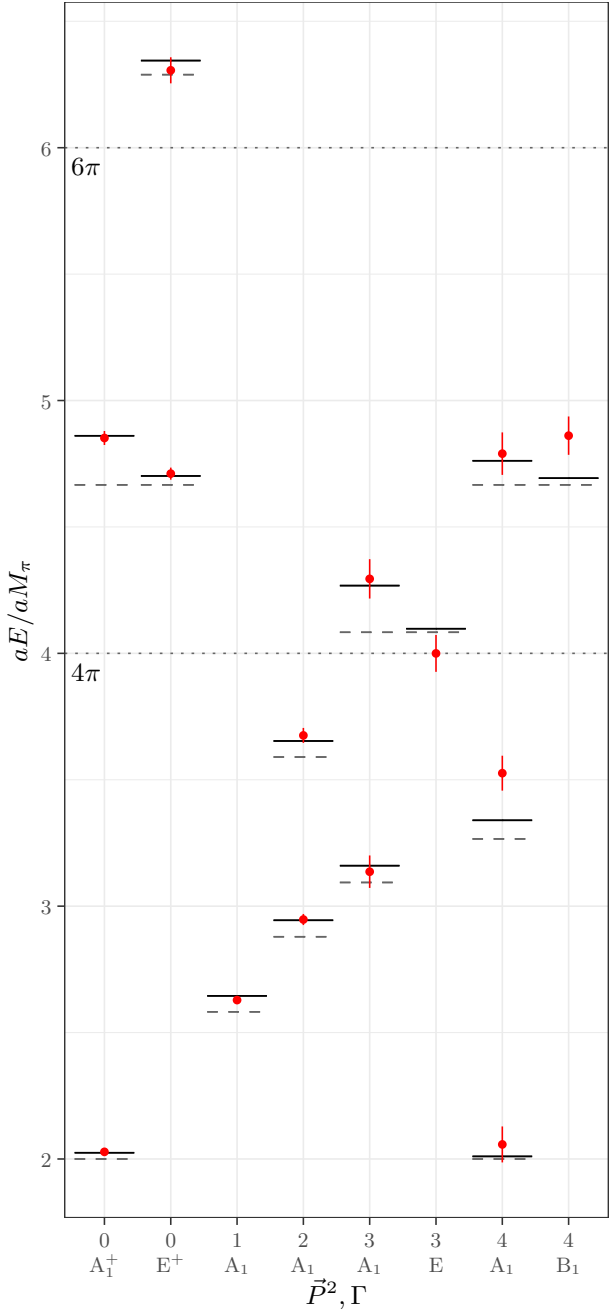
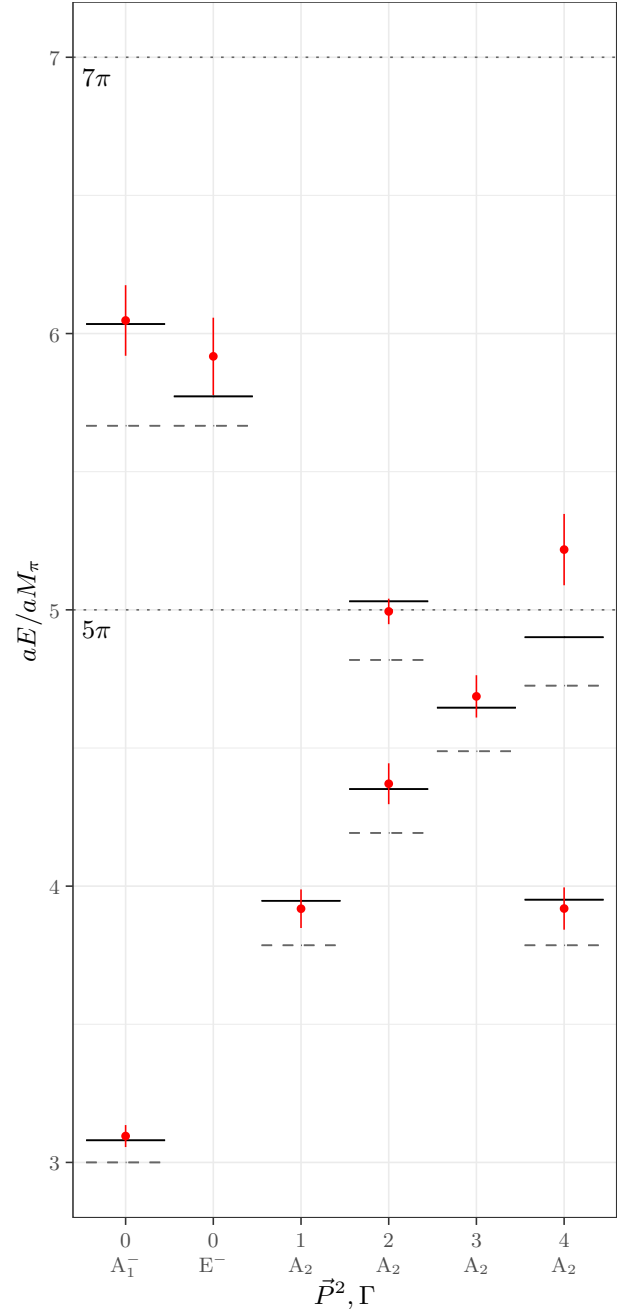
\begin{figure*}[h!]
    \centering
    \subfloat[Two pion channel \label{fig:spectrum-prediction-2pi}]{
        \input{plots/cm_mass_1.tex}
    }
    \hfill
    \subfloat[Three pion channel \label{fig:spectrum-prediction-3pi}]{
        \input{plots/cm_mass_2.tex}
    }
    \caption{%
        The center-of-mass spectrum for two and three pions on the physical
        point ensemble(cA2.09.48). The red data points are the energy
        levels determined from the correlator. The black lines denote the
        prediction from the quantization condition. For the two-pion $A_1$
        levels, and all three-pion levels, we use the fit in
       \Cref{tab:3piadler} in the main text. For the non-$A_1$ two-pion levels, which are
        dominated by $d$-wave interactions, we use the fit in \Cref{tab:dwave} in the main text.
        The short dashed gray lines denote the noninteracting energy levels.
        We also include the relevant inelastic thresholds as long dotted gray
        lines.
    }
    \label{fig:spectrum-prediction}
\end{figure*}

We conclude the discussion by comparing the spectrum from the lattice to the one predicted by the quantization conditions using the best fits. This is shown in \Cref{fig:spectrum-prediction} for the ensemble at the physical point.

%% file: quark-diagrams.tex
\subfloat[C2c  \label{fig:C2c} ]{
 \centering
\begin{minipage}{0.3\linewidth}
 \begin{tikzpicture}
    \coordinate[draw, fill=black!30, circle] (so1) at (0, -0);
    \coordinate[draw, fill=black!30, circle] (si1) at (1, -0);
    \draw[->, >=Latex, out=30, in=150] (so1) to (si1);
    \draw[->, >=Latex, out=210, in=330] (si1) to (so1);
  \end{tikzpicture}
\end{minipage} 
}
\hfill
\subfloat[C4cC  \label{fig:C4cC} ]{
 \centering
\begin{minipage}{0.3\linewidth}
  \begin{tikzpicture}
    \coordinate[draw, fill=black!30, circle] (so1) at (0, 0);
    \coordinate[draw, fill=black!30, circle] (si1) at (1, 0);
    \coordinate[draw, fill=black!30, circle] (so2) at (0, 0.7);
    \coordinate[draw, fill=black!30, circle] (si2) at (1, 0.7);
    \draw[->, >=Latex] (so1) to (si1);
    \draw[->, >=Latex] (si1) to (so2);
    \draw[->, >=Latex] (so2) to (si2);
    \draw[->, >=Latex] (si2) to (so1);
  \end{tikzpicture}
\end{minipage} 
}
\hfill
\subfloat[C4cD  \label{fig:C4cD} ]{
 \centering
\begin{minipage}{0.3\linewidth}
   \begin{tikzpicture}
    \coordinate[draw, fill=black!30, circle] (so1) at (0, -0);
    \coordinate[draw, fill=black!30, circle] (si1) at (1, -0);
    \coordinate[draw, fill=black!30, circle] (so2) at (0, -0.7);
    \coordinate[draw, fill=black!30, circle] (si2) at (1, -0.7);
    \draw[->, >=Latex, out=30, in=150] (so1) to (si1);
    \draw[->, >=Latex, out=210, in=330] (si1) to (so1);
    \draw[->, >=Latex, out=30, in=150] (so2) to (si2);
    \draw[->, >=Latex, out=210, in=330] (si2) to (so2);
  \end{tikzpicture}
\end{minipage} 
}

\vspace{3ex} 

\subfloat[C6cC  \label{fig:C6cC} ]{
 \centering
\begin{minipage}{0.3\linewidth}
   \begin{tikzpicture}
    \coordinate[draw, fill=black!30, circle] (so1) at (0, -0);
    \coordinate[draw, fill=black!30, circle] (si1) at (1, -0);
    \coordinate[draw, fill=black!30, circle] (so2) at (0, -0.7);
    \coordinate[draw, fill=black!30, circle] (si2) at (1, -0.7);
    \coordinate[draw, fill=black!30, circle] (so3) at (0, -1.4);
    \coordinate[draw, fill=black!30, circle] (si3) at (1, -1.4);
    \draw[->, >=Latex] (so1) to (si1);
    \draw[->, >=Latex] (si1) to (so2);
    \draw[->, >=Latex] (so2) to (si2);
    \draw[->, >=Latex] (si2) to (so3);
    \draw[->, >=Latex] (so3) to (si3);
    \draw[->, >=Latex] (si3) to (so1);
  \end{tikzpicture}
\end{minipage} 
}
\hfill
\subfloat[C6cCD  \label{fig:C6cCD} ]{
 \centering
\begin{minipage}{0.3\linewidth}
  \begin{tikzpicture}
    \coordinate[draw, fill=black!30, circle] (so1) at (0, -0);
    \coordinate[draw, fill=black!30, circle] (si1) at (1, -0);
    \coordinate[draw, fill=black!30, circle] (so2) at (0, -0.7);
    \coordinate[draw, fill=black!30, circle] (si2) at (1, -0.7);
    \coordinate[draw, fill=black!30, circle] (so3) at (0, -1.4);
    \coordinate[draw, fill=black!30, circle] (si3) at (1, -1.4);
    \draw[->, >=Latex] (so1) to (si1);
    \draw[->, >=Latex] (si1) to (so2);
    \draw[->, >=Latex] (so2) to (si2);
    \draw[->, >=Latex] (si2) to (so1);
    \draw[->, >=Latex, out=30, in=150] (so3) to (si3);
    \draw[->, >=Latex, out=210, in=330] (si3) to (so3);
  \end{tikzpicture}
\end{minipage} 
}
\hfill
\subfloat[C6cD  \label{fig:C6cD} ]{
 \centering
\begin{minipage}{0.3\linewidth}
  \begin{tikzpicture}
    \coordinate[draw, fill=black!30, circle] (so1) at (0, -0);
    \coordinate[draw, fill=black!30, circle] (si1) at (1, -0);
    \coordinate[draw, fill=black!30, circle] (so2) at (0, -0.7);
    \coordinate[draw, fill=black!30, circle] (si2) at (1, -0.7);
    \coordinate[draw, fill=black!30, circle] (so3) at (0, -1.4);
    \coordinate[draw, fill=black!30, circle] (si3) at (1, -1.4);
    \draw[->, >=Latex, out=30, in=150] (so1) to (si1);
    \draw[->, >=Latex, out=210, in=330] (si1) to (so1);
    \draw[->, >=Latex, out=30, in=150] (so2) to (si2);
    \draw[->, >=Latex, out=210, in=330] (si2) to (so2);
    \draw[->, >=Latex, out=30, in=150] (so3) to (si3);
    \draw[->, >=Latex, out=210, in=330] (si3) to (so3);
  \end{tikzpicture}
\end{minipage} 
}
\caption{Quark contraction diagrams for the one-, two- and three-$\pi^+$
correlation functions needed in this work.}

%% file: irrep_ranges.tex
\begin{table}[h!]
    \subfloat[cA2.09.48, two pions]{
        \begin{tabular}{rll}
            \hline
            $\mathbf{P}^2$ & Irrep & $E / M_\pi$ range \\
            \hline
            0 & $A_1^+$ & [2.03, 4.85]\\
            0 & $E^+$ & [4.71, 6.31]\\
            1 & $A_1$ & [2.63, 6.64]\\
            2 & $A_1$ & [2.95, 5.79]\\
            3 & $A_1$ & [3.14, 4.29]\\
            3 & $E$ & [4.00, 4.00]\\
            4 & $A_1$ & [2.05, 4.79]\\
            4 & $B_1$ & [4.86, 4.86]\\
            \hline
        \end{tabular}
    }
    \subfloat[cA2.09.48, three pions]{
        \begin{tabular}{rll}
            \hline
            $\mathbf{P}^2$ & Irrep & $E / M_\pi$ range \\
            \hline
            0 & $A_1^-$ & [3.09, 6.05]\\
            0 & $E^-$ & [5.91, 5.92]\\
            1 & $A_2$ & [3.92, 3.92]\\
            2 & $A_2$ & [4.37, 4.99]\\
            3 & $A_2$ & [4.69, 6.40]\\
            3 & $E$ & [5.70, 8.06]\\
            4 & $A_2$ & [3.92, 6.42]\\
            4 & $B_2$ & [6.42, 6.42]\\
            \hline
        \end{tabular}
    }

    \subfloat[cA2.30.48, two pions]{
        \begin{tabular}{rll}
            \hline
            $\mathbf{P}^2$ & Irrep & $E / M_\pi$ range \\
            \hline
            0 & $A_1^+$ & [2.01, 3.99]\\
            0 & $E^+$ & [3.09, 3.88]\\
            1 & $A_1$ & [2.29, 4.15]\\
            1 & $B_1$ & [3.28, 3.28]\\
            1 & $B_2$ & [4.09, 4.09]\\
            1 & $E$ & [3.31, 4.02]\\
            2 & $A_1$ & [2.46, 4.26]\\
            2 & $A_2$ & [3.50, 3.50]\\
            2 & $B_1$ & [3.45, 3.45]\\
            2 & $B_2$ & [4.22, 4.22]\\
            3 & $A_1$ & [2.59, 4.46]\\
            3 & $E$ & [2.84, 4.44]\\
            4 & $A_1$ & [2.03, 3.13]\\
            4 & $B_1$ & [3.12, 3.12]\\
            \hline
        \end{tabular}
    }
    \subfloat[cA2.30.48, three pions]{
        \begin{tabular}{rll}
            \hline
            $\mathbf{P}^2$ & Irrep & $E / M_\pi$ range \\
            \hline
            0 & $A_1^-$ & [3.05, 4.26]\\
            0 & $E^-$ & [4.16, 4.16]\\
            1 & $A_2$ & [3.39, 4.66]\\
            1 & $B_2$ & [4.43, 4.59]\\
            1 & $E $& [4.43, 4.43]\\
            2 & $A_1 $& [4.69, 4.85]\\
            2 & $A_2 $& [3.66, 5.01]\\
            2 & $B_1 $& [4.81, 4.90]\\
            2 & $B_2 $& [4.59, 4.90]\\
            3 & $A_1$ & [5.14, 5.14]\\
            3 & $A_2$ & [3.83, 5.32]\\
            3 & $E $& [4.11, 5.18]\\
            4 & $A_2$ & [3.42, 4.67]\\
            4 & $B_2$ & [4.41, 4.56]\\
            4 & $E$ & [4.57, 4.57]\\
            \hline
        \end{tabular}
    }
    \subfloat[cA2.60.32, two pions]{
        \begin{tabular}{rll}
            \hline
            $\mathbf{P}^2$ & Irrep & $E / M_\pi$ range \\
            \hline
            0 & $A_1^+$ & [2.02, 4.20]\\
            0 & $E^+$ & [3.21, 4.08]\\
            1 & $A_1$ & [2.32, 4.38]\\
            1 & $B_1$ & [3.43, 3.43]\\
            1 & $B_2$ & [4.25, 4.26]\\
            1 & $E$ & [3.41, 4.26]\\
            2 & $A_1$ & [2.50, 4.44]\\
            2 & $A_2$ & [3.67, 3.67]\\
            2 & $B_1$ & [3.60, 3.60]\\
            2 & $B_2$ & [4.39, 4.39]\\
            3 & $A_1$ & [2.65, 4.66]\\
            3 & $E$ & [2.94, 4.68]\\
            4 & $A_1$ & [2.02, 3.28]\\
            4 & $B_1$ & [3.21, 3.21]\\
            \hline
        \end{tabular}
    }
    \subfloat[cA2.60.32, three pions]{
        \begin{tabular}{rll}
            \hline
            $\mathbf{P}^2$ & Irrep & $E / M_\pi$ range \\
            \hline
            0 & $A_1^-$ & [3.06, 4.40]\\
            0 & $E^-$ & [4.31, 4.31]\\
            1 & $A_2$ & [3.50, 4.86]\\
            1 & $B_2$ & [4.56, 4.74]\\
            1 & $E$ & [4.56, 4.56]\\
            2 & $A_1$ & [4.84, 5.01]\\
            2 & $A_2$ & [3.72, 5.23]\\
            2 & $B_1$ & [4.98, 5.13]\\
            2 & $B_2$ & [4.79, 5.06]\\
            3 & $A_1$ & [5.32, 5.32]\\
            3 & $A_2$ & [3.94, 5.58]\\
            3 & $E$ & [4.21, 5.42]\\
            4 & $A_2$ & [3.51, 4.86]\\
            4 & $B_2$ & [4.51, 4.71]\\
            4 & $E$ & [4.70, 4.70]\\
            \hline
        \end{tabular}
    }
 \caption{Summary of energy levels included in this work. The
   $E/M_\pi$ range indicates in which energy range the energy levels
   from the various principal correlators in that specific irrep where located.\label{tab:summarylevels}}
\end{table}

\begin{tabular}{llrll}
\hline
\hline
\hline
\hline
\hline
\end{tabular}

%% file: plots/cm_mass_1.tex
\begin{tikzpicture}[x=1pt,y=1pt]
\definecolor{fillColor}{RGB}{255,255,255}
\path[use as bounding box,fill=fillColor,fill opacity=0.00] (0,0) rectangle (238.49,505.89);
\begin{scope}
\path[clip] (  0.00,  0.00) rectangle (238.49,505.89);
\definecolor{drawColor}{RGB}{255,255,255}
\definecolor{fillColor}{RGB}{255,255,255}

\path[draw=drawColor,line width= 0.6pt,line join=round,line cap=round,fill=fillColor] (  0.00,  0.00) rectangle (238.49,505.89);
\end{scope}
\begin{scope}
\path[clip] ( 27.31, 40.19) rectangle (232.99,500.39);
\definecolor{fillColor}{RGB}{255,255,255}

\path[fill=fillColor] ( 27.31, 40.19) rectangle (232.99,500.39);
\definecolor{drawColor}{gray}{0.92}

\path[draw=drawColor,line width= 0.3pt,line join=round] ( 27.31,110.27) --
	(232.99,110.27);

\path[draw=drawColor,line width= 0.3pt,line join=round] ( 27.31,205.97) --
	(232.99,205.97);

\path[draw=drawColor,line width= 0.3pt,line join=round] ( 27.31,301.68) --
	(232.99,301.68);

\path[draw=drawColor,line width= 0.3pt,line join=round] ( 27.31,397.38) --
	(232.99,397.38);

\path[draw=drawColor,line width= 0.3pt,line join=round] ( 27.31,493.09) --
	(232.99,493.09);

\path[draw=drawColor,line width= 0.6pt,line join=round] ( 27.31, 62.41) --
	(232.99, 62.41);

\path[draw=drawColor,line width= 0.6pt,line join=round] ( 27.31,158.12) --
	(232.99,158.12);

\path[draw=drawColor,line width= 0.6pt,line join=round] ( 27.31,253.82) --
	(232.99,253.82);

\path[draw=drawColor,line width= 0.6pt,line join=round] ( 27.31,349.53) --
	(232.99,349.53);

\path[draw=drawColor,line width= 0.6pt,line join=round] ( 27.31,445.23) --
	(232.99,445.23);

\path[draw=drawColor,line width= 0.6pt,line join=round] ( 42.36, 40.19) --
	( 42.36,500.39);

\path[draw=drawColor,line width= 0.6pt,line join=round] ( 67.45, 40.19) --
	( 67.45,500.39);

\path[draw=drawColor,line width= 0.6pt,line join=round] ( 92.53, 40.19) --
	( 92.53,500.39);

\path[draw=drawColor,line width= 0.6pt,line join=round] (117.61, 40.19) --
	(117.61,500.39);

\path[draw=drawColor,line width= 0.6pt,line join=round] (142.69, 40.19) --
	(142.69,500.39);

\path[draw=drawColor,line width= 0.6pt,line join=round] (167.78, 40.19) --
	(167.78,500.39);

\path[draw=drawColor,line width= 0.6pt,line join=round] (192.86, 40.19) --
	(192.86,500.39);

\path[draw=drawColor,line width= 0.6pt,line join=round] (217.94, 40.19) --
	(217.94,500.39);
\definecolor{drawColor}{gray}{0.40}

\path[draw=drawColor,line width= 0.6pt,dash pattern=on 1pt off 3pt ,line join=round] ( 27.31,253.82) -- (232.99,253.82);

\path[draw=drawColor,line width= 0.6pt,dash pattern=on 1pt off 3pt ,line join=round] ( 27.31,445.23) -- (232.99,445.23);
\definecolor{drawColor}{RGB}{0,0,0}

\node[text=drawColor,anchor=base west,inner sep=0pt, outer sep=0pt, scale=  1.10] at ( 29.82,243.35) {$4 \pi$};

\node[text=drawColor,anchor=base west,inner sep=0pt, outer sep=0pt, scale=  1.10] at ( 29.82,434.76) {$6 \pi$};
\definecolor{drawColor}{gray}{0.40}

\path[draw=drawColor,line width= 0.6pt,dash pattern=on 4pt off 4pt ,line join=round] ( 81.24,118.11) --
	(103.82,118.11);

\path[draw=drawColor,line width= 0.6pt,dash pattern=on 4pt off 4pt ,line join=round] ( 92.53,118.11) --
	( 92.53,118.11);

\path[draw=drawColor,line width= 0.6pt,dash pattern=on 4pt off 4pt ,line join=round] ( 81.24,118.11) --
	(103.82,118.11);

\path[draw=drawColor,line width= 0.6pt,dash pattern=on 4pt off 4pt ,line join=round] (106.32,146.54) --
	(128.90,146.54);

\path[draw=drawColor,line width= 0.6pt,dash pattern=on 4pt off 4pt ,line join=round] (117.61,146.54) --
	(117.61,146.54);

\path[draw=drawColor,line width= 0.6pt,dash pattern=on 4pt off 4pt ,line join=round] (106.32,146.54) --
	(128.90,146.54);

\path[draw=drawColor,line width= 0.6pt,dash pattern=on 4pt off 4pt ,line join=round] (106.32,214.58) --
	(128.90,214.58);

\path[draw=drawColor,line width= 0.6pt,dash pattern=on 4pt off 4pt ,line join=round] (117.61,214.58) --
	(117.61,214.58);

\path[draw=drawColor,line width= 0.6pt,dash pattern=on 4pt off 4pt ,line join=round] (106.32,214.58) --
	(128.90,214.58);

\path[draw=drawColor,line width= 0.6pt,dash pattern=on 4pt off 4pt ,line join=round] (131.41,167.09) --
	(153.98,167.09);

\path[draw=drawColor,line width= 0.6pt,dash pattern=on 4pt off 4pt ,line join=round] (142.69,167.09) --
	(142.69,167.09);

\path[draw=drawColor,line width= 0.6pt,dash pattern=on 4pt off 4pt ,line join=round] (131.41,167.09) --
	(153.98,167.09);

\path[draw=drawColor,line width= 0.6pt,dash pattern=on 4pt off 4pt ,line join=round] (131.41,261.80) --
	(153.98,261.80);

\path[draw=drawColor,line width= 0.6pt,dash pattern=on 4pt off 4pt ,line join=round] (142.69,261.80) --
	(142.69,261.80);

\path[draw=drawColor,line width= 0.6pt,dash pattern=on 4pt off 4pt ,line join=round] (131.41,261.80) --
	(153.98,261.80);

\path[draw=drawColor,line width= 0.6pt,dash pattern=on 4pt off 4pt ,line join=round] (181.57, 62.41) --
	(204.15, 62.41);

\path[draw=drawColor,line width= 0.6pt,dash pattern=on 4pt off 4pt ,line join=round] (192.86, 62.41) --
	(192.86, 62.41);

\path[draw=drawColor,line width= 0.6pt,dash pattern=on 4pt off 4pt ,line join=round] (181.57, 62.41) --
	(204.15, 62.41);

\path[draw=drawColor,line width= 0.6pt,dash pattern=on 4pt off 4pt ,line join=round] (181.57,183.57) --
	(204.15,183.57);

\path[draw=drawColor,line width= 0.6pt,dash pattern=on 4pt off 4pt ,line join=round] (192.86,183.57) --
	(192.86,183.57);

\path[draw=drawColor,line width= 0.6pt,dash pattern=on 4pt off 4pt ,line join=round] (181.57,183.57) --
	(204.15,183.57);

\path[draw=drawColor,line width= 0.6pt,dash pattern=on 4pt off 4pt ,line join=round] (181.57,317.60) --
	(204.15,317.60);

\path[draw=drawColor,line width= 0.6pt,dash pattern=on 4pt off 4pt ,line join=round] (192.86,317.60) --
	(192.86,317.60);

\path[draw=drawColor,line width= 0.6pt,dash pattern=on 4pt off 4pt ,line join=round] (181.57,317.60) --
	(204.15,317.60);

\path[draw=drawColor,line width= 0.6pt,dash pattern=on 4pt off 4pt ,line join=round] ( 31.08, 62.41) --
	( 53.65, 62.41);

\path[draw=drawColor,line width= 0.6pt,dash pattern=on 4pt off 4pt ,line join=round] ( 42.36, 62.41) --
	( 42.36, 62.41);

\path[draw=drawColor,line width= 0.6pt,dash pattern=on 4pt off 4pt ,line join=round] ( 31.08, 62.41) --
	( 53.65, 62.41);

\path[draw=drawColor,line width= 0.6pt,dash pattern=on 4pt off 4pt ,line join=round] ( 31.08,317.60) --
	( 53.65,317.60);

\path[draw=drawColor,line width= 0.6pt,dash pattern=on 4pt off 4pt ,line join=round] ( 42.36,317.60) --
	( 42.36,317.60);

\path[draw=drawColor,line width= 0.6pt,dash pattern=on 4pt off 4pt ,line join=round] ( 31.08,317.60) --
	( 53.65,317.60);

\path[draw=drawColor,line width= 0.6pt,dash pattern=on 4pt off 4pt ,line join=round] (206.65,317.60) --
	(229.23,317.60);

\path[draw=drawColor,line width= 0.6pt,dash pattern=on 4pt off 4pt ,line join=round] (217.94,317.60) --
	(217.94,317.60);

\path[draw=drawColor,line width= 0.6pt,dash pattern=on 4pt off 4pt ,line join=round] (206.65,317.60) --
	(229.23,317.60);

\path[draw=drawColor,line width= 0.6pt,dash pattern=on 4pt off 4pt ,line join=round] (156.49,261.80) --
	(179.06,261.80);

\path[draw=drawColor,line width= 0.6pt,dash pattern=on 4pt off 4pt ,line join=round] (167.78,261.80) --
	(167.78,261.80);

\path[draw=drawColor,line width= 0.6pt,dash pattern=on 4pt off 4pt ,line join=round] (156.49,261.80) --
	(179.06,261.80);

\path[draw=drawColor,line width= 0.6pt,dash pattern=on 4pt off 4pt ,line join=round] ( 56.16,317.60) --
	( 78.73,317.60);

\path[draw=drawColor,line width= 0.6pt,dash pattern=on 4pt off 4pt ,line join=round] ( 67.45,317.60) --
	( 67.45,317.60);

\path[draw=drawColor,line width= 0.6pt,dash pattern=on 4pt off 4pt ,line join=round] ( 56.16,317.60) --
	( 78.73,317.60);

\path[draw=drawColor,line width= 0.6pt,dash pattern=on 4pt off 4pt ,line join=round] ( 56.16,472.89) --
	( 78.73,472.89);

\path[draw=drawColor,line width= 0.6pt,dash pattern=on 4pt off 4pt ,line join=round] ( 67.45,472.89) --
	( 67.45,472.89);

\path[draw=drawColor,line width= 0.6pt,dash pattern=on 4pt off 4pt ,line join=round] ( 56.16,472.89) --
	( 78.73,472.89);
\definecolor{drawColor}{RGB}{0,0,0}

\path[draw=drawColor,line width= 0.6pt,line join=round] ( 81.24,124.16) --
	(103.82,124.16);

\path[draw=drawColor,line width= 0.6pt,line join=round] ( 92.53,124.16) --
	( 92.53,124.16);

\path[draw=drawColor,line width= 0.6pt,line join=round] ( 81.24,124.16) --
	(103.82,124.16);

\path[draw=drawColor,line width= 0.6pt,line join=round] (106.32,152.77) --
	(128.90,152.77);

\path[draw=drawColor,line width= 0.6pt,line join=round] (117.61,152.77) --
	(117.61,152.77);

\path[draw=drawColor,line width= 0.6pt,line join=round] (106.32,152.77) --
	(128.90,152.77);

\path[draw=drawColor,line width= 0.6pt,line join=round] (106.32,220.65) --
	(128.90,220.65);

\path[draw=drawColor,line width= 0.6pt,line join=round] (117.61,220.65) --
	(117.61,220.65);

\path[draw=drawColor,line width= 0.6pt,line join=round] (106.32,220.65) --
	(128.90,220.65);

\path[draw=drawColor,line width= 0.6pt,line join=round] (131.41,173.45) --
	(153.98,173.45);

\path[draw=drawColor,line width= 0.6pt,line join=round] (142.69,173.45) --
	(142.69,173.45);

\path[draw=drawColor,line width= 0.6pt,line join=round] (131.41,173.45) --
	(153.98,173.45);

\path[draw=drawColor,line width= 0.6pt,line join=round] (131.41,279.47) --
	(153.98,279.47);

\path[draw=drawColor,line width= 0.6pt,line join=round] (142.69,279.47) --
	(142.69,279.47);

\path[draw=drawColor,line width= 0.6pt,line join=round] (131.41,279.47) --
	(153.98,279.47);

\path[draw=drawColor,line width= 0.6pt,line join=round] (181.57, 63.39) --
	(204.15, 63.39);

\path[draw=drawColor,line width= 0.6pt,line join=round] (192.86, 63.39) --
	(192.86, 63.39);

\path[draw=drawColor,line width= 0.6pt,line join=round] (181.57, 63.39) --
	(204.15, 63.39);

\path[draw=drawColor,line width= 0.6pt,line join=round] (181.57,190.62) --
	(204.15,190.62);

\path[draw=drawColor,line width= 0.6pt,line join=round] (192.86,190.62) --
	(192.86,190.62);

\path[draw=drawColor,line width= 0.6pt,line join=round] (181.57,190.62) --
	(204.15,190.62);

\path[draw=drawColor,line width= 0.6pt,line join=round] (181.57,326.73) --
	(204.15,326.73);

\path[draw=drawColor,line width= 0.6pt,line join=round] (192.86,326.73) --
	(192.86,326.73);

\path[draw=drawColor,line width= 0.6pt,line join=round] (181.57,326.73) --
	(204.15,326.73);

\path[draw=drawColor,line width= 0.6pt,line join=round] ( 31.08, 64.78) --
	( 53.65, 64.78);

\path[draw=drawColor,line width= 0.6pt,line join=round] ( 42.36, 64.78) --
	( 42.36, 64.78);

\path[draw=drawColor,line width= 0.6pt,line join=round] ( 31.08, 64.78) --
	( 53.65, 64.78);

\path[draw=drawColor,line width= 0.6pt,line join=round] ( 31.08,336.19) --
	( 53.65,336.19);

\path[draw=drawColor,line width= 0.6pt,line join=round] ( 42.36,336.19) --
	( 42.36,336.19);

\path[draw=drawColor,line width= 0.6pt,line join=round] ( 31.08,336.19) --
	( 53.65,336.19);

\path[draw=drawColor,line width= 0.6pt,line join=round] (206.65,320.14) --
	(229.23,320.14);

\path[draw=drawColor,line width= 0.6pt,line join=round] (217.94,320.14) --
	(217.94,320.14);

\path[draw=drawColor,line width= 0.6pt,line join=round] (206.65,320.14) --
	(229.23,320.14);

\path[draw=drawColor,line width= 0.6pt,line join=round] (156.49,263.15) --
	(179.06,263.15);

\path[draw=drawColor,line width= 0.6pt,line join=round] (167.78,263.15) --
	(167.78,263.15);

\path[draw=drawColor,line width= 0.6pt,line join=round] (156.49,263.15) --
	(179.06,263.15);

\path[draw=drawColor,line width= 0.6pt,line join=round] ( 56.16,320.98) --
	( 78.73,320.98);

\path[draw=drawColor,line width= 0.6pt,line join=round] ( 67.45,320.98) --
	( 67.45,320.98);

\path[draw=drawColor,line width= 0.6pt,line join=round] ( 56.16,320.98) --
	( 78.73,320.98);

\path[draw=drawColor,line width= 0.6pt,line join=round] ( 56.16,478.22) --
	( 78.73,478.22);

\path[draw=drawColor,line width= 0.6pt,line join=round] ( 67.45,478.22) --
	( 67.45,478.22);

\path[draw=drawColor,line width= 0.6pt,line join=round] ( 56.16,478.22) --
	( 78.73,478.22);
\definecolor{drawColor}{RGB}{255,0,0}

\path[draw=drawColor,line width= 0.6pt,line join=round] ( 42.36, 64.46) -- ( 42.36, 65.76);

\path[draw=drawColor,line width= 0.6pt,line join=round] ( 42.36,332.69) -- ( 42.36,338.00);

\path[draw=drawColor,line width= 0.6pt,line join=round] ( 67.45,319.57) -- ( 67.45,324.13);

\path[draw=drawColor,line width= 0.6pt,line join=round] ( 67.45,469.60) -- ( 67.45,479.47);

\path[draw=drawColor,line width= 0.6pt,line join=round] ( 92.53,121.11) -- ( 92.53,124.03);

\path[draw=drawColor,line width= 0.6pt,line join=round] (117.61,150.96) -- (117.61,155.17);

\path[draw=drawColor,line width= 0.6pt,line join=round] (117.61,219.92) -- (117.61,225.59);

\path[draw=drawColor,line width= 0.6pt,line join=round] (142.69,165.02) -- (142.69,177.31);

\path[draw=drawColor,line width= 0.6pt,line join=round] (142.69,274.57) -- (142.69,289.47);

\path[draw=drawColor,line width= 0.6pt,line join=round] (167.78,246.82) -- (167.78,260.78);

\path[draw=drawColor,line width= 0.6pt,line join=round] (192.86, 61.11) -- (192.86, 74.73);

\path[draw=drawColor,line width= 0.6pt,line join=round] (192.86,201.85) -- (192.86,215.07);

\path[draw=drawColor,line width= 0.6pt,line join=round] (192.86,321.38) -- (192.86,337.46);

\path[draw=drawColor,line width= 0.6pt,line join=round] (217.94,328.97) -- (217.94,343.49);
\definecolor{fillColor}{RGB}{255,0,0}

\path[draw=drawColor,line width= 0.8pt,line join=round,line cap=round,fill=fillColor] ( 42.36, 65.11) circle (  1.25);

\path[draw=drawColor,line width= 0.8pt,line join=round,line cap=round,fill=fillColor] ( 42.36,335.34) circle (  1.25);

\path[draw=drawColor,line width= 0.8pt,line join=round,line cap=round,fill=fillColor] ( 67.45,321.85) circle (  1.25);

\path[draw=drawColor,line width= 0.8pt,line join=round,line cap=round,fill=fillColor] ( 67.45,474.53) circle (  1.25);

\path[draw=drawColor,line width= 0.8pt,line join=round,line cap=round,fill=fillColor] ( 92.53,122.57) circle (  1.25);

\path[draw=drawColor,line width= 0.8pt,line join=round,line cap=round,fill=fillColor] (117.61,153.06) circle (  1.25);

\path[draw=drawColor,line width= 0.8pt,line join=round,line cap=round,fill=fillColor] (117.61,222.75) circle (  1.25);

\path[draw=drawColor,line width= 0.8pt,line join=round,line cap=round,fill=fillColor] (142.69,171.16) circle (  1.25);

\path[draw=drawColor,line width= 0.8pt,line join=round,line cap=round,fill=fillColor] (142.69,282.02) circle (  1.25);

\path[draw=drawColor,line width= 0.8pt,line join=round,line cap=round,fill=fillColor] (167.78,253.80) circle (  1.25);

\path[draw=drawColor,line width= 0.8pt,line join=round,line cap=round,fill=fillColor] (192.86, 67.92) circle (  1.25);

\path[draw=drawColor,line width= 0.8pt,line join=round,line cap=round,fill=fillColor] (192.86,208.46) circle (  1.25);

\path[draw=drawColor,line width= 0.8pt,line join=round,line cap=round,fill=fillColor] (192.86,329.42) circle (  1.25);

\path[draw=drawColor,line width= 0.8pt,line join=round,line cap=round,fill=fillColor] (217.94,336.23) circle (  1.25);
\definecolor{drawColor}{gray}{0.20}

\path[draw=drawColor,line width= 0.6pt,line join=round,line cap=round] ( 27.31, 40.19) rectangle (232.99,500.39);
\end{scope}
\begin{scope}
\path[clip] (  0.00,  0.00) rectangle (238.49,505.89);
\definecolor{drawColor}{gray}{0.30}

\node[text=drawColor,anchor=base east,inner sep=0pt, outer sep=0pt, scale=  0.88] at ( 22.36, 59.38) {2};

\node[text=drawColor,anchor=base east,inner sep=0pt, outer sep=0pt, scale=  0.88] at ( 22.36,155.09) {3};

\node[text=drawColor,anchor=base east,inner sep=0pt, outer sep=0pt, scale=  0.88] at ( 22.36,250.79) {4};

\node[text=drawColor,anchor=base east,inner sep=0pt, outer sep=0pt, scale=  0.88] at ( 22.36,346.50) {5};

\node[text=drawColor,anchor=base east,inner sep=0pt, outer sep=0pt, scale=  0.88] at ( 22.36,442.20) {6};
\end{scope}
\begin{scope}
\path[clip] (  0.00,  0.00) rectangle (238.49,505.89);
\definecolor{drawColor}{gray}{0.20}

\path[draw=drawColor,line width= 0.6pt,line join=round] ( 24.56, 62.41) --
	( 27.31, 62.41);

\path[draw=drawColor,line width= 0.6pt,line join=round] ( 24.56,158.12) --
	( 27.31,158.12);

\path[draw=drawColor,line width= 0.6pt,line join=round] ( 24.56,253.82) --
	( 27.31,253.82);

\path[draw=drawColor,line width= 0.6pt,line join=round] ( 24.56,349.53) --
	( 27.31,349.53);

\path[draw=drawColor,line width= 0.6pt,line join=round] ( 24.56,445.23) --
	( 27.31,445.23);
\end{scope}
\begin{scope}
\path[clip] (  0.00,  0.00) rectangle (238.49,505.89);
\definecolor{drawColor}{gray}{0.20}

\path[draw=drawColor,line width= 0.6pt,line join=round] ( 42.36, 37.44) --
	( 42.36, 40.19);

\path[draw=drawColor,line width= 0.6pt,line join=round] ( 67.45, 37.44) --
	( 67.45, 40.19);

\path[draw=drawColor,line width= 0.6pt,line join=round] ( 92.53, 37.44) --
	( 92.53, 40.19);

\path[draw=drawColor,line width= 0.6pt,line join=round] (117.61, 37.44) --
	(117.61, 40.19);

\path[draw=drawColor,line width= 0.6pt,line join=round] (142.69, 37.44) --
	(142.69, 40.19);

\path[draw=drawColor,line width= 0.6pt,line join=round] (167.78, 37.44) --
	(167.78, 40.19);

\path[draw=drawColor,line width= 0.6pt,line join=round] (192.86, 37.44) --
	(192.86, 40.19);

\path[draw=drawColor,line width= 0.6pt,line join=round] (217.94, 37.44) --
	(217.94, 40.19);
\end{scope}
\begin{scope}
\path[clip] (  0.00,  0.00) rectangle (238.49,505.89);
\definecolor{drawColor}{gray}{0.30}

\node[text=drawColor,anchor=base,inner sep=0pt, outer sep=0pt, scale=  0.88] at ( 42.36, 29.18) {0};

\node[text=drawColor,anchor=base,inner sep=0pt, outer sep=0pt, scale=  0.88] at ( 42.36, 19.68) {$\mathrm A_1^+$};

\node[text=drawColor,anchor=base,inner sep=0pt, outer sep=0pt, scale=  0.88] at ( 67.45, 29.18) {0};

\node[text=drawColor,anchor=base,inner sep=0pt, outer sep=0pt, scale=  0.88] at ( 67.45, 19.68) {$\mathrm E^+$};

\node[text=drawColor,anchor=base,inner sep=0pt, outer sep=0pt, scale=  0.88] at ( 92.53, 29.18) {1};

\node[text=drawColor,anchor=base,inner sep=0pt, outer sep=0pt, scale=  0.88] at ( 92.53, 19.68) {$\mathrm A_1$};

\node[text=drawColor,anchor=base,inner sep=0pt, outer sep=0pt, scale=  0.88] at (117.61, 29.18) {2};

\node[text=drawColor,anchor=base,inner sep=0pt, outer sep=0pt, scale=  0.88] at (117.61, 19.68) {$\mathrm A_1$};

\node[text=drawColor,anchor=base,inner sep=0pt, outer sep=0pt, scale=  0.88] at (142.69, 29.18) {3};

\node[text=drawColor,anchor=base,inner sep=0pt, outer sep=0pt, scale=  0.88] at (142.69, 19.68) {$\mathrm A_1$};

\node[text=drawColor,anchor=base,inner sep=0pt, outer sep=0pt, scale=  0.88] at (167.78, 29.18) {3};

\node[text=drawColor,anchor=base,inner sep=0pt, outer sep=0pt, scale=  0.88] at (167.78, 19.68) {$\mathrm E$};

\node[text=drawColor,anchor=base,inner sep=0pt, outer sep=0pt, scale=  0.88] at (192.86, 29.18) {4};

\node[text=drawColor,anchor=base,inner sep=0pt, outer sep=0pt, scale=  0.88] at (192.86, 19.68) {$\mathrm A_1$};

\node[text=drawColor,anchor=base,inner sep=0pt, outer sep=0pt, scale=  0.88] at (217.94, 29.18) {4};

\node[text=drawColor,anchor=base,inner sep=0pt, outer sep=0pt, scale=  0.88] at (217.94, 19.68) {$\mathrm B_1$};
\end{scope}
\begin{scope}
\path[clip] (  0.00,  0.00) rectangle (238.49,505.89);
\definecolor{drawColor}{RGB}{0,0,0}

\node[text=drawColor,anchor=base,inner sep=0pt, outer sep=0pt, scale=  1.10] at (130.15,  7.64) {$\vec P^2, \Gamma$};
\end{scope}
\begin{scope}
\path[clip] (  0.00,  0.00) rectangle (238.49,505.89);
\definecolor{drawColor}{RGB}{0,0,0}

\node[text=drawColor,rotate= 90.00,anchor=base,inner sep=0pt, outer sep=0pt, scale=  1.10] at ( 13.08,270.29) {$a E / a M_\pi$};
\end{scope}
\end{tikzpicture}

%% file: plots/cm_mass_2.tex
\begin{tikzpicture}[x=1pt,y=1pt]
\definecolor{fillColor}{RGB}{255,255,255}
\path[use as bounding box,fill=fillColor,fill opacity=0.00] (0,0) rectangle (238.49,505.89);
\begin{scope}
\path[clip] (  0.00,  0.00) rectangle (238.49,505.89);
\definecolor{drawColor}{RGB}{255,255,255}
\definecolor{fillColor}{RGB}{255,255,255}

\path[draw=drawColor,line width= 0.6pt,line join=round,line cap=round,fill=fillColor] (  0.00,  0.00) rectangle (238.49,505.89);
\end{scope}
\begin{scope}
\path[clip] ( 27.31, 40.19) rectangle (232.99,500.39);
\definecolor{fillColor}{RGB}{255,255,255}

\path[fill=fillColor] ( 27.31, 40.19) rectangle (232.99,500.39);
\definecolor{drawColor}{gray}{0.92}

\path[draw=drawColor,line width= 0.3pt,line join=round] ( 27.31,113.40) --
	(232.99,113.40);

\path[draw=drawColor,line width= 0.3pt,line join=round] ( 27.31,217.99) --
	(232.99,217.99);

\path[draw=drawColor,line width= 0.3pt,line join=round] ( 27.31,322.59) --
	(232.99,322.59);

\path[draw=drawColor,line width= 0.3pt,line join=round] ( 27.31,427.18) --
	(232.99,427.18);

\path[draw=drawColor,line width= 0.6pt,line join=round] ( 27.31, 61.11) --
	(232.99, 61.11);

\path[draw=drawColor,line width= 0.6pt,line join=round] ( 27.31,165.70) --
	(232.99,165.70);

\path[draw=drawColor,line width= 0.6pt,line join=round] ( 27.31,270.29) --
	(232.99,270.29);

\path[draw=drawColor,line width= 0.6pt,line join=round] ( 27.31,374.88) --
	(232.99,374.88);

\path[draw=drawColor,line width= 0.6pt,line join=round] ( 27.31,479.47) --
	(232.99,479.47);

\path[draw=drawColor,line width= 0.6pt,line join=round] ( 47.22, 40.19) --
	( 47.22,500.39);

\path[draw=drawColor,line width= 0.6pt,line join=round] ( 80.39, 40.19) --
	( 80.39,500.39);

\path[draw=drawColor,line width= 0.6pt,line join=round] (113.57, 40.19) --
	(113.57,500.39);

\path[draw=drawColor,line width= 0.6pt,line join=round] (146.74, 40.19) --
	(146.74,500.39);

\path[draw=drawColor,line width= 0.6pt,line join=round] (179.91, 40.19) --
	(179.91,500.39);

\path[draw=drawColor,line width= 0.6pt,line join=round] (213.09, 40.19) --
	(213.09,500.39);
\definecolor{drawColor}{gray}{0.40}

\path[draw=drawColor,line width= 0.6pt,dash pattern=on 1pt off 3pt ,line join=round] ( 27.31,270.29) -- (232.99,270.29);

\path[draw=drawColor,line width= 0.6pt,dash pattern=on 1pt off 3pt ,line join=round] ( 27.31,479.47) -- (232.99,479.47);
\definecolor{drawColor}{RGB}{0,0,0}

\node[text=drawColor,anchor=base west,inner sep=0pt, outer sep=0pt, scale=  1.10] at ( 30.63,259.55) {$5 \pi$};

\node[text=drawColor,anchor=base west,inner sep=0pt, outer sep=0pt, scale=  1.10] at ( 30.63,468.73) {$7 \pi$};
\definecolor{drawColor}{gray}{0.40}

\path[draw=drawColor,line width= 0.6pt,dash pattern=on 4pt off 4pt ,line join=round] ( 32.29, 61.11) --
	( 62.15, 61.11);

\path[draw=drawColor,line width= 0.6pt,dash pattern=on 4pt off 4pt ,line join=round] ( 47.22, 61.11) --
	( 47.22, 61.11);

\path[draw=drawColor,line width= 0.6pt,dash pattern=on 4pt off 4pt ,line join=round] ( 32.29, 61.11) --
	( 62.15, 61.11);

\path[draw=drawColor,line width= 0.6pt,dash pattern=on 4pt off 4pt ,line join=round] ( 32.29,339.99) --
	( 62.15,339.99);

\path[draw=drawColor,line width= 0.6pt,dash pattern=on 4pt off 4pt ,line join=round] ( 47.22,339.99) --
	( 47.22,339.99);

\path[draw=drawColor,line width= 0.6pt,dash pattern=on 4pt off 4pt ,line join=round] ( 32.29,339.99) --
	( 62.15,339.99);

\path[draw=drawColor,line width= 0.6pt,dash pattern=on 4pt off 4pt ,line join=round] ( 98.64,143.30) --
	(128.49,143.30);

\path[draw=drawColor,line width= 0.6pt,dash pattern=on 4pt off 4pt ,line join=round] (113.57,143.30) --
	(113.57,143.30);

\path[draw=drawColor,line width= 0.6pt,dash pattern=on 4pt off 4pt ,line join=round] ( 98.64,143.30) --
	(128.49,143.30);

\path[draw=drawColor,line width= 0.6pt,dash pattern=on 4pt off 4pt ,line join=round] (131.81,185.84) --
	(161.67,185.84);

\path[draw=drawColor,line width= 0.6pt,dash pattern=on 4pt off 4pt ,line join=round] (146.74,185.84) --
	(146.74,185.84);

\path[draw=drawColor,line width= 0.6pt,dash pattern=on 4pt off 4pt ,line join=round] (131.81,185.84) --
	(161.67,185.84);

\path[draw=drawColor,line width= 0.6pt,dash pattern=on 4pt off 4pt ,line join=round] (131.81,251.33) --
	(161.67,251.33);

\path[draw=drawColor,line width= 0.6pt,dash pattern=on 4pt off 4pt ,line join=round] (146.74,251.33) --
	(146.74,251.33);

\path[draw=drawColor,line width= 0.6pt,dash pattern=on 4pt off 4pt ,line join=round] (131.81,251.33) --
	(161.67,251.33);

\path[draw=drawColor,line width= 0.6pt,dash pattern=on 4pt off 4pt ,line join=round] (164.98,216.75) --
	(194.84,216.75);

\path[draw=drawColor,line width= 0.6pt,dash pattern=on 4pt off 4pt ,line join=round] (179.91,216.75) --
	(179.91,216.75);

\path[draw=drawColor,line width= 0.6pt,dash pattern=on 4pt off 4pt ,line join=round] (164.98,216.75) --
	(194.84,216.75);

\path[draw=drawColor,line width= 0.6pt,dash pattern=on 4pt off 4pt ,line join=round] (198.16,143.30) --
	(228.01,143.30);

\path[draw=drawColor,line width= 0.6pt,dash pattern=on 4pt off 4pt ,line join=round] (213.09,143.30) --
	(213.09,143.30);

\path[draw=drawColor,line width= 0.6pt,dash pattern=on 4pt off 4pt ,line join=round] (198.16,143.30) --
	(228.01,143.30);

\path[draw=drawColor,line width= 0.6pt,dash pattern=on 4pt off 4pt ,line join=round] (198.16,241.60) --
	(228.01,241.60);

\path[draw=drawColor,line width= 0.6pt,dash pattern=on 4pt off 4pt ,line join=round] (213.09,241.60) --
	(213.09,241.60);

\path[draw=drawColor,line width= 0.6pt,dash pattern=on 4pt off 4pt ,line join=round] (198.16,241.60) --
	(228.01,241.60);

\path[draw=drawColor,line width= 0.6pt,dash pattern=on 4pt off 4pt ,line join=round] ( 65.46,339.99) --
	( 95.32,339.99);

\path[draw=drawColor,line width= 0.6pt,dash pattern=on 4pt off 4pt ,line join=round] ( 80.39,339.99) --
	( 80.39,339.99);

\path[draw=drawColor,line width= 0.6pt,dash pattern=on 4pt off 4pt ,line join=round] ( 65.46,339.99) --
	( 95.32,339.99);
\definecolor{drawColor}{RGB}{0,0,0}

\path[draw=drawColor,line width= 0.6pt,line join=round] ( 32.29, 69.48) --
	( 62.15, 69.48);

\path[draw=drawColor,line width= 0.6pt,line join=round] ( 47.22, 69.48) --
	( 47.22, 69.48);

\path[draw=drawColor,line width= 0.6pt,line join=round] ( 32.29, 69.48) --
	( 62.15, 69.48);

\path[draw=drawColor,line width= 0.6pt,line join=round] ( 32.29,378.48) --
	( 62.15,378.48);

\path[draw=drawColor,line width= 0.6pt,line join=round] ( 47.22,378.48) --
	( 47.22,378.48);

\path[draw=drawColor,line width= 0.6pt,line join=round] ( 32.29,378.48) --
	( 62.15,378.48);

\path[draw=drawColor,line width= 0.6pt,line join=round] ( 98.64,160.16) --
	(128.49,160.16);

\path[draw=drawColor,line width= 0.6pt,line join=round] (113.57,160.16) --
	(113.57,160.16);

\path[draw=drawColor,line width= 0.6pt,line join=round] ( 98.64,160.16) --
	(128.49,160.16);

\path[draw=drawColor,line width= 0.6pt,line join=round] (131.81,202.44) --
	(161.67,202.44);

\path[draw=drawColor,line width= 0.6pt,line join=round] (146.74,202.44) --
	(146.74,202.44);

\path[draw=drawColor,line width= 0.6pt,line join=round] (131.81,202.44) --
	(161.67,202.44);

\path[draw=drawColor,line width= 0.6pt,line join=round] (131.81,273.51) --
	(161.67,273.51);

\path[draw=drawColor,line width= 0.6pt,line join=round] (146.74,273.51) --
	(146.74,273.51);

\path[draw=drawColor,line width= 0.6pt,line join=round] (131.81,273.51) --
	(161.67,273.51);

\path[draw=drawColor,line width= 0.6pt,line join=round] (164.98,233.27) --
	(194.84,233.27);

\path[draw=drawColor,line width= 0.6pt,line join=round] (179.91,233.27) --
	(179.91,233.27);

\path[draw=drawColor,line width= 0.6pt,line join=round] (164.98,233.27) --
	(194.84,233.27);

\path[draw=drawColor,line width= 0.6pt,line join=round] (198.16,160.58) --
	(228.01,160.58);

\path[draw=drawColor,line width= 0.6pt,line join=round] (213.09,160.58) --
	(213.09,160.58);

\path[draw=drawColor,line width= 0.6pt,line join=round] (198.16,160.58) --
	(228.01,160.58);

\path[draw=drawColor,line width= 0.6pt,line join=round] (198.16,259.91) --
	(228.01,259.91);

\path[draw=drawColor,line width= 0.6pt,line join=round] (213.09,259.91) --
	(213.09,259.91);

\path[draw=drawColor,line width= 0.6pt,line join=round] (198.16,259.91) --
	(228.01,259.91);

\path[draw=drawColor,line width= 0.6pt,line join=round] ( 65.46,351.09) --
	( 95.32,351.09);

\path[draw=drawColor,line width= 0.6pt,line join=round] ( 80.39,351.09) --
	( 80.39,351.09);

\path[draw=drawColor,line width= 0.6pt,line join=round] ( 65.46,351.09) --
	( 95.32,351.09);
\definecolor{drawColor}{RGB}{255,0,0}

\path[draw=drawColor,line width= 0.6pt,line join=round] ( 47.22, 66.93) -- ( 47.22, 75.25);

\path[draw=drawColor,line width= 0.6pt,line join=round] ( 47.22,366.47) -- ( 47.22,393.19);

\path[draw=drawColor,line width= 0.6pt,line join=round] ( 80.39,351.62) -- ( 80.39,380.88);

\path[draw=drawColor,line width= 0.6pt,line join=round] (113.57,149.85) -- (113.57,164.45);

\path[draw=drawColor,line width= 0.6pt,line join=round] (146.74,196.72) -- (146.74,212.20);

\path[draw=drawColor,line width= 0.6pt,line join=round] (146.74,264.86) -- (146.74,274.52);

\path[draw=drawColor,line width= 0.6pt,line join=round] (179.91,229.52) -- (179.91,245.54);

\path[draw=drawColor,line width= 0.6pt,line join=round] (213.09,149.21) -- (213.09,165.22);

\path[draw=drawColor,line width= 0.6pt,line join=round] (213.09,279.59) -- (213.09,306.60);
\definecolor{fillColor}{RGB}{255,0,0}

\path[draw=drawColor,line width= 0.8pt,line join=round,line cap=round,fill=fillColor] ( 47.22, 71.09) circle (  1.25);

\path[draw=drawColor,line width= 0.8pt,line join=round,line cap=round,fill=fillColor] ( 47.22,379.83) circle (  1.25);

\path[draw=drawColor,line width= 0.8pt,line join=round,line cap=round,fill=fillColor] ( 80.39,366.25) circle (  1.25);

\path[draw=drawColor,line width= 0.8pt,line join=round,line cap=round,fill=fillColor] (113.57,157.15) circle (  1.25);

\path[draw=drawColor,line width= 0.8pt,line join=round,line cap=round,fill=fillColor] (146.74,204.46) circle (  1.25);

\path[draw=drawColor,line width= 0.8pt,line join=round,line cap=round,fill=fillColor] (146.74,269.69) circle (  1.25);

\path[draw=drawColor,line width= 0.8pt,line join=round,line cap=round,fill=fillColor] (179.91,237.53) circle (  1.25);

\path[draw=drawColor,line width= 0.8pt,line join=round,line cap=round,fill=fillColor] (213.09,157.22) circle (  1.25);

\path[draw=drawColor,line width= 0.8pt,line join=round,line cap=round,fill=fillColor] (213.09,293.10) circle (  1.25);
\definecolor{drawColor}{gray}{0.20}

\path[draw=drawColor,line width= 0.6pt,line join=round,line cap=round] ( 27.31, 40.19) rectangle (232.99,500.39);
\end{scope}
\begin{scope}
\path[clip] (  0.00,  0.00) rectangle (238.49,505.89);
\definecolor{drawColor}{gray}{0.30}

\node[text=drawColor,anchor=base east,inner sep=0pt, outer sep=0pt, scale=  0.88] at ( 22.36, 58.08) {3};

\node[text=drawColor,anchor=base east,inner sep=0pt, outer sep=0pt, scale=  0.88] at ( 22.36,162.67) {4};

\node[text=drawColor,anchor=base east,inner sep=0pt, outer sep=0pt, scale=  0.88] at ( 22.36,267.26) {5};

\node[text=drawColor,anchor=base east,inner sep=0pt, outer sep=0pt, scale=  0.88] at ( 22.36,371.85) {6};

\node[text=drawColor,anchor=base east,inner sep=0pt, outer sep=0pt, scale=  0.88] at ( 22.36,476.44) {7};
\end{scope}
\begin{scope}
\path[clip] (  0.00,  0.00) rectangle (238.49,505.89);
\definecolor{drawColor}{gray}{0.20}

\path[draw=drawColor,line width= 0.6pt,line join=round] ( 24.56, 61.11) --
	( 27.31, 61.11);

\path[draw=drawColor,line width= 0.6pt,line join=round] ( 24.56,165.70) --
	( 27.31,165.70);

\path[draw=drawColor,line width= 0.6pt,line join=round] ( 24.56,270.29) --
	( 27.31,270.29);

\path[draw=drawColor,line width= 0.6pt,line join=round] ( 24.56,374.88) --
	( 27.31,374.88);

\path[draw=drawColor,line width= 0.6pt,line join=round] ( 24.56,479.47) --
	( 27.31,479.47);
\end{scope}
\begin{scope}
\path[clip] (  0.00,  0.00) rectangle (238.49,505.89);
\definecolor{drawColor}{gray}{0.20}

\path[draw=drawColor,line width= 0.6pt,line join=round] ( 47.22, 37.44) --
	( 47.22, 40.19);

\path[draw=drawColor,line width= 0.6pt,line join=round] ( 80.39, 37.44) --
	( 80.39, 40.19);

\path[draw=drawColor,line width= 0.6pt,line join=round] (113.57, 37.44) --
	(113.57, 40.19);

\path[draw=drawColor,line width= 0.6pt,line join=round] (146.74, 37.44) --
	(146.74, 40.19);

\path[draw=drawColor,line width= 0.6pt,line join=round] (179.91, 37.44) --
	(179.91, 40.19);

\path[draw=drawColor,line width= 0.6pt,line join=round] (213.09, 37.44) --
	(213.09, 40.19);
\end{scope}
\begin{scope}
\path[clip] (  0.00,  0.00) rectangle (238.49,505.89);
\definecolor{drawColor}{gray}{0.30}

\node[text=drawColor,anchor=base,inner sep=0pt, outer sep=0pt, scale=  0.88] at ( 47.22, 29.18) {0};

\node[text=drawColor,anchor=base,inner sep=0pt, outer sep=0pt, scale=  0.88] at ( 47.22, 19.68) {$\mathrm A_1^-$};

\node[text=drawColor,anchor=base,inner sep=0pt, outer sep=0pt, scale=  0.88] at ( 80.39, 29.18) {0};

\node[text=drawColor,anchor=base,inner sep=0pt, outer sep=0pt, scale=  0.88] at ( 80.39, 19.68) {$\mathrm E^-$};

\node[text=drawColor,anchor=base,inner sep=0pt, outer sep=0pt, scale=  0.88] at (113.57, 29.18) {1};

\node[text=drawColor,anchor=base,inner sep=0pt, outer sep=0pt, scale=  0.88] at (113.57, 19.68) {$\mathrm A_2$};

\node[text=drawColor,anchor=base,inner sep=0pt, outer sep=0pt, scale=  0.88] at (146.74, 29.18) {2};

\node[text=drawColor,anchor=base,inner sep=0pt, outer sep=0pt, scale=  0.88] at (146.74, 19.68) {$\mathrm A_2$};

\node[text=drawColor,anchor=base,inner sep=0pt, outer sep=0pt, scale=  0.88] at (179.91, 29.18) {3};

\node[text=drawColor,anchor=base,inner sep=0pt, outer sep=0pt, scale=  0.88] at (179.91, 19.68) {$\mathrm A_2$};

\node[text=drawColor,anchor=base,inner sep=0pt, outer sep=0pt, scale=  0.88] at (213.09, 29.18) {4};

\node[text=drawColor,anchor=base,inner sep=0pt, outer sep=0pt, scale=  0.88] at (213.09, 19.68) {$\mathrm A_2$};
\end{scope}
\begin{scope}
\path[clip] (  0.00,  0.00) rectangle (238.49,505.89);
\definecolor{drawColor}{RGB}{0,0,0}

\node[text=drawColor,anchor=base,inner sep=0pt, outer sep=0pt, scale=  1.10] at (130.15,  7.64) {$\vec P^2, \Gamma$};
\end{scope}
\begin{scope}
\path[clip] (  0.00,  0.00) rectangle (238.49,505.89);
\definecolor{drawColor}{RGB}{0,0,0}

\node[text=drawColor,rotate= 90.00,anchor=base,inner sep=0pt, outer sep=0pt, scale=  1.10] at ( 13.08,270.29) {$a E / a M_\pi$};
\end{scope}
\end{tikzpicture}

%% file: main.bbl
\begin{thebibliography}{120}%
\makeatletter
\providecommand \@ifxundefined [1]{%
 \@ifx{#1\undefined}
}%
\providecommand \@ifnum [1]{%
 \ifnum #1\expandafter \@firstoftwo
 \else \expandafter \@secondoftwo
 \fi
}%
\providecommand \@ifx [1]{%
 \ifx #1\expandafter \@firstoftwo
 \else \expandafter \@secondoftwo
 \fi
}%
\providecommand \natexlab [1]{#1}%
\providecommand \enquote  [1]{``#1''}%
\providecommand \bibnamefont  [1]{#1}%
\providecommand \bibfnamefont [1]{#1}%
\providecommand \citenamefont [1]{#1}%
\providecommand \href@noop [0]{\@secondoftwo}%
\providecommand \href [0]{\begingroup \@sanitize@url \@href}%
\providecommand \@href[1]{\@@startlink{#1}\@@href}%
\providecommand \@@href[1]{\endgroup#1\@@endlink}%
\providecommand \@sanitize@url [0]{\catcode `\\12\catcode `\$12\catcode
  `\&12\catcode `\#12\catcode `\^12\catcode `\_12\catcode `\%12\relax}%
\providecommand \@@startlink[1]{}%
\providecommand \@@endlink[0]{}%
\providecommand \url  [0]{\begingroup\@sanitize@url \@url }%
\providecommand \@url [1]{\endgroup\@href {#1}{\urlprefix }}%
\providecommand \urlprefix  [0]{URL }%
\providecommand \Eprint [0]{\href }%
\providecommand \doibase [0]{http://dx.doi.org/}%
\providecommand \selectlanguage [0]{\@gobble}%
\providecommand \bibinfo  [0]{\@secondoftwo}%
\providecommand \bibfield  [0]{\@secondoftwo}%
\providecommand \translation [1]{[#1]}%
\providecommand \BibitemOpen [0]{}%
\providecommand \bibitemStop [0]{}%
\providecommand \bibitemNoStop [0]{.\EOS\space}%
\providecommand \EOS [0]{\spacefactor3000\relax}%
\providecommand \BibitemShut  [1]{\csname bibitem#1\endcsname}%
\let\auto@bib@innerbib\@empty
\bibitem [{\citenamefont {Tanabashi}\ \emph {et~al.}(2018)\citenamefont
  {Tanabashi} \emph {et~al.}}]{PhysRevD.98.030001}%
  \BibitemOpen
  \bibfield  {author} {\bibinfo {author} {\bibfnamefont {M.}~\bibnamefont
  {Tanabashi}} \emph {et~al.} (\bibinfo {collaboration} {PDG}),\ }\bibfield
  {title} {\enquote {\bibinfo {title} {Review of particle physics},}\ }\href
  {\doibase 10.1103/PhysRevD.98.030001} {\bibfield  {journal} {\bibinfo
  {journal} {Phys. Rev. D}\ }\textbf {\bibinfo {volume} {98}},\ \bibinfo
  {pages} {030001} (\bibinfo {year} {2018})}\BibitemShut {NoStop}%
\bibitem [{\citenamefont {Roper}(1964)}]{Roper}%
  \BibitemOpen
  \bibfield  {author} {\bibinfo {author} {\bibfnamefont {L.~David}\
  \bibnamefont {Roper}},\ }\bibfield  {title} {\enquote {\bibinfo {title}
  {Evidence for a ${P}_{11}$ pion-nucleon resonance at 556 {MeV}},}\ }\href
  {\doibase 10.1103/PhysRevLett.12.340} {\bibfield  {journal} {\bibinfo
  {journal} {Phys. Rev. Lett.}\ }\textbf {\bibinfo {volume} {12}},\ \bibinfo
  {pages} {340--342} (\bibinfo {year} {1964})}\BibitemShut {NoStop}%
\bibitem [{\citenamefont {Helmes}\ \emph {et~al.}(2015)\citenamefont {Helmes},
  \citenamefont {Jost}, \citenamefont {Knippschild}, \citenamefont {Liu},
  \citenamefont {Liu}, \citenamefont {Liu}, \citenamefont {Urbach},
  \citenamefont {Ueding}, \citenamefont {Wang},\ and\ \citenamefont
  {Werner}}]{Helmes:2015gla}%
  \BibitemOpen
  \bibfield  {author} {\bibinfo {author} {\bibfnamefont {C.}~\bibnamefont
  {Helmes}}, \bibinfo {author} {\bibfnamefont {C.}~\bibnamefont {Jost}},
  \bibinfo {author} {\bibfnamefont {B.}~\bibnamefont {Knippschild}}, \bibinfo
  {author} {\bibfnamefont {C.}~\bibnamefont {Liu}}, \bibinfo {author}
  {\bibfnamefont {J.}~\bibnamefont {Liu}}, \bibinfo {author} {\bibfnamefont
  {L.}~\bibnamefont {Liu}}, \bibinfo {author} {\bibfnamefont {C.}~\bibnamefont
  {Urbach}}, \bibinfo {author} {\bibfnamefont {M.}~\bibnamefont {Ueding}},
  \bibinfo {author} {\bibfnamefont {Z.}~\bibnamefont {Wang}}, \ and\ \bibinfo
  {author} {\bibfnamefont {M.}~\bibnamefont {Werner}} (\bibinfo {collaboration}
  {ETM}),\ }\bibfield  {title} {\enquote {\bibinfo {title} {{Hadron-hadron
  interactions from N$_{f}$ = 2 + 1 + 1 lattice QCD: isospin-2 $\pi\pi$
  scattering length}},}\ }\href {\doibase 10.1007/JHEP09(2015)109} {\bibfield
  {journal} {\bibinfo  {journal} {JHEP}\ }\textbf {\bibinfo {volume} {09}},\
  \bibinfo {pages} {109} (\bibinfo {year} {2015})},\ \Eprint
  {http://arxiv.org/abs/1506.00408} {arXiv:1506.00408 [hep-lat]} \BibitemShut
  {NoStop}%
\bibitem [{\citenamefont {Lüscher}(1986)}]{Luscher:1986pf}%
  \BibitemOpen
  \bibfield  {author} {\bibinfo {author} {\bibfnamefont {M.}~\bibnamefont
  {Lüscher}},\ }\bibfield  {title} {\enquote {\bibinfo {title} {{Volume
  Dependence of the Energy Spectrum in Massive Quantum Field Theories. 2.
  Scattering States}},}\ }\href {\doibase 10.1007/BF01211097} {\bibfield
  {journal} {\bibinfo  {journal} {Commun. Math. Phys.}\ }\textbf {\bibinfo
  {volume} {105}},\ \bibinfo {pages} {153--188} (\bibinfo {year}
  {1986})}\BibitemShut {NoStop}%
\bibitem [{\citenamefont {Lüscher}(1991)}]{Luscher:1990ux}%
  \BibitemOpen
  \bibfield  {author} {\bibinfo {author} {\bibfnamefont {Martin}\ \bibnamefont
  {Lüscher}},\ }\bibfield  {title} {\enquote {\bibinfo {title} {{Two particle
  states on a torus and their relation to the scattering matrix}},}\ }\href
  {\doibase 10.1016/0550-3213(91)90366-6} {\bibfield  {journal} {\bibinfo
  {journal} {Nucl. Phys.}\ }\textbf {\bibinfo {volume} {B354}},\ \bibinfo
  {pages} {531--578} (\bibinfo {year} {1991})}\BibitemShut {NoStop}%
\bibitem [{\citenamefont {Lüscher}\ and\ \citenamefont
  {Wolff}(1990)}]{Luscher:1990ck}%
  \BibitemOpen
  \bibfield  {author} {\bibinfo {author} {\bibfnamefont {Martin}\ \bibnamefont
  {Lüscher}}\ and\ \bibinfo {author} {\bibfnamefont {Ulli}\ \bibnamefont
  {Wolff}},\ }\bibfield  {title} {\enquote {\bibinfo {title} {{How to Calculate
  the Elastic Scattering Matrix in Two-dimensional Quantum Field Theories by
  Numerical Simulation}},}\ }\href {\doibase 10.1016/0550-3213(90)90540-T}
  {\bibfield  {journal} {\bibinfo  {journal} {Nucl. Phys.}\ }\textbf {\bibinfo
  {volume} {B339}},\ \bibinfo {pages} {222--252} (\bibinfo {year}
  {1990})}\BibitemShut {NoStop}%
\bibitem [{\citenamefont {Rummukainen}\ and\ \citenamefont
  {Gottlieb}(1995)}]{Kari:1995}%
  \BibitemOpen
  \bibfield  {author} {\bibinfo {author} {\bibfnamefont {K.}~\bibnamefont
  {Rummukainen}}\ and\ \bibinfo {author} {\bibfnamefont {Steven~A.}\
  \bibnamefont {Gottlieb}},\ }\bibfield  {title} {\enquote {\bibinfo {title}
  {{Resonance scattering phase shifts on a nonrest frame lattice}},}\ }\href
  {\doibase 10.1016/0550-3213(95)00313-H} {\bibfield  {journal} {\bibinfo
  {journal} {Nucl.Phys.}\ }\textbf {\bibinfo {volume} {B450}},\ \bibinfo
  {pages} {397--436} (\bibinfo {year} {1995})},\ \Eprint
  {http://arxiv.org/abs/hep-lat/9503028} {arXiv:hep-lat/9503028} \BibitemShut
  {NoStop}%
\bibitem [{\citenamefont {Kim}\ \emph {et~al.}(2005)\citenamefont {Kim},
  \citenamefont {Sachrajda},\ and\ \citenamefont {Sharpe}}]{Kim:2005gf}%
  \BibitemOpen
  \bibfield  {author} {\bibinfo {author} {\bibfnamefont {C.~h.}\ \bibnamefont
  {Kim}}, \bibinfo {author} {\bibfnamefont {C.~T.}\ \bibnamefont {Sachrajda}},
  \ and\ \bibinfo {author} {\bibfnamefont {Stephen~R.}\ \bibnamefont
  {Sharpe}},\ }\bibfield  {title} {\enquote {\bibinfo {title} {{Finite-volume
  effects for two-hadron states in moving frames}},}\ }\href {\doibase
  10.1016/j.nuclphysb.2005.08.029} {\bibfield  {journal} {\bibinfo  {journal}
  {Nucl. Phys.}\ }\textbf {\bibinfo {volume} {B727}},\ \bibinfo {pages}
  {218--243} (\bibinfo {year} {2005})},\ \Eprint
  {http://arxiv.org/abs/hep-lat/0507006} {arXiv:hep-lat/0507006 [hep-lat]}
  \BibitemShut {NoStop}%
\bibitem [{\citenamefont {He}\ \emph {et~al.}(2005)\citenamefont {He},
  \citenamefont {Feng},\ and\ \citenamefont {Liu}}]{He:2005ey}%
  \BibitemOpen
  \bibfield  {author} {\bibinfo {author} {\bibfnamefont {Song}\ \bibnamefont
  {He}}, \bibinfo {author} {\bibfnamefont {Xu}~\bibnamefont {Feng}}, \ and\
  \bibinfo {author} {\bibfnamefont {Chuan}\ \bibnamefont {Liu}},\ }\bibfield
  {title} {\enquote {\bibinfo {title} {{Two particle states and the S-matrix
  elements in multi-channel scattering}},}\ }\href {\doibase
  10.1088/1126-6708/2005/07/011} {\bibfield  {journal} {\bibinfo  {journal}
  {JHEP}\ }\textbf {\bibinfo {volume} {07}},\ \bibinfo {pages} {011} (\bibinfo
  {year} {2005})},\ \Eprint {http://arxiv.org/abs/hep-lat/0504019}
  {arXiv:hep-lat/0504019 [hep-lat]} \BibitemShut {NoStop}%
\bibitem [{\citenamefont {Bernard}\ \emph {et~al.}(2011)\citenamefont
  {Bernard}, \citenamefont {Lage}, \citenamefont {Meißner},\ and\
  \citenamefont {Rusetsky}}]{Bernard:2010}%
  \BibitemOpen
  \bibfield  {author} {\bibinfo {author} {\bibfnamefont {V.}~\bibnamefont
  {Bernard}}, \bibinfo {author} {\bibfnamefont {M.}~\bibnamefont {Lage}},
  \bibinfo {author} {\bibfnamefont {U.-G.}\ \bibnamefont {Meißner}}, \ and\
  \bibinfo {author} {\bibfnamefont {A.}~\bibnamefont {Rusetsky}},\ }\href
  {\doibase 10.1007/JHEP01(2011)019} {\bibfield  {journal} {\bibinfo  {journal}
  {JHEP}\ }\textbf {\bibinfo {volume} {1101}},\ \bibinfo {pages} {019}
  (\bibinfo {year} {2011})},\ \Eprint {http://arxiv.org/abs/1010.6018}
  {arXiv:1010.6018 [hep-lat]} \BibitemShut {NoStop}%
\bibitem [{\citenamefont {Hansen}\ and\ \citenamefont
  {Sharpe}(2012)}]{Hansen:2012tf}%
  \BibitemOpen
  \bibfield  {author} {\bibinfo {author} {\bibfnamefont {Maxwell~T.}\
  \bibnamefont {Hansen}}\ and\ \bibinfo {author} {\bibfnamefont {Stephen~R.}\
  \bibnamefont {Sharpe}},\ }\bibfield  {title} {\enquote {\bibinfo {title}
  {{Multiple-channel generalization of Lellouch-Luscher formula}},}\ }\href
  {\doibase 10.1103/PhysRevD.86.016007} {\bibfield  {journal} {\bibinfo
  {journal} {Phys.Rev.}\ }\textbf {\bibinfo {volume} {D86}},\ \bibinfo {pages}
  {016007} (\bibinfo {year} {2012})},\ \Eprint {http://arxiv.org/abs/1204.0826}
  {arXiv:1204.0826 [hep-lat]} \BibitemShut {NoStop}%
\bibitem [{\citenamefont {Brice\~no}\ and\ \citenamefont
  {Davoudi}(2013)}]{Briceno:2012yi}%
  \BibitemOpen
  \bibfield  {author} {\bibinfo {author} {\bibfnamefont {Raul~A.}\ \bibnamefont
  {Brice\~no}}\ and\ \bibinfo {author} {\bibfnamefont {Zohreh}\ \bibnamefont
  {Davoudi}},\ }\bibfield  {title} {\enquote {\bibinfo {title} {{Moving
  multichannel systems in a finite volume with application to proton-proton
  fusion}},}\ }\href {\doibase 10.1103/PhysRevD.88.094507} {\bibfield
  {journal} {\bibinfo  {journal} {Phys. Rev.}\ }\textbf {\bibinfo {volume}
  {D88}},\ \bibinfo {pages} {094507} (\bibinfo {year} {2013})},\ \Eprint
  {http://arxiv.org/abs/1204.1110} {arXiv:1204.1110 [hep-lat]} \BibitemShut
  {NoStop}%
\bibitem [{\citenamefont {Brice\~no}(2014)}]{Briceno:2014oea}%
  \BibitemOpen
  \bibfield  {author} {\bibinfo {author} {\bibfnamefont {Raul~A.}\ \bibnamefont
  {Brice\~no}},\ }\bibfield  {title} {\enquote {\bibinfo {title} {{Two-particle
  multichannel systems in a finite volume with arbitrary spin}},}\ }\href
  {\doibase 10.1103/PhysRevD.89.074507} {\bibfield  {journal} {\bibinfo
  {journal} {Phys. Rev.}\ }\textbf {\bibinfo {volume} {D89}},\ \bibinfo {pages}
  {074507} (\bibinfo {year} {2014})},\ \Eprint {http://arxiv.org/abs/1401.3312}
  {arXiv:1401.3312 [hep-lat]} \BibitemShut {NoStop}%
\bibitem [{\citenamefont {Romero-López}\ \emph
  {et~al.}(2018{\natexlab{a}})\citenamefont {Romero-López}, \citenamefont
  {Rusetsky},\ and\ \citenamefont {Urbach}}]{Romero-Lopez:2018zyy}%
  \BibitemOpen
  \bibfield  {author} {\bibinfo {author} {\bibfnamefont {F.}~\bibnamefont
  {Romero-López}}, \bibinfo {author} {\bibfnamefont {A.}~\bibnamefont
  {Rusetsky}}, \ and\ \bibinfo {author} {\bibfnamefont {C.}~\bibnamefont
  {Urbach}},\ }\bibfield  {title} {\enquote {\bibinfo {title} {{Vector particle
  scattering on the lattice}},}\ }\href {\doibase 10.1103/PhysRevD.98.014503}
  {\bibfield  {journal} {\bibinfo  {journal} {Phys. Rev.}\ }\textbf {\bibinfo
  {volume} {D98}},\ \bibinfo {pages} {014503} (\bibinfo {year}
  {2018}{\natexlab{a}})},\ \Eprint {http://arxiv.org/abs/1802.03458}
  {arXiv:1802.03458 [hep-lat]} \BibitemShut {NoStop}%
\bibitem [{\citenamefont {Luu}\ and\ \citenamefont
  {Savage}(2011)}]{Luu:2011ep}%
  \BibitemOpen
  \bibfield  {author} {\bibinfo {author} {\bibfnamefont {Thomas}\ \bibnamefont
  {Luu}}\ and\ \bibinfo {author} {\bibfnamefont {Martin~J.}\ \bibnamefont
  {Savage}},\ }\bibfield  {title} {\enquote {\bibinfo {title} {{Extracting
  Scattering Phase-Shifts in Higher Partial-Waves from Lattice QCD
  Calculations}},}\ }\href {\doibase 10.1103/PhysRevD.83.114508} {\bibfield
  {journal} {\bibinfo  {journal} {Phys. Rev.}\ }\textbf {\bibinfo {volume}
  {D83}},\ \bibinfo {pages} {114508} (\bibinfo {year} {2011})},\ \Eprint
  {http://arxiv.org/abs/1101.3347} {arXiv:1101.3347 [hep-lat]} \BibitemShut
  {NoStop}%
\bibitem [{\citenamefont {Göckeler}\ \emph {et~al.}(2012)\citenamefont
  {Göckeler}, \citenamefont {Horsley}, \citenamefont {Lage}, \citenamefont
  {Meißner}, \citenamefont {Rakow}, \citenamefont {Rusetsky}, \citenamefont
  {Schierholz},\ and\ \citenamefont {Zanotti}}]{Gockeler:2012yj}%
  \BibitemOpen
  \bibfield  {author} {\bibinfo {author} {\bibfnamefont {M.}~\bibnamefont
  {Göckeler}}, \bibinfo {author} {\bibfnamefont {R.}~\bibnamefont {Horsley}},
  \bibinfo {author} {\bibfnamefont {M.}~\bibnamefont {Lage}}, \bibinfo {author}
  {\bibfnamefont {U.~G.}\ \bibnamefont {Meißner}}, \bibinfo {author}
  {\bibfnamefont {P.~E.~L.}\ \bibnamefont {Rakow}}, \bibinfo {author}
  {\bibfnamefont {A.}~\bibnamefont {Rusetsky}}, \bibinfo {author}
  {\bibfnamefont {G.}~\bibnamefont {Schierholz}}, \ and\ \bibinfo {author}
  {\bibfnamefont {J.~M.}\ \bibnamefont {Zanotti}},\ }\bibfield  {title}
  {\enquote {\bibinfo {title} {{Scattering phases for meson and baryon
  resonances on general moving-frame lattices}},}\ }\href {\doibase
  10.1103/PhysRevD.86.094513} {\bibfield  {journal} {\bibinfo  {journal} {Phys.
  Rev.}\ }\textbf {\bibinfo {volume} {D86}},\ \bibinfo {pages} {094513}
  (\bibinfo {year} {2012})},\ \Eprint {http://arxiv.org/abs/1206.4141}
  {arXiv:1206.4141 [hep-lat]} \BibitemShut {NoStop}%
\bibitem [{\citenamefont {Feng}\ \emph {et~al.}(2010)\citenamefont {Feng},
  \citenamefont {Jansen},\ and\ \citenamefont {Renner}}]{Feng:2009ij}%
  \BibitemOpen
  \bibfield  {author} {\bibinfo {author} {\bibfnamefont {Xu}~\bibnamefont
  {Feng}}, \bibinfo {author} {\bibfnamefont {Karl}\ \bibnamefont {Jansen}}, \
  and\ \bibinfo {author} {\bibfnamefont {Dru~B.}\ \bibnamefont {Renner}},\
  }\bibfield  {title} {\enquote {\bibinfo {title} {{The pi+ pi+ scattering
  length from maximally twisted mass lattice QCD}},}\ }\href {\doibase
  10.1016/j.physletb.2010.01.018} {\bibfield  {journal} {\bibinfo  {journal}
  {Phys. Lett.}\ }\textbf {\bibinfo {volume} {B684}},\ \bibinfo {pages}
  {268--274} (\bibinfo {year} {2010})},\ \Eprint
  {http://arxiv.org/abs/0909.3255} {arXiv:0909.3255 [hep-lat]} \BibitemShut
  {NoStop}%
\bibitem [{\citenamefont {Lage}\ \emph {et~al.}(2009)\citenamefont {Lage},
  \citenamefont {Meißner},\ and\ \citenamefont {Rusetsky}}]{Lage:2009zv}%
  \BibitemOpen
  \bibfield  {author} {\bibinfo {author} {\bibfnamefont {Michael}\ \bibnamefont
  {Lage}}, \bibinfo {author} {\bibfnamefont {Ulf-G.}\ \bibnamefont {Meißner}},
  \ and\ \bibinfo {author} {\bibfnamefont {Akaki}\ \bibnamefont {Rusetsky}},\
  }\bibfield  {title} {\enquote {\bibinfo {title} {{A Method to measure the
  antikaon-nucleon scattering length in lattice QCD}},}\ }\href {\doibase
  10.1016/j.physletb.2009.10.055} {\bibfield  {journal} {\bibinfo  {journal}
  {Phys. Lett.}\ }\textbf {\bibinfo {volume} {B681}},\ \bibinfo {pages}
  {439--443} (\bibinfo {year} {2009})},\ \Eprint
  {http://arxiv.org/abs/0905.0069} {arXiv:0905.0069 [hep-lat]} \BibitemShut
  {NoStop}%
\bibitem [{\citenamefont {Wilson}\ \emph {et~al.}(2015)\citenamefont {Wilson},
  \citenamefont {Brice{\~n}o}, \citenamefont {Dudek}, \citenamefont {Edwards},\
  and\ \citenamefont {Thomas}}]{Wilson:2015dqa}%
  \BibitemOpen
  \bibfield  {author} {\bibinfo {author} {\bibfnamefont {David~J.}\
  \bibnamefont {Wilson}}, \bibinfo {author} {\bibfnamefont {Raul~A.}\
  \bibnamefont {Brice{\~n}o}}, \bibinfo {author} {\bibfnamefont {Jozef~J.}\
  \bibnamefont {Dudek}}, \bibinfo {author} {\bibfnamefont {Robert~G.}\
  \bibnamefont {Edwards}}, \ and\ \bibinfo {author} {\bibfnamefont
  {Christopher~E.}\ \bibnamefont {Thomas}},\ }\bibfield  {title} {\enquote
  {\bibinfo {title} {{Coupled $\pi\pi, K\bar{K}$ scattering in $P$-wave and the
  $\rho$ resonance from lattice QCD}},}\ }\href {\doibase
  10.1103/PhysRevD.92.094502} {\bibfield  {journal} {\bibinfo  {journal} {Phys.
  Rev.}\ }\textbf {\bibinfo {volume} {D92}},\ \bibinfo {pages} {094502}
  (\bibinfo {year} {2015})},\ \Eprint {http://arxiv.org/abs/1507.02599}
  {arXiv:1507.02599 [hep-ph]} \BibitemShut {NoStop}%
\bibitem [{\citenamefont {Brice{\~n}o}\ \emph {et~al.}(2017)\citenamefont
  {Brice{\~n}o}, \citenamefont {Dudek}, \citenamefont {Edwards},\ and\
  \citenamefont {Wilson}}]{Briceno:2016mjc}%
  \BibitemOpen
  \bibfield  {author} {\bibinfo {author} {\bibfnamefont {Raul~A.}\ \bibnamefont
  {Brice{\~n}o}}, \bibinfo {author} {\bibfnamefont {Jozef~J.}\ \bibnamefont
  {Dudek}}, \bibinfo {author} {\bibfnamefont {Robert~G.}\ \bibnamefont
  {Edwards}}, \ and\ \bibinfo {author} {\bibfnamefont {David~J.}\ \bibnamefont
  {Wilson}},\ }\bibfield  {title} {\enquote {\bibinfo {title} {{Isoscalar
  $\pi\pi$ scattering and the $\sigma$ meson resonance from QCD}},}\ }\href
  {\doibase 10.1103/PhysRevLett.118.022002} {\bibfield  {journal} {\bibinfo
  {journal} {Phys. Rev. Lett.}\ }\textbf {\bibinfo {volume} {118}},\ \bibinfo
  {pages} {022002} (\bibinfo {year} {2017})},\ \Eprint
  {http://arxiv.org/abs/1607.05900} {arXiv:1607.05900 [hep-ph]} \BibitemShut
  {NoStop}%
\bibitem [{\citenamefont {Brett}\ \emph {et~al.}(2018)\citenamefont {Brett},
  \citenamefont {Bulava}, \citenamefont {Fallica}, \citenamefont {Hanlon},
  \citenamefont {Höz},\ and\ \citenamefont {Morningstar}}]{Brett:2018jqw}%
  \BibitemOpen
  \bibfield  {author} {\bibinfo {author} {\bibfnamefont {Ruair}\ \bibnamefont
  {Brett}}, \bibinfo {author} {\bibfnamefont {John}\ \bibnamefont {Bulava}},
  \bibinfo {author} {\bibfnamefont {Jacob}\ \bibnamefont {Fallica}}, \bibinfo
  {author} {\bibfnamefont {Andrew}\ \bibnamefont {Hanlon}}, \bibinfo {author}
  {\bibfnamefont {Ben}\ \bibnamefont {Höz}}, \ and\ \bibinfo {author}
  {\bibfnamefont {Colin}\ \bibnamefont {Morningstar}},\ }\bibfield  {title}
  {\enquote {\bibinfo {title} {{Determination of $s$- and $p$-wave $I=1/2$
  $K\pi$ scattering amplitudes in $N_{\mathrm{f}}=2+1$ lattice QCD}},}\ }\href
  {\doibase 10.1016/j.nuclphysb.2018.05.008} {\bibfield  {journal} {\bibinfo
  {journal} {Nucl. Phys.}\ }\textbf {\bibinfo {volume} {B932}},\ \bibinfo
  {pages} {29--51} (\bibinfo {year} {2018})},\ \Eprint
  {http://arxiv.org/abs/1802.03100} {arXiv:1802.03100 [hep-lat]} \BibitemShut
  {NoStop}%
\bibitem [{\citenamefont {Andersen}\ \emph {et~al.}(2018)\citenamefont
  {Andersen}, \citenamefont {Bulava}, \citenamefont {Hörz},\ and\
  \citenamefont {Morningstar}}]{Andersen:2017una}%
  \BibitemOpen
  \bibfield  {author} {\bibinfo {author} {\bibfnamefont {Christian~Walther}\
  \bibnamefont {Andersen}}, \bibinfo {author} {\bibfnamefont {John}\
  \bibnamefont {Bulava}}, \bibinfo {author} {\bibfnamefont {Ben}\ \bibnamefont
  {Hörz}}, \ and\ \bibinfo {author} {\bibfnamefont {Colin}\ \bibnamefont
  {Morningstar}},\ }\bibfield  {title} {\enquote {\bibinfo {title} {{Elastic
  $I=3/2 p$-wave nucleon-pion scattering amplitude and the $\Delta$(1232)
  resonance from N$_f$=2+1 lattice QCD}},}\ }\href {\doibase
  10.1103/PhysRevD.97.014506} {\bibfield  {journal} {\bibinfo  {journal} {Phys.
  Rev.}\ }\textbf {\bibinfo {volume} {D97}},\ \bibinfo {pages} {014506}
  (\bibinfo {year} {2018})},\ \Eprint {http://arxiv.org/abs/1710.01557}
  {arXiv:1710.01557 [hep-lat]} \BibitemShut {NoStop}%
\bibitem [{\citenamefont {Guo}\ \emph {et~al.}(2018)\citenamefont {Guo},
  \citenamefont {Alexandru}, \citenamefont {Molina}, \citenamefont {Mai},\ and\
  \citenamefont {Döring}}]{Guo:2018zss}%
  \BibitemOpen
  \bibfield  {author} {\bibinfo {author} {\bibfnamefont {Dehua}\ \bibnamefont
  {Guo}}, \bibinfo {author} {\bibfnamefont {Andrei}\ \bibnamefont {Alexandru}},
  \bibinfo {author} {\bibfnamefont {Raquel}\ \bibnamefont {Molina}}, \bibinfo
  {author} {\bibfnamefont {Maxim}\ \bibnamefont {Mai}}, \ and\ \bibinfo
  {author} {\bibfnamefont {Michael}\ \bibnamefont {Döring}},\ }\bibfield
  {title} {\enquote {\bibinfo {title} {{Extraction of isoscalar $\pi\pi$
  phase-shifts from lattice QCD}},}\ }\href {\doibase
  10.1103/PhysRevD.98.014507} {\bibfield  {journal} {\bibinfo  {journal} {Phys.
  Rev.}\ }\textbf {\bibinfo {volume} {D98}},\ \bibinfo {pages} {014507}
  (\bibinfo {year} {2018})},\ \Eprint {http://arxiv.org/abs/1803.02897}
  {arXiv:1803.02897 [hep-lat]} \BibitemShut {NoStop}%
\bibitem [{\citenamefont {Andersen}\ \emph {et~al.}(2019)\citenamefont
  {Andersen}, \citenamefont {Bulava}, \citenamefont {Hörz},\ and\
  \citenamefont {Morningstar}}]{Andersen:2018mau}%
  \BibitemOpen
  \bibfield  {author} {\bibinfo {author} {\bibfnamefont {Christian}\
  \bibnamefont {Andersen}}, \bibinfo {author} {\bibfnamefont {John}\
  \bibnamefont {Bulava}}, \bibinfo {author} {\bibfnamefont {Ben}\ \bibnamefont
  {Hörz}}, \ and\ \bibinfo {author} {\bibfnamefont {Colin}\ \bibnamefont
  {Morningstar}},\ }\bibfield  {title} {\enquote {\bibinfo {title} {{The $I=1$
  pion-pion scattering amplitude and timelike pion form factor from $N_{\rm f}
  = 2+1$ lattice QCD}},}\ }\href {\doibase 10.1016/j.nuclphysb.2018.12.018}
  {\bibfield  {journal} {\bibinfo  {journal} {Nucl. Phys.}\ }\textbf {\bibinfo
  {volume} {B939}},\ \bibinfo {pages} {145--173} (\bibinfo {year} {2019})},\
  \Eprint {http://arxiv.org/abs/1808.05007} {arXiv:1808.05007 [hep-lat]}
  \BibitemShut {NoStop}%
\bibitem [{\citenamefont {Dudek}\ \emph {et~al.}(2014)\citenamefont {Dudek},
  \citenamefont {Edwards}, \citenamefont {Thomas},\ and\ \citenamefont
  {Wilson}}]{Dudek:2014qha}%
  \BibitemOpen
  \bibfield  {author} {\bibinfo {author} {\bibfnamefont {Jozef~J.}\
  \bibnamefont {Dudek}}, \bibinfo {author} {\bibfnamefont {Robert~G.}\
  \bibnamefont {Edwards}}, \bibinfo {author} {\bibfnamefont {Christopher~E.}\
  \bibnamefont {Thomas}}, \ and\ \bibinfo {author} {\bibfnamefont {David~J.}\
  \bibnamefont {Wilson}} (\bibinfo {collaboration} {Hadron Spectrum}),\
  }\bibfield  {title} {\enquote {\bibinfo {title} {{Resonances in coupled $\pi
  K -\eta K$ scattering from quantum chromodynamics}},}\ }\href {\doibase
  10.1103/PhysRevLett.113.182001} {\bibfield  {journal} {\bibinfo  {journal}
  {Phys. Rev. Lett.}\ }\textbf {\bibinfo {volume} {113}},\ \bibinfo {pages}
  {182001} (\bibinfo {year} {2014})},\ \Eprint {http://arxiv.org/abs/1406.4158}
  {arXiv:1406.4158 [hep-ph]} \BibitemShut {NoStop}%
\bibitem [{\citenamefont {Dudek}\ \emph {et~al.}(2016)\citenamefont {Dudek},
  \citenamefont {Edwards},\ and\ \citenamefont {Wilson}}]{Dudek:2016cru}%
  \BibitemOpen
  \bibfield  {author} {\bibinfo {author} {\bibfnamefont {Jozef~J.}\
  \bibnamefont {Dudek}}, \bibinfo {author} {\bibfnamefont {Robert~G.}\
  \bibnamefont {Edwards}}, \ and\ \bibinfo {author} {\bibfnamefont {David~J.}\
  \bibnamefont {Wilson}} (\bibinfo {collaboration} {Hadron Spectrum}),\
  }\bibfield  {title} {\enquote {\bibinfo {title} {{An $a_0$ resonance in
  strongly coupled $\pi \eta$, $K\overline{K}$ scattering from lattice QCD}},}\
  }\href {\doibase 10.1103/PhysRevD.93.094506} {\bibfield  {journal} {\bibinfo
  {journal} {Phys. Rev.}\ }\textbf {\bibinfo {volume} {D93}},\ \bibinfo {pages}
  {094506} (\bibinfo {year} {2016})},\ \Eprint
  {http://arxiv.org/abs/1602.05122} {arXiv:1602.05122 [hep-ph]} \BibitemShut
  {NoStop}%
\bibitem [{\citenamefont {Woss}\ \emph {et~al.}(2018)\citenamefont {Woss},
  \citenamefont {Thomas}, \citenamefont {Dudek}, \citenamefont {Edwards},\ and\
  \citenamefont {Wilson}}]{Woss:2018irj}%
  \BibitemOpen
  \bibfield  {author} {\bibinfo {author} {\bibfnamefont {Antoni}\ \bibnamefont
  {Woss}}, \bibinfo {author} {\bibfnamefont {Christopher~E.}\ \bibnamefont
  {Thomas}}, \bibinfo {author} {\bibfnamefont {Jozef~J.}\ \bibnamefont
  {Dudek}}, \bibinfo {author} {\bibfnamefont {Robert~G.}\ \bibnamefont
  {Edwards}}, \ and\ \bibinfo {author} {\bibfnamefont {David~J.}\ \bibnamefont
  {Wilson}},\ }\bibfield  {title} {\enquote {\bibinfo {title}
  {{Dynamically-coupled partial-waves in $\rho\pi$ isospin-2 scattering from
  lattice QCD}},}\ }\href {\doibase 10.1007/JHEP07(2018)043} {\bibfield
  {journal} {\bibinfo  {journal} {JHEP}\ }\textbf {\bibinfo {volume} {07}},\
  \bibinfo {pages} {043} (\bibinfo {year} {2018})},\ \Eprint
  {http://arxiv.org/abs/1802.05580} {arXiv:1802.05580 [hep-lat]} \BibitemShut
  {NoStop}%
\bibitem [{\citenamefont {Woss}\ \emph {et~al.}(2019)\citenamefont {Woss},
  \citenamefont {Thomas}, \citenamefont {Dudek}, \citenamefont {Edwards},\ and\
  \citenamefont {Wilson}}]{Woss:2019hse}%
  \BibitemOpen
  \bibfield  {author} {\bibinfo {author} {\bibfnamefont {Antoni~J.}\
  \bibnamefont {Woss}}, \bibinfo {author} {\bibfnamefont {Christopher~E.}\
  \bibnamefont {Thomas}}, \bibinfo {author} {\bibfnamefont {Jozef~J.}\
  \bibnamefont {Dudek}}, \bibinfo {author} {\bibfnamefont {Robert~G.}\
  \bibnamefont {Edwards}}, \ and\ \bibinfo {author} {\bibfnamefont {David~J.}\
  \bibnamefont {Wilson}},\ }\bibfield  {title} {\enquote {\bibinfo {title}
  {{The $b_1$ resonance in coupled $\pi\omega$, $\pi\phi$ scattering from
  lattice QCD}},}\ }\href@noop {} {\  (\bibinfo {year} {2019})},\ \Eprint
  {http://arxiv.org/abs/1904.04136} {arXiv:1904.04136 [hep-lat]} \BibitemShut
  {NoStop}%
\bibitem [{\citenamefont {Helmes}\ \emph {et~al.}(2018)\citenamefont {Helmes},
  \citenamefont {Jost}, \citenamefont {Knippschild}, \citenamefont {Kostrzewa},
  \citenamefont {Liu}, \citenamefont {Pittler}, \citenamefont {Urbach},\ and\
  \citenamefont {Werner}}]{Helmes:2018nug}%
  \BibitemOpen
  \bibfield  {author} {\bibinfo {author} {\bibfnamefont {Christopher}\
  \bibnamefont {Helmes}}, \bibinfo {author} {\bibfnamefont {Christian}\
  \bibnamefont {Jost}}, \bibinfo {author} {\bibfnamefont {Bastian}\
  \bibnamefont {Knippschild}}, \bibinfo {author} {\bibfnamefont {Bartosz}\
  \bibnamefont {Kostrzewa}}, \bibinfo {author} {\bibfnamefont {Liuming}\
  \bibnamefont {Liu}}, \bibinfo {author} {\bibfnamefont {Ferenc}\ \bibnamefont
  {Pittler}}, \bibinfo {author} {\bibfnamefont {Carsten}\ \bibnamefont
  {Urbach}}, \ and\ \bibinfo {author} {\bibfnamefont {Markus}\ \bibnamefont
  {Werner}} (\bibinfo {collaboration} {ETM}),\ }\bibfield  {title} {\enquote
  {\bibinfo {title} {{Hadron-Hadron Interactions from $N_f=2+1+1$ Lattice QCD:
  $I=3/2$ $\pi K$ Scattering Length}},}\ }\href {\doibase
  10.1103/PhysRevD.98.114511} {\bibfield  {journal} {\bibinfo  {journal} {Phys.
  Rev.}\ }\textbf {\bibinfo {volume} {D98}},\ \bibinfo {pages} {114511}
  (\bibinfo {year} {2018})},\ \Eprint {http://arxiv.org/abs/1809.08886}
  {arXiv:1809.08886 [hep-lat]} \BibitemShut {NoStop}%
\bibitem [{\citenamefont {Liu}\ \emph {et~al.}(2017)\citenamefont {Liu} \emph
  {et~al.}}]{Liu:2016cba}%
  \BibitemOpen
  \bibfield  {author} {\bibinfo {author} {\bibfnamefont {L.}~\bibnamefont
  {Liu}} \emph {et~al.},\ }\bibfield  {title} {\enquote {\bibinfo {title}
  {{Isospin-0 $\pi\pi$ s-wave scattering length from twisted mass lattice
  QCD}},}\ }\href {\doibase 10.1103/PhysRevD.96.054516} {\bibfield  {journal}
  {\bibinfo  {journal} {Phys. Rev.}\ }\textbf {\bibinfo {volume} {D96}},\
  \bibinfo {pages} {054516} (\bibinfo {year} {2017})},\ \Eprint
  {http://arxiv.org/abs/1612.02061} {arXiv:1612.02061 [hep-lat]} \BibitemShut
  {NoStop}%
\bibitem [{\citenamefont {Helmes}\ \emph {et~al.}(2017)\citenamefont {Helmes},
  \citenamefont {Jost}, \citenamefont {Knippschild}, \citenamefont {Kostrzewa},
  \citenamefont {Liu}, \citenamefont {Urbach},\ and\ \citenamefont
  {Werner}}]{Helmes:2017smr}%
  \BibitemOpen
  \bibfield  {author} {\bibinfo {author} {\bibfnamefont {Christopher}\
  \bibnamefont {Helmes}}, \bibinfo {author} {\bibfnamefont {Christian}\
  \bibnamefont {Jost}}, \bibinfo {author} {\bibfnamefont {Bastian}\
  \bibnamefont {Knippschild}}, \bibinfo {author} {\bibfnamefont {Bartosz}\
  \bibnamefont {Kostrzewa}}, \bibinfo {author} {\bibfnamefont {Liuming}\
  \bibnamefont {Liu}}, \bibinfo {author} {\bibfnamefont {Carsten}\ \bibnamefont
  {Urbach}}, \ and\ \bibinfo {author} {\bibfnamefont {Markus}\ \bibnamefont
  {Werner}},\ }\bibfield  {title} {\enquote {\bibinfo {title} {{Hadron-Hadron
  Interactions from $N_f=2+1+1$ lattice QCD: Isospin-1 $KK$ scattering
  length}},}\ }\href {\doibase 10.1103/PhysRevD.96.034510} {\bibfield
  {journal} {\bibinfo  {journal} {Phys. Rev.}\ }\textbf {\bibinfo {volume}
  {D96}},\ \bibinfo {pages} {034510} (\bibinfo {year} {2017})},\ \Eprint
  {http://arxiv.org/abs/1703.04737} {arXiv:1703.04737 [hep-lat]} \BibitemShut
  {NoStop}%
\bibitem [{\citenamefont {Werner}\ \emph {et~al.}(2020)\citenamefont {Werner}
  \emph {et~al.}}]{Werner:2019hxc}%
  \BibitemOpen
  \bibfield  {author} {\bibinfo {author} {\bibfnamefont {Markus}\ \bibnamefont
  {Werner}} \emph {et~al.},\ }\bibfield  {title} {\enquote {\bibinfo {title}
  {{Hadron-Hadron Interactions from $N_f=2+1+1$ Lattice QCD: The
  $\rho$-resonance}},}\ }\href {\doibase 10.1140/epja/s10050-020-00057-4}
  {\bibfield  {journal} {\bibinfo  {journal} {Eur. Phys. J. A}\ }\textbf
  {\bibinfo {volume} {56}},\ \bibinfo {pages} {61} (\bibinfo {year} {2020})},\
  \Eprint {http://arxiv.org/abs/1907.01237} {arXiv:1907.01237 [hep-lat]}
  \BibitemShut {NoStop}%
\bibitem [{\citenamefont {Culver}\ \emph {et~al.}(2019)\citenamefont {Culver},
  \citenamefont {Mai}, \citenamefont {Alexandru}, \citenamefont {Döring},\
  and\ \citenamefont {Lee}}]{Culver:2019qtx}%
  \BibitemOpen
  \bibfield  {author} {\bibinfo {author} {\bibfnamefont {C.}~\bibnamefont
  {Culver}}, \bibinfo {author} {\bibfnamefont {M.}~\bibnamefont {Mai}},
  \bibinfo {author} {\bibfnamefont {A.}~\bibnamefont {Alexandru}}, \bibinfo
  {author} {\bibfnamefont {M.}~\bibnamefont {Döring}}, \ and\ \bibinfo
  {author} {\bibfnamefont {F.~X.}\ \bibnamefont {Lee}},\ }\bibfield  {title}
  {\enquote {\bibinfo {title} {{Pion scattering in the isospin I=2 channel from
  elongated lattices}},}\ }\href@noop {} {\  (\bibinfo {year} {2019})},\
  \Eprint {http://arxiv.org/abs/1905.10202} {arXiv:1905.10202 [hep-lat]}
  \BibitemShut {NoStop}%
\bibitem [{\citenamefont {Mai}\ \emph {et~al.}(2019)\citenamefont {Mai},
  \citenamefont {Culver}, \citenamefont {Alexandru}, \citenamefont {Döring},\
  and\ \citenamefont {Lee}}]{Mai:2019pqr}%
  \BibitemOpen
  \bibfield  {author} {\bibinfo {author} {\bibfnamefont {Maxim}\ \bibnamefont
  {Mai}}, \bibinfo {author} {\bibfnamefont {Chris}\ \bibnamefont {Culver}},
  \bibinfo {author} {\bibfnamefont {Andrei}\ \bibnamefont {Alexandru}},
  \bibinfo {author} {\bibfnamefont {Michael}\ \bibnamefont {Döring}}, \ and\
  \bibinfo {author} {\bibfnamefont {Frank~X.}\ \bibnamefont {Lee}},\ }\bibfield
   {title} {\enquote {\bibinfo {title} {{A cross-channel study of pion
  scattering from lattice QCD}},}\ }\href@noop {} {\  (\bibinfo {year}
  {2019})},\ \Eprint {http://arxiv.org/abs/1908.01847} {arXiv:1908.01847
  [hep-lat]} \BibitemShut {NoStop}%
\bibitem [{\citenamefont {Doring}\ \emph {et~al.}(2012)\citenamefont {Doring},
  \citenamefont {Meißner}, \citenamefont {Oset},\ and\ \citenamefont
  {Rusetsky}}]{Doring:2012eu}%
  \BibitemOpen
  \bibfield  {author} {\bibinfo {author} {\bibfnamefont {M.}~\bibnamefont
  {Doring}}, \bibinfo {author} {\bibfnamefont {U.~G.}\ \bibnamefont
  {Meißner}}, \bibinfo {author} {\bibfnamefont {E.}~\bibnamefont {Oset}}, \
  and\ \bibinfo {author} {\bibfnamefont {A.}~\bibnamefont {Rusetsky}},\
  }\bibfield  {title} {\enquote {\bibinfo {title} {{Scalar mesons moving in a
  finite volume and the role of partial wave mixing}},}\ }\href {\doibase
  10.1140/epja/i2012-12114-6} {\bibfield  {journal} {\bibinfo  {journal} {Eur.
  Phys. J.}\ }\textbf {\bibinfo {volume} {A48}},\ \bibinfo {pages} {114}
  (\bibinfo {year} {2012})},\ \Eprint {http://arxiv.org/abs/1205.4838}
  {arXiv:1205.4838 [hep-lat]} \BibitemShut {NoStop}%
\bibitem [{\citenamefont {Fischer}\ \emph
  {et~al.}(2020{\natexlab{a}})\citenamefont {Fischer}, \citenamefont
  {Kostrzewa}, \citenamefont {Mai}, \citenamefont {Petschlies}, \citenamefont
  {Pittler}, \citenamefont {Ueding}, \citenamefont {Urbach},\ and\
  \citenamefont {Werner}}]{Fischer:2020fvl}%
  \BibitemOpen
  \bibfield  {author} {\bibinfo {author} {\bibfnamefont {Matthias}\
  \bibnamefont {Fischer}}, \bibinfo {author} {\bibfnamefont {Bartosz}\
  \bibnamefont {Kostrzewa}}, \bibinfo {author} {\bibfnamefont {Maxim}\
  \bibnamefont {Mai}}, \bibinfo {author} {\bibfnamefont {Marcus}\ \bibnamefont
  {Petschlies}}, \bibinfo {author} {\bibfnamefont {Ferenc}\ \bibnamefont
  {Pittler}}, \bibinfo {author} {\bibfnamefont {Martin}\ \bibnamefont
  {Ueding}}, \bibinfo {author} {\bibfnamefont {Carsten}\ \bibnamefont
  {Urbach}}, \ and\ \bibinfo {author} {\bibfnamefont {Markus}\ \bibnamefont
  {Werner}} (\bibinfo {collaboration} {ETM}),\ }\bibfield  {title} {\enquote
  {\bibinfo {title} {{The $\rho$-resonance with physical pion mass from $N_f=2$
  lattice QCD}},}\ }\href@noop {} {\  (\bibinfo {year} {2020}{\natexlab{a}})},\
  \Eprint {http://arxiv.org/abs/2006.13805} {arXiv:2006.13805 [hep-lat]}
  \BibitemShut {NoStop}%
\bibitem [{\citenamefont {Woss}\ \emph {et~al.}(2020)\citenamefont {Woss},
  \citenamefont {Wilson},\ and\ \citenamefont {Dudek}}]{Woss:2020cmp}%
  \BibitemOpen
  \bibfield  {author} {\bibinfo {author} {\bibfnamefont {Antoni~J.}\
  \bibnamefont {Woss}}, \bibinfo {author} {\bibfnamefont {David~J.}\
  \bibnamefont {Wilson}}, \ and\ \bibinfo {author} {\bibfnamefont {Jozef~J.}\
  \bibnamefont {Dudek}} (\bibinfo {collaboration} {Hadron Spectrum}),\
  }\bibfield  {title} {\enquote {\bibinfo {title} {{Efficient solution of the
  multichannel Lüscher determinant condition through eigenvalue
  decomposition}},}\ }\href {\doibase 10.1103/PhysRevD.101.114505} {\bibfield
  {journal} {\bibinfo  {journal} {Phys. Rev. D}\ }\textbf {\bibinfo {volume}
  {101}},\ \bibinfo {pages} {114505} (\bibinfo {year} {2020})},\ \Eprint
  {http://arxiv.org/abs/2001.08474} {arXiv:2001.08474 [hep-lat]} \BibitemShut
  {NoStop}%
\bibitem [{\citenamefont {Bulava}\ \emph {et~al.}(2016)\citenamefont {Bulava},
  \citenamefont {Fahy}, \citenamefont {Hörz}, \citenamefont {Juge},
  \citenamefont {Morningstar},\ and\ \citenamefont {Wong}}]{Bulava:2016mks}%
  \BibitemOpen
  \bibfield  {author} {\bibinfo {author} {\bibfnamefont {John}\ \bibnamefont
  {Bulava}}, \bibinfo {author} {\bibfnamefont {Brendan}\ \bibnamefont {Fahy}},
  \bibinfo {author} {\bibfnamefont {Ben}\ \bibnamefont {Hörz}}, \bibinfo
  {author} {\bibfnamefont {Keisuke~J.}\ \bibnamefont {Juge}}, \bibinfo {author}
  {\bibfnamefont {Colin}\ \bibnamefont {Morningstar}}, \ and\ \bibinfo {author}
  {\bibfnamefont {Chik~Him}\ \bibnamefont {Wong}},\ }\bibfield  {title}
  {\enquote {\bibinfo {title} {{$I=1$ and $I=2$ $\pi-\pi$ scattering phase
  shifts from $N_{\mathrm{f}} = 2+1$ lattice QCD}},}\ }\href {\doibase
  10.1016/j.nuclphysb.2016.07.024} {\bibfield  {journal} {\bibinfo  {journal}
  {Nucl. Phys.}\ }\textbf {\bibinfo {volume} {B910}},\ \bibinfo {pages}
  {842--867} (\bibinfo {year} {2016})},\ \Eprint
  {http://arxiv.org/abs/1604.05593} {arXiv:1604.05593 [hep-lat]} \BibitemShut
  {NoStop}%
\bibitem [{\citenamefont {Rendon}\ \emph {et~al.}(2020)\citenamefont {Rendon},
  \citenamefont {Leskovec}, \citenamefont {Meinel}, \citenamefont {Negele},
  \citenamefont {Paul}, \citenamefont {Petschlies}, \citenamefont {Pochinsky},
  \citenamefont {Silvi},\ and\ \citenamefont {Syritsyn}}]{Rendon:2020rtw}%
  \BibitemOpen
  \bibfield  {author} {\bibinfo {author} {\bibfnamefont {Gumaro}\ \bibnamefont
  {Rendon}}, \bibinfo {author} {\bibfnamefont {Luka}\ \bibnamefont {Leskovec}},
  \bibinfo {author} {\bibfnamefont {Stefan}\ \bibnamefont {Meinel}}, \bibinfo
  {author} {\bibfnamefont {John}\ \bibnamefont {Negele}}, \bibinfo {author}
  {\bibfnamefont {Srijit}\ \bibnamefont {Paul}}, \bibinfo {author}
  {\bibfnamefont {Marcus}\ \bibnamefont {Petschlies}}, \bibinfo {author}
  {\bibfnamefont {Andrew}\ \bibnamefont {Pochinsky}}, \bibinfo {author}
  {\bibfnamefont {Giorgio}\ \bibnamefont {Silvi}}, \ and\ \bibinfo {author}
  {\bibfnamefont {Sergey}\ \bibnamefont {Syritsyn}},\ }\bibfield  {title}
  {\enquote {\bibinfo {title} {{$I=1/2$ $S$-wave and $P$-wave $K\pi$ scattering
  and the $\kappa$ and $K^*$ resonances from lattice QCD}},}\ }\href@noop {} {\
   (\bibinfo {year} {2020})},\ \Eprint {http://arxiv.org/abs/2006.14035}
  {arXiv:2006.14035 [hep-lat]} \BibitemShut {NoStop}%
\bibitem [{\citenamefont {Alexandrou}\ \emph {et~al.}(2017)\citenamefont
  {Alexandrou}, \citenamefont {Leskovec}, \citenamefont {Meinel}, \citenamefont
  {Negele}, \citenamefont {Paul}, \citenamefont {Petschlies}, \citenamefont
  {Pochinsky}, \citenamefont {Rendon},\ and\ \citenamefont
  {Syritsyn}}]{Alexandrou:2017mpi}%
  \BibitemOpen
  \bibfield  {author} {\bibinfo {author} {\bibfnamefont {Constantia}\
  \bibnamefont {Alexandrou}}, \bibinfo {author} {\bibfnamefont {Luka}\
  \bibnamefont {Leskovec}}, \bibinfo {author} {\bibfnamefont {Stefan}\
  \bibnamefont {Meinel}}, \bibinfo {author} {\bibfnamefont {John}\ \bibnamefont
  {Negele}}, \bibinfo {author} {\bibfnamefont {Srijit}\ \bibnamefont {Paul}},
  \bibinfo {author} {\bibfnamefont {Marcus}\ \bibnamefont {Petschlies}},
  \bibinfo {author} {\bibfnamefont {Andrew}\ \bibnamefont {Pochinsky}},
  \bibinfo {author} {\bibfnamefont {Gumaro}\ \bibnamefont {Rendon}}, \ and\
  \bibinfo {author} {\bibfnamefont {Sergey}\ \bibnamefont {Syritsyn}},\
  }\bibfield  {title} {\enquote {\bibinfo {title} {{$P$-wave $\pi\pi$
  scattering and the $\rho$ resonance from lattice QCD}},}\ }\href {\doibase
  10.1103/PhysRevD.96.034525} {\bibfield  {journal} {\bibinfo  {journal} {Phys.
  Rev. D}\ }\textbf {\bibinfo {volume} {96}},\ \bibinfo {pages} {034525}
  (\bibinfo {year} {2017})},\ \Eprint {http://arxiv.org/abs/1704.05439}
  {arXiv:1704.05439 [hep-lat]} \BibitemShut {NoStop}%
\bibitem [{\citenamefont {Brice{\~n}o}\ \emph {et~al.}(2018)\citenamefont
  {Brice{\~n}o}, \citenamefont {Dudek},\ and\ \citenamefont
  {Young}}]{Briceno:2017max}%
  \BibitemOpen
  \bibfield  {author} {\bibinfo {author} {\bibfnamefont {Raul~A.}\ \bibnamefont
  {Brice{\~n}o}}, \bibinfo {author} {\bibfnamefont {Jozef~J.}\ \bibnamefont
  {Dudek}}, \ and\ \bibinfo {author} {\bibfnamefont {Ross~D.}\ \bibnamefont
  {Young}},\ }\bibfield  {title} {\enquote {\bibinfo {title} {{Scattering
  processes and resonances from lattice QCD}},}\ }\href {\doibase
  10.1103/RevModPhys.90.025001} {\bibfield  {journal} {\bibinfo  {journal}
  {Rev. Mod. Phys.}\ }\textbf {\bibinfo {volume} {90}},\ \bibinfo {pages}
  {025001} (\bibinfo {year} {2018})},\ \Eprint
  {http://arxiv.org/abs/1706.06223} {arXiv:1706.06223 [hep-lat]} \BibitemShut
  {NoStop}%
\bibitem [{\citenamefont {Hansen}\ and\ \citenamefont
  {Sharpe}(2014)}]{Hansen:2014eka}%
  \BibitemOpen
  \bibfield  {author} {\bibinfo {author} {\bibfnamefont {Maxwell~T.}\
  \bibnamefont {Hansen}}\ and\ \bibinfo {author} {\bibfnamefont {Stephen~R.}\
  \bibnamefont {Sharpe}},\ }\bibfield  {title} {\enquote {\bibinfo {title}
  {{Relativistic, model-independent, three-particle quantization condition}},}\
  }\href {\doibase 10.1103/PhysRevD.90.116003} {\bibfield  {journal} {\bibinfo
  {journal} {Phys. Rev. D}\ }\textbf {\bibinfo {volume} {90}},\ \bibinfo
  {pages} {116003} (\bibinfo {year} {2014})},\ \Eprint
  {http://arxiv.org/abs/1408.5933} {arXiv:1408.5933 [hep-lat]} \BibitemShut
  {NoStop}%
\bibitem [{\citenamefont {Hansen}\ and\ \citenamefont
  {Sharpe}(2015)}]{Hansen:2015zga}%
  \BibitemOpen
  \bibfield  {author} {\bibinfo {author} {\bibfnamefont {Maxwell~T.}\
  \bibnamefont {Hansen}}\ and\ \bibinfo {author} {\bibfnamefont {Stephen~R.}\
  \bibnamefont {Sharpe}},\ }\bibfield  {title} {\enquote {\bibinfo {title}
  {{Expressing the three-particle finite-volume spectrum in terms of the
  three-to-three scattering amplitude}},}\ }\href {\doibase
  10.1103/PhysRevD.92.114509} {\bibfield  {journal} {\bibinfo  {journal} {Phys.
  Rev. D}\ }\textbf {\bibinfo {volume} {92}},\ \bibinfo {pages} {114509}
  (\bibinfo {year} {2015})},\ \Eprint {http://arxiv.org/abs/1504.04248}
  {arXiv:1504.04248 [hep-lat]} \BibitemShut {NoStop}%
\bibitem [{\citenamefont {Hansen}\ and\ \citenamefont
  {Sharpe}(2016{\natexlab{a}})}]{Hansen:2015zta}%
  \BibitemOpen
  \bibfield  {author} {\bibinfo {author} {\bibfnamefont {Maxwell~T.}\
  \bibnamefont {Hansen}}\ and\ \bibinfo {author} {\bibfnamefont {Stephen~R.}\
  \bibnamefont {Sharpe}},\ }\bibfield  {title} {\enquote {\bibinfo {title}
  {{Perturbative results for two and three particle threshold energies in
  finite volume}},}\ }\href {\doibase 10.1103/PhysRevD.93.014506} {\bibfield
  {journal} {\bibinfo  {journal} {Phys. Rev. D}\ }\textbf {\bibinfo {volume}
  {93}},\ \bibinfo {pages} {014506} (\bibinfo {year} {2016}{\natexlab{a}})},\
  \Eprint {http://arxiv.org/abs/1509.07929} {arXiv:1509.07929 [hep-lat]}
  \BibitemShut {NoStop}%
\bibitem [{\citenamefont {Hansen}\ and\ \citenamefont
  {Sharpe}(2016{\natexlab{b}})}]{Hansen:2016fzj}%
  \BibitemOpen
  \bibfield  {author} {\bibinfo {author} {\bibfnamefont {Maxwell~T.}\
  \bibnamefont {Hansen}}\ and\ \bibinfo {author} {\bibfnamefont {Stephen~R.}\
  \bibnamefont {Sharpe}},\ }\bibfield  {title} {\enquote {\bibinfo {title}
  {{Threshold expansion of the three-particle quantization condition}},}\
  }\href {\doibase 10.1103/PhysRevD.93.096006} {\bibfield  {journal} {\bibinfo
  {journal} {Phys. Rev. D}\ }\textbf {\bibinfo {volume} {93}},\ \bibinfo
  {pages} {096006} (\bibinfo {year} {2016}{\natexlab{b}})},\ \bibinfo {note}
  {[Erratum: Phys.Rev.D 96, 039901 (2017)]},\ \Eprint
  {http://arxiv.org/abs/1602.00324} {arXiv:1602.00324 [hep-lat]} \BibitemShut
  {NoStop}%
\bibitem [{\citenamefont {Briceño}\ \emph {et~al.}(2017)\citenamefont
  {Briceño}, \citenamefont {Hansen},\ and\ \citenamefont
  {Sharpe}}]{Briceno:2017tce}%
  \BibitemOpen
  \bibfield  {author} {\bibinfo {author} {\bibfnamefont {Raúl~A.}\
  \bibnamefont {Briceño}}, \bibinfo {author} {\bibfnamefont {Maxwell~T.}\
  \bibnamefont {Hansen}}, \ and\ \bibinfo {author} {\bibfnamefont {Stephen~R.}\
  \bibnamefont {Sharpe}},\ }\bibfield  {title} {\enquote {\bibinfo {title}
  {{Relating the finite-volume spectrum and the two-and-three-particle $S$
  matrix for relativistic systems of identical scalar particles}},}\ }\href
  {\doibase 10.1103/PhysRevD.95.074510} {\bibfield  {journal} {\bibinfo
  {journal} {Phys. Rev. D}\ }\textbf {\bibinfo {volume} {95}},\ \bibinfo
  {pages} {074510} (\bibinfo {year} {2017})},\ \Eprint
  {http://arxiv.org/abs/1701.07465} {arXiv:1701.07465 [hep-lat]} \BibitemShut
  {NoStop}%
\bibitem [{\citenamefont {Briceño}\ \emph {et~al.}(2018)\citenamefont
  {Briceño}, \citenamefont {Hansen},\ and\ \citenamefont
  {Sharpe}}]{Briceno:2018mlh}%
  \BibitemOpen
  \bibfield  {author} {\bibinfo {author} {\bibfnamefont {Raúl~A.}\
  \bibnamefont {Briceño}}, \bibinfo {author} {\bibfnamefont {Maxwell~T.}\
  \bibnamefont {Hansen}}, \ and\ \bibinfo {author} {\bibfnamefont {Stephen~R.}\
  \bibnamefont {Sharpe}},\ }\bibfield  {title} {\enquote {\bibinfo {title}
  {{Numerical study of the relativistic three-body quantization condition in
  the isotropic approximation}},}\ }\href {\doibase 10.1103/PhysRevD.98.014506}
  {\bibfield  {journal} {\bibinfo  {journal} {Phys. Rev. D}\ }\textbf {\bibinfo
  {volume} {98}},\ \bibinfo {pages} {014506} (\bibinfo {year} {2018})},\
  \Eprint {http://arxiv.org/abs/1803.04169} {arXiv:1803.04169 [hep-lat]}
  \BibitemShut {NoStop}%
\bibitem [{\citenamefont {Brice\~no}\ \emph {et~al.}(2019)\citenamefont
  {Brice\~no}, \citenamefont {Hansen},\ and\ \citenamefont
  {Sharpe}}]{Briceno:2018aml}%
  \BibitemOpen
  \bibfield  {author} {\bibinfo {author} {\bibfnamefont {Raúl~A.}\
  \bibnamefont {Brice\~no}}, \bibinfo {author} {\bibfnamefont {Maxwell~T.}\
  \bibnamefont {Hansen}}, \ and\ \bibinfo {author} {\bibfnamefont {Stephen~R.}\
  \bibnamefont {Sharpe}},\ }\bibfield  {title} {\enquote {\bibinfo {title}
  {{Three-particle systems with resonant subprocesses in a finite volume}},}\
  }\href {\doibase 10.1103/PhysRevD.99.014516} {\bibfield  {journal} {\bibinfo
  {journal} {Phys. Rev.}\ }\textbf {\bibinfo {volume} {D99}},\ \bibinfo {pages}
  {014516} (\bibinfo {year} {2019})},\ \Eprint
  {http://arxiv.org/abs/1810.01429} {arXiv:1810.01429 [hep-lat]} \BibitemShut
  {NoStop}%
\bibitem [{\citenamefont {Blanton}\ \emph {et~al.}(2019)\citenamefont
  {Blanton}, \citenamefont {Romero-L\'opez},\ and\ \citenamefont
  {Sharpe}}]{Blanton:2019igq}%
  \BibitemOpen
  \bibfield  {author} {\bibinfo {author} {\bibfnamefont {Tyler~D.}\
  \bibnamefont {Blanton}}, \bibinfo {author} {\bibfnamefont {Fernando}\
  \bibnamefont {Romero-L\'opez}}, \ and\ \bibinfo {author} {\bibfnamefont
  {Stephen~R.}\ \bibnamefont {Sharpe}},\ }\bibfield  {title} {\enquote
  {\bibinfo {title} {{Implementing the three-particle quantization condition
  including higher partial waves}},}\ }\href {\doibase 10.1007/JHEP03(2019)106}
  {\bibfield  {journal} {\bibinfo  {journal} {JHEP}\ }\textbf {\bibinfo
  {volume} {03}},\ \bibinfo {pages} {106} (\bibinfo {year} {2019})},\ \Eprint
  {http://arxiv.org/abs/1901.07095} {arXiv:1901.07095 [hep-lat]} \BibitemShut
  {NoStop}%
\bibitem [{\citenamefont {Romero-López}\ \emph {et~al.}(2019)\citenamefont
  {Romero-López}, \citenamefont {Sharpe}, \citenamefont {Blanton},
  \citenamefont {Briceño},\ and\ \citenamefont
  {Hansen}}]{Romero-Lopez:2019qrt}%
  \BibitemOpen
  \bibfield  {author} {\bibinfo {author} {\bibfnamefont {Fernando}\
  \bibnamefont {Romero-López}}, \bibinfo {author} {\bibfnamefont {Stephen~R.}\
  \bibnamefont {Sharpe}}, \bibinfo {author} {\bibfnamefont {Tyler~D.}\
  \bibnamefont {Blanton}}, \bibinfo {author} {\bibfnamefont {Raúl~A.}\
  \bibnamefont {Briceño}}, \ and\ \bibinfo {author} {\bibfnamefont
  {Maxwell~T.}\ \bibnamefont {Hansen}},\ }\bibfield  {title} {\enquote
  {\bibinfo {title} {{Numerical exploration of three relativistic particles in
  a finite volume including two-particle resonances and bound states}},}\
  }\href {\doibase 10.1007/JHEP10(2019)007} {\bibfield  {journal} {\bibinfo
  {journal} {JHEP}\ }\textbf {\bibinfo {volume} {10}},\ \bibinfo {pages} {007}
  (\bibinfo {year} {2019})},\ \Eprint {http://arxiv.org/abs/1908.02411}
  {arXiv:1908.02411 [hep-lat]} \BibitemShut {NoStop}%
\bibitem [{\citenamefont {Hansen}\ \emph {et~al.}(2020)\citenamefont {Hansen},
  \citenamefont {Romero-López},\ and\ \citenamefont
  {Sharpe}}]{Hansen:2020zhy}%
  \BibitemOpen
  \bibfield  {author} {\bibinfo {author} {\bibfnamefont {Maxwell~T.}\
  \bibnamefont {Hansen}}, \bibinfo {author} {\bibfnamefont {Fernando}\
  \bibnamefont {Romero-López}}, \ and\ \bibinfo {author} {\bibfnamefont
  {Stephen~R.}\ \bibnamefont {Sharpe}},\ }\bibfield  {title} {\enquote
  {\bibinfo {title} {{Generalizing the relativistic quantization condition to
  include all three-pion isospin channels}},}\ }\href {\doibase
  10.1007/JHEP07(2020)047} {\bibfield  {journal} {\bibinfo  {journal} {JHEP}\
  }\textbf {\bibinfo {volume} {20}},\ \bibinfo {pages} {047} (\bibinfo {year}
  {2020})},\ \Eprint {http://arxiv.org/abs/2003.10974} {arXiv:2003.10974
  [hep-lat]} \BibitemShut {NoStop}%
\bibitem [{\citenamefont {Blanton}\ and\ \citenamefont
  {Sharpe}(2020{\natexlab{a}})}]{Blanton:2020jnm}%
  \BibitemOpen
  \bibfield  {author} {\bibinfo {author} {\bibfnamefont {Tyler~D.}\
  \bibnamefont {Blanton}}\ and\ \bibinfo {author} {\bibfnamefont {Stephen~R.}\
  \bibnamefont {Sharpe}},\ }\bibfield  {title} {\enquote {\bibinfo {title}
  {{Equivalence of relativistic three-particle quantization conditions}},}\
  }\href@noop {} {\  (\bibinfo {year} {2020}{\natexlab{a}})},\ \Eprint
  {http://arxiv.org/abs/2007.16190} {arXiv:2007.16190 [hep-lat]} \BibitemShut
  {NoStop}%
\bibitem [{\citenamefont {Blanton}\ and\ \citenamefont
  {Sharpe}(2020{\natexlab{b}})}]{Blanton:2020gha}%
  \BibitemOpen
  \bibfield  {author} {\bibinfo {author} {\bibfnamefont {Tyler~D.}\
  \bibnamefont {Blanton}}\ and\ \bibinfo {author} {\bibfnamefont {Stephen~R.}\
  \bibnamefont {Sharpe}},\ }\bibfield  {title} {\enquote {\bibinfo {title}
  {{Alternative derivation of the relativistic, three-particle quantization
  condition}},}\ }\href@noop {} {\  (\bibinfo {year} {2020}{\natexlab{b}})},\
  \Eprint {http://arxiv.org/abs/2007.16188} {arXiv:2007.16188 [hep-lat]}
  \BibitemShut {NoStop}%
\bibitem [{\citenamefont {Polejaeva}\ and\ \citenamefont
  {Rusetsky}(2012)}]{Polejaeva:2012ut}%
  \BibitemOpen
  \bibfield  {author} {\bibinfo {author} {\bibfnamefont {K.}~\bibnamefont
  {Polejaeva}}\ and\ \bibinfo {author} {\bibfnamefont {A.}~\bibnamefont
  {Rusetsky}},\ }\bibfield  {title} {\enquote {\bibinfo {title} {{Three
  particles in a finite volume}},}\ }\href {\doibase
  10.1140/epja/i2012-12067-8} {\bibfield  {journal} {\bibinfo  {journal} {Eur.
  Phys. J. A}\ }\textbf {\bibinfo {volume} {48}},\ \bibinfo {pages} {67}
  (\bibinfo {year} {2012})},\ \Eprint {http://arxiv.org/abs/1203.1241}
  {arXiv:1203.1241 [hep-lat]} \BibitemShut {NoStop}%
\bibitem [{\citenamefont {Meißner}\ \emph {et~al.}(2015)\citenamefont
  {Meißner}, \citenamefont {R{\'i}os},\ and\ \citenamefont
  {Rusetsky}}]{Meissner:2014dea}%
  \BibitemOpen
  \bibfield  {author} {\bibinfo {author} {\bibfnamefont {Ulf-G.}\ \bibnamefont
  {Meißner}}, \bibinfo {author} {\bibfnamefont {Guillermo}\ \bibnamefont
  {R{\'i}os}}, \ and\ \bibinfo {author} {\bibfnamefont {Akaki}\ \bibnamefont
  {Rusetsky}},\ }\bibfield  {title} {\enquote {\bibinfo {title} {{Spectrum of
  three-body bound states in a finite volume}},}\ }\href {\doibase
  10.1103/PhysRevLett.117.069902, 10.1103/PhysRevLett.114.091602} {\bibfield
  {journal} {\bibinfo  {journal} {Phys. Rev. Lett.}\ }\textbf {\bibinfo
  {volume} {114}},\ \bibinfo {pages} {091602} (\bibinfo {year} {2015})},\
  \bibinfo {note} {[Erratum: Phys. Rev. Lett.117,no.6,069902(2016)]},\ \Eprint
  {http://arxiv.org/abs/1412.4969} {arXiv:1412.4969 [hep-lat]} \BibitemShut
  {NoStop}%
\bibitem [{\citenamefont {Hammer}\ \emph
  {et~al.}(2017{\natexlab{a}})\citenamefont {Hammer}, \citenamefont {Pang},\
  and\ \citenamefont {Rusetsky}}]{Hammer:2017uqm}%
  \BibitemOpen
  \bibfield  {author} {\bibinfo {author} {\bibfnamefont {Hans-Werner}\
  \bibnamefont {Hammer}}, \bibinfo {author} {\bibfnamefont {Jin-Yi}\
  \bibnamefont {Pang}}, \ and\ \bibinfo {author} {\bibfnamefont
  {A.}~\bibnamefont {Rusetsky}},\ }\bibfield  {title} {\enquote {\bibinfo
  {title} {{Three-particle quantization condition in a finite volume: 1. The
  role of the three-particle force}},}\ }\href {\doibase
  10.1007/JHEP09(2017)109} {\bibfield  {journal} {\bibinfo  {journal} {JHEP}\
  }\textbf {\bibinfo {volume} {09}},\ \bibinfo {pages} {109} (\bibinfo {year}
  {2017}{\natexlab{a}})},\ \Eprint {http://arxiv.org/abs/1706.07700}
  {arXiv:1706.07700 [hep-lat]} \BibitemShut {NoStop}%
\bibitem [{\citenamefont {Hammer}\ \emph
  {et~al.}(2017{\natexlab{b}})\citenamefont {Hammer}, \citenamefont {Pang},\
  and\ \citenamefont {Rusetsky}}]{Hammer:2017kms}%
  \BibitemOpen
  \bibfield  {author} {\bibinfo {author} {\bibfnamefont {H.~W.}\ \bibnamefont
  {Hammer}}, \bibinfo {author} {\bibfnamefont {J.~Y.}\ \bibnamefont {Pang}}, \
  and\ \bibinfo {author} {\bibfnamefont {A.}~\bibnamefont {Rusetsky}},\
  }\bibfield  {title} {\enquote {\bibinfo {title} {{Three particle quantization
  condition in a finite volume: 2. general formalism and the analysis of
  data}},}\ }\href {\doibase 10.1007/JHEP10(2017)115} {\bibfield  {journal}
  {\bibinfo  {journal} {JHEP}\ }\textbf {\bibinfo {volume} {10}},\ \bibinfo
  {pages} {115} (\bibinfo {year} {2017}{\natexlab{b}})},\ \Eprint
  {http://arxiv.org/abs/1707.02176} {arXiv:1707.02176 [hep-lat]} \BibitemShut
  {NoStop}%
\bibitem [{\citenamefont {{D\"oring}}\ \emph {et~al.}(2018)\citenamefont
  {{D\"oring}}, \citenamefont {Hammer}, \citenamefont {Mai}, \citenamefont
  {Pang}, \citenamefont {Rusetsky},\ and\ \citenamefont {Wu}}]{Doring:2018xxx}%
  \BibitemOpen
  \bibfield  {author} {\bibinfo {author} {\bibfnamefont {M.}~\bibnamefont
  {{D\"oring}}}, \bibinfo {author} {\bibfnamefont {H.~W.}\ \bibnamefont
  {Hammer}}, \bibinfo {author} {\bibfnamefont {M.}~\bibnamefont {Mai}},
  \bibinfo {author} {\bibfnamefont {J.~Y}\ \bibnamefont {Pang}}, \bibinfo
  {author} {\bibfnamefont {A.}~\bibnamefont {Rusetsky}}, \ and\ \bibinfo
  {author} {\bibfnamefont {J.}~\bibnamefont {Wu}},\ }\bibfield  {title}
  {\enquote {\bibinfo {title} {{Three-body spectrum in a finite volume: the
  role of cubic symmetry}},}\ }\href {\doibase 10.1103/PhysRevD.97.114508}
  {\bibfield  {journal} {\bibinfo  {journal} {Phys. Rev.}\ }\textbf {\bibinfo
  {volume} {D97}},\ \bibinfo {pages} {114508} (\bibinfo {year} {2018})},\
  \Eprint {http://arxiv.org/abs/1802.03362} {arXiv:1802.03362 [hep-lat]}
  \BibitemShut {NoStop}%
\bibitem [{\citenamefont {Pang}\ \emph {et~al.}(2019)\citenamefont {Pang},
  \citenamefont {Wu}, \citenamefont {Hammer}, \citenamefont {Meißner},\ and\
  \citenamefont {Rusetsky}}]{Pang:2019dfe}%
  \BibitemOpen
  \bibfield  {author} {\bibinfo {author} {\bibfnamefont {Jin-Yi}\ \bibnamefont
  {Pang}}, \bibinfo {author} {\bibfnamefont {Jia-Jun}\ \bibnamefont {Wu}},
  \bibinfo {author} {\bibfnamefont {H.~W.}\ \bibnamefont {Hammer}}, \bibinfo
  {author} {\bibfnamefont {Ulf-G.}\ \bibnamefont {Meißner}}, \ and\ \bibinfo
  {author} {\bibfnamefont {Akaki}\ \bibnamefont {Rusetsky}},\ }\bibfield
  {title} {\enquote {\bibinfo {title} {{Energy shift of the three-particle
  system in a finite volume}},}\ }\href {\doibase 10.1103/PhysRevD.99.074513}
  {\bibfield  {journal} {\bibinfo  {journal} {Phys. Rev.}\ }\textbf {\bibinfo
  {volume} {D99}},\ \bibinfo {pages} {074513} (\bibinfo {year} {2019})},\
  \Eprint {http://arxiv.org/abs/1902.01111} {arXiv:1902.01111 [hep-lat]}
  \BibitemShut {NoStop}%
\bibitem [{\citenamefont {Mai}\ and\ \citenamefont
  {{D\"oring}}(2017)}]{Mai:2017bge}%
  \BibitemOpen
  \bibfield  {author} {\bibinfo {author} {\bibfnamefont {M.}~\bibnamefont
  {Mai}}\ and\ \bibinfo {author} {\bibfnamefont {M.}~\bibnamefont
  {{D\"oring}}},\ }\bibfield  {title} {\enquote {\bibinfo {title} {{Three-body
  Unitarity in the Finite Volume}},}\ }\href {\doibase
  10.1140/epja/i2017-12440-1} {\bibfield  {journal} {\bibinfo  {journal} {Eur.
  Phys. J.}\ }\textbf {\bibinfo {volume} {A53}},\ \bibinfo {pages} {240}
  (\bibinfo {year} {2017})},\ \Eprint {http://arxiv.org/abs/1709.08222}
  {arXiv:1709.08222 [hep-lat]} \BibitemShut {NoStop}%
\bibitem [{\citenamefont {Mai}\ and\ \citenamefont
  {{D\"oring}}(2019)}]{Mai:2018djl}%
  \BibitemOpen
  \bibfield  {author} {\bibinfo {author} {\bibfnamefont {Maxim}\ \bibnamefont
  {Mai}}\ and\ \bibinfo {author} {\bibfnamefont {Michael}\ \bibnamefont
  {{D\"oring}}},\ }\bibfield  {title} {\enquote {\bibinfo {title}
  {{Finite-Volume Spectrum of $\pi^+\pi^+$ and $\pi^+\pi^+\pi^+$ Systems}},}\
  }\href {\doibase 10.1103/PhysRevLett.122.062503} {\bibfield  {journal}
  {\bibinfo  {journal} {Phys. Rev. Lett.}\ }\textbf {\bibinfo {volume} {122}},\
  \bibinfo {pages} {062503} (\bibinfo {year} {2019})},\ \Eprint
  {http://arxiv.org/abs/1807.04746} {arXiv:1807.04746 [hep-lat]} \BibitemShut
  {NoStop}%
\bibitem [{\citenamefont {Klos}\ \emph {et~al.}(2018)\citenamefont {Klos},
  \citenamefont {König}, \citenamefont {Hammer}, \citenamefont {Lynn},\ and\
  \citenamefont {Schwenk}}]{Klos:2018sen}%
  \BibitemOpen
  \bibfield  {author} {\bibinfo {author} {\bibfnamefont {P.}~\bibnamefont
  {Klos}}, \bibinfo {author} {\bibfnamefont {S.}~\bibnamefont {König}},
  \bibinfo {author} {\bibfnamefont {H.~W.}\ \bibnamefont {Hammer}}, \bibinfo
  {author} {\bibfnamefont {J.~E.}\ \bibnamefont {Lynn}}, \ and\ \bibinfo
  {author} {\bibfnamefont {A.}~\bibnamefont {Schwenk}},\ }\bibfield  {title}
  {\enquote {\bibinfo {title} {{Signatures of few-body resonances in finite
  volume}},}\ }\href {\doibase 10.1103/PhysRevC.98.034004} {\bibfield
  {journal} {\bibinfo  {journal} {Phys. Rev.}\ }\textbf {\bibinfo {volume}
  {C98}},\ \bibinfo {pages} {034004} (\bibinfo {year} {2018})},\ \Eprint
  {http://arxiv.org/abs/1805.02029} {arXiv:1805.02029 [nucl-th]} \BibitemShut
  {NoStop}%
\bibitem [{\citenamefont {Guo}\ and\ \citenamefont
  {Gasparian}(2017)}]{Guo:2017ism}%
  \BibitemOpen
  \bibfield  {author} {\bibinfo {author} {\bibfnamefont {Peng}\ \bibnamefont
  {Guo}}\ and\ \bibinfo {author} {\bibfnamefont {Vladimir}\ \bibnamefont
  {Gasparian}},\ }\bibfield  {title} {\enquote {\bibinfo {title} {{An solvable
  three-body model in finite volume}},}\ }\href {\doibase
  10.1016/j.physletb.2017.10.009} {\bibfield  {journal} {\bibinfo  {journal}
  {Phys. Lett.}\ }\textbf {\bibinfo {volume} {B774}},\ \bibinfo {pages}
  {441--445} (\bibinfo {year} {2017})},\ \Eprint
  {http://arxiv.org/abs/1701.00438} {arXiv:1701.00438 [hep-lat]} \BibitemShut
  {NoStop}%
\bibitem [{\citenamefont {Jackura}\ \emph {et~al.}(2019)\citenamefont
  {Jackura}, \citenamefont {Fernández-Ramírez}, \citenamefont {Mathieu},
  \citenamefont {Mikhasenko}, \citenamefont {Nys}, \citenamefont {Pilloni},
  \citenamefont {Salda\~na}, \citenamefont {Sherrill},\ and\ \citenamefont
  {Szczepaniak}}]{Jackura:2018xnx}%
  \BibitemOpen
  \bibfield  {author} {\bibinfo {author} {\bibfnamefont {A.}~\bibnamefont
  {Jackura}}, \bibinfo {author} {\bibfnamefont {C.}~\bibnamefont
  {Fernández-Ramírez}}, \bibinfo {author} {\bibfnamefont {V.}~\bibnamefont
  {Mathieu}}, \bibinfo {author} {\bibfnamefont {M.}~\bibnamefont {Mikhasenko}},
  \bibinfo {author} {\bibfnamefont {J.}~\bibnamefont {Nys}}, \bibinfo {author}
  {\bibfnamefont {A.}~\bibnamefont {Pilloni}}, \bibinfo {author} {\bibfnamefont
  {K.}~\bibnamefont {Salda\~na}}, \bibinfo {author} {\bibfnamefont
  {N.}~\bibnamefont {Sherrill}}, \ and\ \bibinfo {author} {\bibfnamefont
  {A.~P.}\ \bibnamefont {Szczepaniak}} (\bibinfo {collaboration} {JPAC}),\
  }\bibfield  {title} {\enquote {\bibinfo {title} {{Phenomenology of
  Relativistic $\mathbf{3} \to \mathbf{3}$ Reaction Amplitudes within the
  Isobar Approximation}},}\ }\href {\doibase 10.1140/epjc/s10052-019-6566-1}
  {\bibfield  {journal} {\bibinfo  {journal} {Eur. Phys. J.}\ }\textbf
  {\bibinfo {volume} {C79}},\ \bibinfo {pages} {56} (\bibinfo {year} {2019})},\
  \Eprint {http://arxiv.org/abs/1809.10523} {arXiv:1809.10523 [hep-ph]}
  \BibitemShut {NoStop}%
\bibitem [{\citenamefont {Hansen}\ and\ \citenamefont
  {Sharpe}(2019)}]{Hansen:2019nir}%
  \BibitemOpen
  \bibfield  {author} {\bibinfo {author} {\bibfnamefont {Maxwell~T.}\
  \bibnamefont {Hansen}}\ and\ \bibinfo {author} {\bibfnamefont {Stephen~R.}\
  \bibnamefont {Sharpe}},\ }\bibfield  {title} {\enquote {\bibinfo {title}
  {{Lattice QCD and Three-particle Decays of Resonances}},}\ }\href@noop {} {\
  (\bibinfo {year} {2019})},\ \Eprint {http://arxiv.org/abs/1901.00483}
  {arXiv:1901.00483 [hep-lat]} \BibitemShut {NoStop}%
\bibitem [{\citenamefont {Hörz}\ and\ \citenamefont
  {Hanlon}(2019)}]{Horz:2019rrn}%
  \BibitemOpen
  \bibfield  {author} {\bibinfo {author} {\bibfnamefont {Ben}\ \bibnamefont
  {Hörz}}\ and\ \bibinfo {author} {\bibfnamefont {Andrew}\ \bibnamefont
  {Hanlon}},\ }\bibfield  {title} {\enquote {\bibinfo {title} {{Two- and
  three-pion finite-volume spectra at maximal isospin from lattice QCD}},}\
  }\href {\doibase 10.1103/PhysRevLett.123.142002} {\bibfield  {journal}
  {\bibinfo  {journal} {Phys. Rev. Lett.}\ }\textbf {\bibinfo {volume} {123}},\
  \bibinfo {pages} {142002} (\bibinfo {year} {2019})},\ \Eprint
  {http://arxiv.org/abs/1905.04277} {arXiv:1905.04277 [hep-lat]} \BibitemShut
  {NoStop}%
\bibitem [{\citenamefont {Culver}\ \emph {et~al.}(2020)\citenamefont {Culver},
  \citenamefont {Mai}, \citenamefont {Brett}, \citenamefont {Alexandru},\ and\
  \citenamefont {Döring}}]{Culver:2019vvu}%
  \BibitemOpen
  \bibfield  {author} {\bibinfo {author} {\bibfnamefont {Chris}\ \bibnamefont
  {Culver}}, \bibinfo {author} {\bibfnamefont {Maxim}\ \bibnamefont {Mai}},
  \bibinfo {author} {\bibfnamefont {Ruairí}\ \bibnamefont {Brett}}, \bibinfo
  {author} {\bibfnamefont {Andrei}\ \bibnamefont {Alexandru}}, \ and\ \bibinfo
  {author} {\bibfnamefont {Michael}\ \bibnamefont {Döring}},\ }\bibfield
  {title} {\enquote {\bibinfo {title} {{Three body spectrum from lattice
  QCD}},}\ }\href {\doibase 10.1103/PhysRevD.101.114507} {\bibfield  {journal}
  {\bibinfo  {journal} {Phys. Rev. D}\ }\textbf {\bibinfo {volume} {101}},\
  \bibinfo {pages} {114507} (\bibinfo {year} {2020})},\ \Eprint
  {http://arxiv.org/abs/1911.09047} {arXiv:1911.09047 [hep-lat]} \BibitemShut
  {NoStop}%
\bibitem [{\citenamefont {Mai}\ \emph {et~al.}(2020)\citenamefont {Mai},
  \citenamefont {Döring}, \citenamefont {Culver},\ and\ \citenamefont
  {Alexandru}}]{Mai:2019fba}%
  \BibitemOpen
  \bibfield  {author} {\bibinfo {author} {\bibfnamefont {M.}~\bibnamefont
  {Mai}}, \bibinfo {author} {\bibfnamefont {M.}~\bibnamefont {Döring}},
  \bibinfo {author} {\bibfnamefont {C.}~\bibnamefont {Culver}}, \ and\ \bibinfo
  {author} {\bibfnamefont {A.}~\bibnamefont {Alexandru}},\ }\bibfield  {title}
  {\enquote {\bibinfo {title} {{Three-body unitarity versus finite-volume
  $\pi^+\pi^+\pi^+$ spectrum from lattice QCD}},}\ }\href {\doibase
  10.1103/PhysRevD.101.054510} {\bibfield  {journal} {\bibinfo  {journal}
  {Phys. Rev. D}\ }\textbf {\bibinfo {volume} {101}},\ \bibinfo {pages}
  {054510} (\bibinfo {year} {2020})},\ \Eprint
  {http://arxiv.org/abs/1909.05749} {arXiv:1909.05749 [hep-lat]} \BibitemShut
  {NoStop}%
\bibitem [{\citenamefont {Blanton}\ \emph {et~al.}(2020)\citenamefont
  {Blanton}, \citenamefont {Romero-López},\ and\ \citenamefont
  {Sharpe}}]{Blanton:2019vdk}%
  \BibitemOpen
  \bibfield  {author} {\bibinfo {author} {\bibfnamefont {Tyler~D.}\
  \bibnamefont {Blanton}}, \bibinfo {author} {\bibfnamefont {Fernando}\
  \bibnamefont {Romero-López}}, \ and\ \bibinfo {author} {\bibfnamefont
  {Stephen~R.}\ \bibnamefont {Sharpe}},\ }\bibfield  {title} {\enquote
  {\bibinfo {title} {{$I=3$ Three-Pion Scattering Amplitude from Lattice
  QCD}},}\ }\href {\doibase 10.1103/PhysRevLett.124.032001} {\bibfield
  {journal} {\bibinfo  {journal} {Phys. Rev. Lett.}\ }\textbf {\bibinfo
  {volume} {124}},\ \bibinfo {pages} {032001} (\bibinfo {year} {2020})},\
  \Eprint {http://arxiv.org/abs/1909.02973} {arXiv:1909.02973 [hep-lat]}
  \BibitemShut {NoStop}%
\bibitem [{\citenamefont {Guo}\ and\ \citenamefont {Long}(2020)}]{Guo:2020kph}%
  \BibitemOpen
  \bibfield  {author} {\bibinfo {author} {\bibfnamefont {Peng}\ \bibnamefont
  {Guo}}\ and\ \bibinfo {author} {\bibfnamefont {Bingwei}\ \bibnamefont
  {Long}},\ }\bibfield  {title} {\enquote {\bibinfo {title} {{Multi- $\pi^+$
  systems in a finite volume}},}\ }\href {\doibase 10.1103/PhysRevD.101.094510}
  {\bibfield  {journal} {\bibinfo  {journal} {Phys. Rev. D}\ }\textbf {\bibinfo
  {volume} {101}},\ \bibinfo {pages} {094510} (\bibinfo {year} {2020})},\
  \Eprint {http://arxiv.org/abs/2002.09266} {arXiv:2002.09266 [hep-lat]}
  \BibitemShut {NoStop}%
\bibitem [{\citenamefont {Guo}(2020)}]{Guo:2020spn}%
  \BibitemOpen
  \bibfield  {author} {\bibinfo {author} {\bibfnamefont {Peng}\ \bibnamefont
  {Guo}},\ }\bibfield  {title} {\enquote {\bibinfo {title} {{Modeling few-body
  resonances in finite volume}},}\ }\href@noop {} {\  (\bibinfo {year}
  {2020})},\ \Eprint {http://arxiv.org/abs/2007.12790} {arXiv:2007.12790
  [hep-lat]} \BibitemShut {NoStop}%
\bibitem [{\citenamefont {Beane}\ \emph {et~al.}(2007)\citenamefont {Beane},
  \citenamefont {Detmold},\ and\ \citenamefont {Savage}}]{Beane:2007qr}%
  \BibitemOpen
  \bibfield  {author} {\bibinfo {author} {\bibfnamefont {Silas~R.}\
  \bibnamefont {Beane}}, \bibinfo {author} {\bibfnamefont {William}\
  \bibnamefont {Detmold}}, \ and\ \bibinfo {author} {\bibfnamefont {Martin~J.}\
  \bibnamefont {Savage}},\ }\bibfield  {title} {\enquote {\bibinfo {title}
  {{n-Boson Energies at Finite Volume and Three-Boson Interactions}},}\ }\href
  {\doibase 10.1103/PhysRevD.76.074507} {\bibfield  {journal} {\bibinfo
  {journal} {Phys. Rev.}\ }\textbf {\bibinfo {volume} {D76}},\ \bibinfo {pages}
  {074507} (\bibinfo {year} {2007})},\ \Eprint {http://arxiv.org/abs/0707.1670}
  {arXiv:0707.1670 [hep-lat]} \BibitemShut {NoStop}%
\bibitem [{\citenamefont {Detmold}\ \emph {et~al.}(2008)\citenamefont
  {Detmold}, \citenamefont {Savage}, \citenamefont {Torok}, \citenamefont
  {Beane}, \citenamefont {Luu}, \citenamefont {Orginos},\ and\ \citenamefont
  {Parreno}}]{Detmold:2008fn}%
  \BibitemOpen
  \bibfield  {author} {\bibinfo {author} {\bibfnamefont {William}\ \bibnamefont
  {Detmold}}, \bibinfo {author} {\bibfnamefont {Martin~J.}\ \bibnamefont
  {Savage}}, \bibinfo {author} {\bibfnamefont {Aaron}\ \bibnamefont {Torok}},
  \bibinfo {author} {\bibfnamefont {Silas~R.}\ \bibnamefont {Beane}}, \bibinfo
  {author} {\bibfnamefont {Thomas~C.}\ \bibnamefont {Luu}}, \bibinfo {author}
  {\bibfnamefont {Kostas}\ \bibnamefont {Orginos}}, \ and\ \bibinfo {author}
  {\bibfnamefont {Assumpta}\ \bibnamefont {Parreno}},\ }\bibfield  {title}
  {\enquote {\bibinfo {title} {{Multi-Pion States in Lattice QCD and the
  Charged-Pion Condensate}},}\ }\href {\doibase 10.1103/PhysRevD.78.014507}
  {\bibfield  {journal} {\bibinfo  {journal} {Phys. Rev.}\ }\textbf {\bibinfo
  {volume} {D78}},\ \bibinfo {pages} {014507} (\bibinfo {year} {2008})},\
  \Eprint {http://arxiv.org/abs/0803.2728} {arXiv:0803.2728 [hep-lat]}
  \BibitemShut {NoStop}%
\bibitem [{\citenamefont {Romero-López}\ \emph
  {et~al.}(2018{\natexlab{b}})\citenamefont {Romero-López}, \citenamefont
  {Rusetsky},\ and\ \citenamefont {Urbach}}]{Romero-Lopez:2018rcb}%
  \BibitemOpen
  \bibfield  {author} {\bibinfo {author} {\bibfnamefont {Fernando}\
  \bibnamefont {Romero-López}}, \bibinfo {author} {\bibfnamefont {Akaki}\
  \bibnamefont {Rusetsky}}, \ and\ \bibinfo {author} {\bibfnamefont {Carsten}\
  \bibnamefont {Urbach}},\ }\bibfield  {title} {\enquote {\bibinfo {title}
  {{Two- and three-body interactions in $\varphi ^4$ theory from lattice
  simulations}},}\ }\href {\doibase 10.1140/epjc/s10052-018-6325-8} {\bibfield
  {journal} {\bibinfo  {journal} {Eur. Phys. J.}\ }\textbf {\bibinfo {volume}
  {C78}},\ \bibinfo {pages} {846} (\bibinfo {year} {2018}{\natexlab{b}})},\
  \Eprint {http://arxiv.org/abs/1806.02367} {arXiv:1806.02367 [hep-lat]}
  \BibitemShut {NoStop}%
\bibitem [{\citenamefont {Beane}\ \emph {et~al.}(2020)\citenamefont {Beane}
  \emph {et~al.}}]{Beane:2020ycc}%
  \BibitemOpen
  \bibfield  {author} {\bibinfo {author} {\bibfnamefont {S.R.}\ \bibnamefont
  {Beane}} \emph {et~al.},\ }\bibfield  {title} {\enquote {\bibinfo {title}
  {{Charged multi-hadron systems in lattice QCD+QED}},}\ }\href@noop {} {\
  (\bibinfo {year} {2020})},\ \Eprint {http://arxiv.org/abs/2003.12130}
  {arXiv:2003.12130 [hep-lat]} \BibitemShut {NoStop}%
\bibitem [{\citenamefont {Dudek}\ \emph {et~al.}(2012)\citenamefont {Dudek},
  \citenamefont {Edwards},\ and\ \citenamefont {Thomas}}]{Dudek:2012gj}%
  \BibitemOpen
  \bibfield  {author} {\bibinfo {author} {\bibfnamefont {Jozef~J.}\
  \bibnamefont {Dudek}}, \bibinfo {author} {\bibfnamefont {Robert~G.}\
  \bibnamefont {Edwards}}, \ and\ \bibinfo {author} {\bibfnamefont
  {Christopher~E.}\ \bibnamefont {Thomas}},\ }\bibfield  {title} {\enquote
  {\bibinfo {title} {{S and D-wave phase shifts in isospin-2 pi pi scattering
  from lattice QCD}},}\ }\href {\doibase 10.1103/PhysRevD.86.034031} {\bibfield
   {journal} {\bibinfo  {journal} {Phys. Rev.}\ }\textbf {\bibinfo {volume}
  {D86}},\ \bibinfo {pages} {034031} (\bibinfo {year} {2012})},\ \Eprint
  {http://arxiv.org/abs/1203.6041} {arXiv:1203.6041 [hep-ph]} \BibitemShut
  {NoStop}%
\bibitem [{\citenamefont {Gasser}\ and\ \citenamefont
  {Leutwyler}(1987{\natexlab{a}})}]{Gasser:1986vb}%
  \BibitemOpen
  \bibfield  {author} {\bibinfo {author} {\bibfnamefont {J.}~\bibnamefont
  {Gasser}}\ and\ \bibinfo {author} {\bibfnamefont {H.}~\bibnamefont
  {Leutwyler}},\ }\bibfield  {title} {\enquote {\bibinfo {title} {{Light Quarks
  at Low Temperatures}},}\ }\href {\doibase 10.1016/0370-2693(87)90492-8}
  {\bibfield  {journal} {\bibinfo  {journal} {Phys. Lett. B}\ }\textbf
  {\bibinfo {volume} {184}},\ \bibinfo {pages} {83--88} (\bibinfo {year}
  {1987}{\natexlab{a}})}\BibitemShut {NoStop}%
\bibitem [{\citenamefont {Gasser}\ and\ \citenamefont
  {Leutwyler}(1987{\natexlab{b}})}]{Gasser:1987ah}%
  \BibitemOpen
  \bibfield  {author} {\bibinfo {author} {\bibfnamefont {J.}~\bibnamefont
  {Gasser}}\ and\ \bibinfo {author} {\bibfnamefont {H.}~\bibnamefont
  {Leutwyler}},\ }\bibfield  {title} {\enquote {\bibinfo {title}
  {{Thermodynamics of Chiral Symmetry}},}\ }\href {\doibase
  10.1016/0370-2693(87)91652-2} {\bibfield  {journal} {\bibinfo  {journal}
  {Phys. Lett. B}\ }\textbf {\bibinfo {volume} {188}},\ \bibinfo {pages}
  {477--481} (\bibinfo {year} {1987}{\natexlab{b}})}\BibitemShut {NoStop}%
\bibitem [{\citenamefont {Gasser}\ and\ \citenamefont
  {Leutwyler}(1988)}]{Gasser:1987zq}%
  \BibitemOpen
  \bibfield  {author} {\bibinfo {author} {\bibfnamefont {J.}~\bibnamefont
  {Gasser}}\ and\ \bibinfo {author} {\bibfnamefont {H.}~\bibnamefont
  {Leutwyler}},\ }\bibfield  {title} {\enquote {\bibinfo {title}
  {{Spontaneously Broken Symmetries: Effective Lagrangians at Finite
  Volume}},}\ }\href {\doibase 10.1016/0550-3213(88)90107-1} {\bibfield
  {journal} {\bibinfo  {journal} {Nucl. Phys. B}\ }\textbf {\bibinfo {volume}
  {307}},\ \bibinfo {pages} {763--778} (\bibinfo {year} {1988})}\BibitemShut
  {NoStop}%
\bibitem [{\citenamefont {Abdel-Rehim}\ \emph {et~al.}(2017)\citenamefont
  {Abdel-Rehim} \emph {et~al.}}]{Abdel-Rehim:2015pwa}%
  \BibitemOpen
  \bibfield  {author} {\bibinfo {author} {\bibfnamefont {A.}~\bibnamefont
  {Abdel-Rehim}} \emph {et~al.} (\bibinfo {collaboration} {ETM}),\ }\bibfield
  {title} {\enquote {\bibinfo {title} {{First physics results at the physical
  pion mass from $N_f=2$ Wilson twisted mass fermions at maximal twist}},}\
  }\href {\doibase 10.1103/PhysRevD.95.094515} {\bibfield  {journal} {\bibinfo
  {journal} {Phys. Rev.}\ }\textbf {\bibinfo {volume} {D95}},\ \bibinfo {pages}
  {094515} (\bibinfo {year} {2017})},\ \Eprint
  {http://arxiv.org/abs/1507.05068} {arXiv:1507.05068 [hep-lat]} \BibitemShut
  {NoStop}%
\bibitem [{\citenamefont {Iwasaki}(1985)}]{Iwasaki:1985we}%
  \BibitemOpen
  \bibfield  {author} {\bibinfo {author} {\bibfnamefont {Y.}~\bibnamefont
  {Iwasaki}},\ }\bibfield  {title} {\enquote {\bibinfo {title}
  {{Renormalization Group Analysis of Lattice Theories and Improved Lattice
  Action: Two-Dimensional Nonlinear O(N) Sigma Model}},}\ }\href {\doibase
  10.1016/0550-3213(85)90606-6} {\bibfield  {journal} {\bibinfo  {journal}
  {Nucl. Phys. B}\ }\textbf {\bibinfo {volume} {258}},\ \bibinfo {pages}
  {141--156} (\bibinfo {year} {1985})}\BibitemShut {NoStop}%
\bibitem [{\citenamefont {Frezzotti}\ \emph {et~al.}(2001)\citenamefont
  {Frezzotti}, \citenamefont {Grassi}, \citenamefont {Sint},\ and\
  \citenamefont {Weisz}}]{Frezzotti:2000nk}%
  \BibitemOpen
  \bibfield  {author} {\bibinfo {author} {\bibfnamefont {Roberto}\ \bibnamefont
  {Frezzotti}}, \bibinfo {author} {\bibfnamefont {Pietro~Antonio}\ \bibnamefont
  {Grassi}}, \bibinfo {author} {\bibfnamefont {Stefan}\ \bibnamefont {Sint}}, \
  and\ \bibinfo {author} {\bibfnamefont {Peter}\ \bibnamefont {Weisz}}
  (\bibinfo {collaboration} {Alpha}),\ }\bibfield  {title} {\enquote {\bibinfo
  {title} {{Lattice QCD with a chirally twisted mass term}},}\ }\href@noop {}
  {\bibfield  {journal} {\bibinfo  {journal} {JHEP}\ }\textbf {\bibinfo
  {volume} {08}},\ \bibinfo {pages} {058} (\bibinfo {year} {2001})},\ \Eprint
  {http://arxiv.org/abs/hep-lat/0101001} {arXiv:hep-lat/0101001} \BibitemShut
  {NoStop}%
\bibitem [{\citenamefont {Frezzotti}\ and\ \citenamefont
  {Rossi}(2004)}]{Frezzotti:2003ni}%
  \BibitemOpen
  \bibfield  {author} {\bibinfo {author} {\bibfnamefont {R.}~\bibnamefont
  {Frezzotti}}\ and\ \bibinfo {author} {\bibfnamefont {G.~C.}\ \bibnamefont
  {Rossi}},\ }\bibfield  {title} {\enquote {\bibinfo {title} {Chirally
  improving {Wilson} fermions. {I}: {O(a)} improvement},}\ }\href@noop {}
  {\bibfield  {journal} {\bibinfo  {journal} {JHEP}\ }\textbf {\bibinfo
  {volume} {08}},\ \bibinfo {pages} {007} (\bibinfo {year} {2004})},\ \Eprint
  {http://arxiv.org/abs/hep-lat/0306014} {hep-lat/0306014} \BibitemShut
  {NoStop}%
\bibitem [{\citenamefont {Buchoff}\ \emph {et~al.}(2009)\citenamefont
  {Buchoff}, \citenamefont {Chen},\ and\ \citenamefont
  {Walker-Loud}}]{Buchoff:2008hh}%
  \BibitemOpen
  \bibfield  {author} {\bibinfo {author} {\bibfnamefont {Michael~I.}\
  \bibnamefont {Buchoff}}, \bibinfo {author} {\bibfnamefont {Jiunn-Wei}\
  \bibnamefont {Chen}}, \ and\ \bibinfo {author} {\bibfnamefont {Andre}\
  \bibnamefont {Walker-Loud}},\ }\bibfield  {title} {\enquote {\bibinfo {title}
  {{pi-pi Scattering in Twisted Mass Chiral Perturbation Theory}},}\ }\href
  {\doibase 10.1103/PhysRevD.79.074503} {\bibfield  {journal} {\bibinfo
  {journal} {Phys. Rev. D}\ }\textbf {\bibinfo {volume} {79}},\ \bibinfo
  {pages} {074503} (\bibinfo {year} {2009})},\ \Eprint
  {http://arxiv.org/abs/0810.2464} {arXiv:0810.2464 [hep-lat]} \BibitemShut
  {NoStop}%
\bibitem [{\citenamefont {Peardon}\ \emph {et~al.}(2009)\citenamefont
  {Peardon}, \citenamefont {Bulava}, \citenamefont {Foley}, \citenamefont
  {Morningstar}, \citenamefont {Dudek}, \citenamefont {Edwards}, \citenamefont
  {Joo}, \citenamefont {Lin}, \citenamefont {Richards},\ and\ \citenamefont
  {Juge}}]{Peardon:2009gh}%
  \BibitemOpen
  \bibfield  {author} {\bibinfo {author} {\bibfnamefont {Michael}\ \bibnamefont
  {Peardon}}, \bibinfo {author} {\bibfnamefont {John}\ \bibnamefont {Bulava}},
  \bibinfo {author} {\bibfnamefont {Justin}\ \bibnamefont {Foley}}, \bibinfo
  {author} {\bibfnamefont {Colin}\ \bibnamefont {Morningstar}}, \bibinfo
  {author} {\bibfnamefont {Jozef}\ \bibnamefont {Dudek}}, \bibinfo {author}
  {\bibfnamefont {Robert~G.}\ \bibnamefont {Edwards}}, \bibinfo {author}
  {\bibfnamefont {Balint}\ \bibnamefont {Joo}}, \bibinfo {author}
  {\bibfnamefont {Huey-Wen}\ \bibnamefont {Lin}}, \bibinfo {author}
  {\bibfnamefont {David~G.}\ \bibnamefont {Richards}}, \ and\ \bibinfo {author}
  {\bibfnamefont {Keisuke~Jimmy}\ \bibnamefont {Juge}} (\bibinfo
  {collaboration} {Hadron Spectrum}),\ }\bibfield  {title} {\enquote {\bibinfo
  {title} {{A Novel quark-field creation operator construction for hadronic
  physics in lattice QCD}},}\ }\href {\doibase 10.1103/PhysRevD.80.054506}
  {\bibfield  {journal} {\bibinfo  {journal} {Phys. Rev. D}\ }\textbf {\bibinfo
  {volume} {80}},\ \bibinfo {pages} {054506} (\bibinfo {year} {2009})},\
  \Eprint {http://arxiv.org/abs/0905.2160} {arXiv:0905.2160 [hep-lat]}
  \BibitemShut {NoStop}%
\bibitem [{\citenamefont {Morningstar}\ \emph {et~al.}(2011)\citenamefont
  {Morningstar}, \citenamefont {Bulava}, \citenamefont {Foley}, \citenamefont
  {Juge}, \citenamefont {Lenkner}, \citenamefont {Peardon},\ and\ \citenamefont
  {Wong}}]{Morningstar:2011ka}%
  \BibitemOpen
  \bibfield  {author} {\bibinfo {author} {\bibfnamefont {Colin}\ \bibnamefont
  {Morningstar}}, \bibinfo {author} {\bibfnamefont {John}\ \bibnamefont
  {Bulava}}, \bibinfo {author} {\bibfnamefont {Justin}\ \bibnamefont {Foley}},
  \bibinfo {author} {\bibfnamefont {Keisuke~J.}\ \bibnamefont {Juge}}, \bibinfo
  {author} {\bibfnamefont {David}\ \bibnamefont {Lenkner}}, \bibinfo {author}
  {\bibfnamefont {Mike}\ \bibnamefont {Peardon}}, \ and\ \bibinfo {author}
  {\bibfnamefont {Chik~Him}\ \bibnamefont {Wong}},\ }\bibfield  {title}
  {\enquote {\bibinfo {title} {{Improved stochastic estimation of quark
  propagation with Laplacian Heaviside smearing in lattice QCD}},}\ }\href
  {\doibase 10.1103/PhysRevD.83.114505} {\bibfield  {journal} {\bibinfo
  {journal} {Phys. Rev. D}\ }\textbf {\bibinfo {volume} {83}},\ \bibinfo
  {pages} {114505} (\bibinfo {year} {2011})},\ \Eprint
  {http://arxiv.org/abs/1104.3870} {arXiv:1104.3870 [hep-lat]} \BibitemShut
  {NoStop}%
\bibitem [{\citenamefont {Dimopoulos}\ \emph {et~al.}(2019)\citenamefont
  {Dimopoulos} \emph {et~al.}}]{Dimopoulos:2018xkm}%
  \BibitemOpen
  \bibfield  {author} {\bibinfo {author} {\bibfnamefont {Petros}\ \bibnamefont
  {Dimopoulos}} \emph {et~al.},\ }\bibfield  {title} {\enquote {\bibinfo
  {title} {{Topological susceptibility and $\eta'$ meson mass from $N_f=2$
  lattice QCD at the physical point}},}\ }\href {\doibase
  10.1103/PhysRevD.99.034511} {\bibfield  {journal} {\bibinfo  {journal} {Phys.
  Rev.}\ }\textbf {\bibinfo {volume} {D99}},\ \bibinfo {pages} {034511}
  (\bibinfo {year} {2019})},\ \Eprint {http://arxiv.org/abs/1812.08787}
  {arXiv:1812.08787 [hep-lat]} \BibitemShut {NoStop}%
\bibitem [{\citenamefont {Michael}\ and\ \citenamefont
  {Teasdale}(1983)}]{Michael:1982gb}%
  \BibitemOpen
  \bibfield  {author} {\bibinfo {author} {\bibfnamefont {Christopher}\
  \bibnamefont {Michael}}\ and\ \bibinfo {author} {\bibfnamefont
  {I.}~\bibnamefont {Teasdale}},\ }\bibfield  {title} {\enquote {\bibinfo
  {title} {{Extracting Glueball Masses From Lattice {QCD}}},}\ }\href {\doibase
  10.1016/0550-3213(83)90674-0} {\bibfield  {journal} {\bibinfo  {journal}
  {Nucl. Phys.}\ }\textbf {\bibinfo {volume} {B215}},\ \bibinfo {pages}
  {433--446} (\bibinfo {year} {1983})}\BibitemShut {NoStop}%
\bibitem [{\citenamefont {Blossier}\ \emph {et~al.}(2009)\citenamefont
  {Blossier}, \citenamefont {Della~Morte}, \citenamefont {von Hippel},
  \citenamefont {Mendes},\ and\ \citenamefont {Sommer}}]{Blossier:2009kd}%
  \BibitemOpen
  \bibfield  {author} {\bibinfo {author} {\bibfnamefont {Benoit}\ \bibnamefont
  {Blossier}}, \bibinfo {author} {\bibfnamefont {Michele}\ \bibnamefont
  {Della~Morte}}, \bibinfo {author} {\bibfnamefont {Georg}\ \bibnamefont {von
  Hippel}}, \bibinfo {author} {\bibfnamefont {Tereza}\ \bibnamefont {Mendes}},
  \ and\ \bibinfo {author} {\bibfnamefont {Rainer}\ \bibnamefont {Sommer}},\
  }\bibfield  {title} {\enquote {\bibinfo {title} {{On the generalized
  eigenvalue method for energies and matrix elements in lattice field
  theory}},}\ }\href {\doibase 10.1088/1126-6708/2009/04/094} {\bibfield
  {journal} {\bibinfo  {journal} {JHEP}\ }\textbf {\bibinfo {volume} {04}},\
  \bibinfo {pages} {094} (\bibinfo {year} {2009})},\ \Eprint
  {http://arxiv.org/abs/0902.1265} {arXiv:0902.1265 [hep-lat]} \BibitemShut
  {NoStop}%
\bibitem [{\citenamefont {Fischer}\ \emph
  {et~al.}(2020{\natexlab{b}})\citenamefont {Fischer}, \citenamefont
  {Kostrzewa}, \citenamefont {Ostmeyer}, \citenamefont {Ottnad}, \citenamefont
  {Ueding},\ and\ \citenamefont {Urbach}}]{Fischer:2020bgv}%
  \BibitemOpen
  \bibfield  {author} {\bibinfo {author} {\bibfnamefont {Matthias}\
  \bibnamefont {Fischer}}, \bibinfo {author} {\bibfnamefont {Bartosz}\
  \bibnamefont {Kostrzewa}}, \bibinfo {author} {\bibfnamefont {Johann}\
  \bibnamefont {Ostmeyer}}, \bibinfo {author} {\bibfnamefont {Konstantin}\
  \bibnamefont {Ottnad}}, \bibinfo {author} {\bibfnamefont {Martin}\
  \bibnamefont {Ueding}}, \ and\ \bibinfo {author} {\bibfnamefont {Carsten}\
  \bibnamefont {Urbach}},\ }\bibfield  {title} {\enquote {\bibinfo {title} {{On
  the generalised eigenvalue method and its relation to Prony and generalised
  pencil of function methods}},}\ }\href@noop {} {\  (\bibinfo {year}
  {2020}{\natexlab{b}})},\ \Eprint {http://arxiv.org/abs/2004.10472}
  {arXiv:2004.10472 [hep-lat]} \BibitemShut {NoStop}%
\bibitem [{\citenamefont
  {Ueding}({\natexlab{a}})}]{Nf2-3pi-I3-scattering-data}%
  \BibitemOpen
  \bibfield  {author} {\bibinfo {author} {\bibfnamefont {Martin}\ \bibnamefont
  {Ueding}},\ }\href {https://github.com/HISKP-LQCD/Nf2-3pi-I3-scattering-data}
  {\enquote {\bibinfo {title} {{$N_\mathrm f = 2$ three pion $I = 3$ scattering
  data repository}},}\ } ({\natexlab{a}})\BibitemShut {NoStop}%
\bibitem [{\citenamefont {Kaminski}\ \emph {et~al.}(2008)\citenamefont
  {Kaminski}, \citenamefont {Pelaez},\ and\ \citenamefont
  {Yndurain}}]{Kaminski:2006qe}%
  \BibitemOpen
  \bibfield  {author} {\bibinfo {author} {\bibfnamefont {R.}~\bibnamefont
  {Kaminski}}, \bibinfo {author} {\bibfnamefont {J.R.}\ \bibnamefont {Pelaez}},
  \ and\ \bibinfo {author} {\bibfnamefont {F.J.}\ \bibnamefont {Yndurain}},\
  }\bibfield  {title} {\enquote {\bibinfo {title} {{The Pion-pion scattering
  amplitude. III. Improving the analysis with forward dispersion relations and
  Roy equations}},}\ }\href {\doibase 10.1103/PhysRevD.77.054015} {\bibfield
  {journal} {\bibinfo  {journal} {Phys. Rev. D}\ }\textbf {\bibinfo {volume}
  {77}},\ \bibinfo {pages} {054015} (\bibinfo {year} {2008})},\ \Eprint
  {http://arxiv.org/abs/0710.1150} {arXiv:0710.1150 [hep-ph]} \BibitemShut
  {NoStop}%
\bibitem [{\citenamefont {Colangelo}\ \emph {et~al.}(2001)\citenamefont
  {Colangelo}, \citenamefont {Gasser},\ and\ \citenamefont
  {Leutwyler}}]{Colangelo:2001df}%
  \BibitemOpen
  \bibfield  {author} {\bibinfo {author} {\bibfnamefont {G.}~\bibnamefont
  {Colangelo}}, \bibinfo {author} {\bibfnamefont {J.}~\bibnamefont {Gasser}}, \
  and\ \bibinfo {author} {\bibfnamefont {H.}~\bibnamefont {Leutwyler}},\
  }\bibfield  {title} {\enquote {\bibinfo {title} {{$\pi \pi$ scattering}},}\
  }\href {\doibase 10.1016/S0550-3213(01)00147-X} {\bibfield  {journal}
  {\bibinfo  {journal} {Nucl. Phys. B}\ }\textbf {\bibinfo {volume} {603}},\
  \bibinfo {pages} {125--179} (\bibinfo {year} {2001})},\ \Eprint
  {http://arxiv.org/abs/hep-ph/0103088} {arXiv:hep-ph/0103088} \BibitemShut
  {NoStop}%
\bibitem [{\citenamefont {Adler}(1965)}]{adlerref}%
  \BibitemOpen
  \bibfield  {author} {\bibinfo {author} {\bibfnamefont {Stephen~L.}\
  \bibnamefont {Adler}},\ }\bibfield  {title} {\enquote {\bibinfo {title}
  {Consistency conditions on the strong interactions implied by a partially
  conserved axial-vector current},}\ }\href {\doibase
  10.1103/PhysRev.137.B1022} {\bibfield  {journal} {\bibinfo  {journal} {Phys.
  Rev.}\ }\textbf {\bibinfo {volume} {137}},\ \bibinfo {pages} {B1022--B1033}
  (\bibinfo {year} {1965})}\BibitemShut {NoStop}%
\bibitem [{\citenamefont {Yndurain}(2002)}]{Yndurain:2002ud}%
  \BibitemOpen
  \bibfield  {author} {\bibinfo {author} {\bibfnamefont {F.J.}\ \bibnamefont
  {Yndurain}},\ }\bibfield  {title} {\enquote {\bibinfo {title} {{Low-energy
  pion physics}},}\ }\href@noop {} {\  (\bibinfo {year} {2002})},\ \Eprint
  {http://arxiv.org/abs/hep-ph/0212282} {arXiv:hep-ph/0212282} \BibitemShut
  {NoStop}%
\bibitem [{\citenamefont {Pelaez}\ and\ \citenamefont
  {Yndurain}(2005)}]{Pelaez:2004vs}%
  \BibitemOpen
  \bibfield  {author} {\bibinfo {author} {\bibfnamefont {J.R.}\ \bibnamefont
  {Pelaez}}\ and\ \bibinfo {author} {\bibfnamefont {F.J.}\ \bibnamefont
  {Yndurain}},\ }\bibfield  {title} {\enquote {\bibinfo {title} {{The Pion-pion
  scattering amplitude}},}\ }\href {\doibase 10.1103/PhysRevD.71.074016}
  {\bibfield  {journal} {\bibinfo  {journal} {Phys. Rev. D}\ }\textbf {\bibinfo
  {volume} {71}},\ \bibinfo {pages} {074016} (\bibinfo {year} {2005})},\
  \Eprint {http://arxiv.org/abs/hep-ph/0411334} {arXiv:hep-ph/0411334}
  \BibitemShut {NoStop}%
\bibitem [{\citenamefont {Caprini}\ \emph {et~al.}(2012)\citenamefont
  {Caprini}, \citenamefont {Colangelo},\ and\ \citenamefont
  {Leutwyler}}]{Caprini:2011ky}%
  \BibitemOpen
  \bibfield  {author} {\bibinfo {author} {\bibfnamefont {I.}~\bibnamefont
  {Caprini}}, \bibinfo {author} {\bibfnamefont {G.}~\bibnamefont {Colangelo}},
  \ and\ \bibinfo {author} {\bibfnamefont {H.}~\bibnamefont {Leutwyler}},\
  }\bibfield  {title} {\enquote {\bibinfo {title} {{Regge analysis of the pi pi
  scattering amplitude}},}\ }\href {\doibase 10.1140/epjc/s10052-012-1860-1}
  {\bibfield  {journal} {\bibinfo  {journal} {Eur. Phys. J. C}\ }\textbf
  {\bibinfo {volume} {72}},\ \bibinfo {pages} {1860} (\bibinfo {year}
  {2012})},\ \Eprint {http://arxiv.org/abs/1111.7160} {arXiv:1111.7160
  [hep-ph]} \BibitemShut {NoStop}%
\bibitem [{\citenamefont {Albaladejo}\ and\ \citenamefont
  {Oller}(2012)}]{Albaladejo:2012te}%
  \BibitemOpen
  \bibfield  {author} {\bibinfo {author} {\bibfnamefont {M.}~\bibnamefont
  {Albaladejo}}\ and\ \bibinfo {author} {\bibfnamefont {J.~A.}\ \bibnamefont
  {Oller}},\ }\bibfield  {title} {\enquote {\bibinfo {title} {{On the size of
  the sigma meson and its nature}},}\ }\href {\doibase
  10.1103/PhysRevD.86.034003} {\bibfield  {journal} {\bibinfo  {journal} {Phys.
  Rev. D}\ }\textbf {\bibinfo {volume} {86}},\ \bibinfo {pages} {034003}
  (\bibinfo {year} {2012})},\ \Eprint {http://arxiv.org/abs/1205.6606}
  {arXiv:1205.6606 [hep-ph]} \BibitemShut {NoStop}%
\bibitem [{\citenamefont {Yamazaki}\ \emph {et~al.}(2004)\citenamefont
  {Yamazaki} \emph {et~al.}}]{Yamazaki:2004qb}%
  \BibitemOpen
  \bibfield  {author} {\bibinfo {author} {\bibfnamefont {T.}~\bibnamefont
  {Yamazaki}} \emph {et~al.} (\bibinfo {collaboration} {CP-PACS}),\ }\bibfield
  {title} {\enquote {\bibinfo {title} {{I = 2 pi pi scattering phase shift with
  two flavors of O(a) improved dynamical quarks}},}\ }\href {\doibase
  10.1103/PhysRevD.70.074513} {\bibfield  {journal} {\bibinfo  {journal} {Phys.
  Rev. D}\ }\textbf {\bibinfo {volume} {70}},\ \bibinfo {pages} {074513}
  (\bibinfo {year} {2004})},\ \Eprint {http://arxiv.org/abs/hep-lat/0402025}
  {arXiv:hep-lat/0402025} \BibitemShut {NoStop}%
\bibitem [{\citenamefont {Beane}\ \emph {et~al.}(2006)\citenamefont {Beane},
  \citenamefont {Bedaque}, \citenamefont {Orginos},\ and\ \citenamefont
  {Savage}}]{Beane:2005rj}%
  \BibitemOpen
  \bibfield  {author} {\bibinfo {author} {\bibfnamefont {Silas~R.}\
  \bibnamefont {Beane}}, \bibinfo {author} {\bibfnamefont {Paulo~F.}\
  \bibnamefont {Bedaque}}, \bibinfo {author} {\bibfnamefont {Kostas}\
  \bibnamefont {Orginos}}, \ and\ \bibinfo {author} {\bibfnamefont {Martin~J.}\
  \bibnamefont {Savage}} (\bibinfo {collaboration} {NPLQCD}),\ }\bibfield
  {title} {\enquote {\bibinfo {title} {{I = 2 pi-pi scattering from
  fully-dynamical mixed-action lattice QCD}},}\ }\href {\doibase
  10.1103/PhysRevD.73.054503} {\bibfield  {journal} {\bibinfo  {journal} {Phys.
  Rev. D}\ }\textbf {\bibinfo {volume} {73}},\ \bibinfo {pages} {054503}
  (\bibinfo {year} {2006})},\ \Eprint {http://arxiv.org/abs/hep-lat/0506013}
  {arXiv:hep-lat/0506013} \BibitemShut {NoStop}%
\bibitem [{\citenamefont {Beane}\ \emph {et~al.}(2008)\citenamefont {Beane},
  \citenamefont {Luu}, \citenamefont {Orginos}, \citenamefont {Parreno},
  \citenamefont {Savage}, \citenamefont {Torok},\ and\ \citenamefont
  {Walker-Loud}}]{Beane:2007xs}%
  \BibitemOpen
  \bibfield  {author} {\bibinfo {author} {\bibfnamefont {Silas~R.}\
  \bibnamefont {Beane}}, \bibinfo {author} {\bibfnamefont {Thomas~C.}\
  \bibnamefont {Luu}}, \bibinfo {author} {\bibfnamefont {Kostas}\ \bibnamefont
  {Orginos}}, \bibinfo {author} {\bibfnamefont {Assumpta}\ \bibnamefont
  {Parreno}}, \bibinfo {author} {\bibfnamefont {Martin~J.}\ \bibnamefont
  {Savage}}, \bibinfo {author} {\bibfnamefont {Aaron}\ \bibnamefont {Torok}}, \
  and\ \bibinfo {author} {\bibfnamefont {Andre}\ \bibnamefont {Walker-Loud}},\
  }\bibfield  {title} {\enquote {\bibinfo {title} {{Precise Determination of
  the I=2 pi pi Scattering Length from Mixed-Action Lattice QCD}},}\ }\href
  {\doibase 10.1103/PhysRevD.77.014505} {\bibfield  {journal} {\bibinfo
  {journal} {Phys. Rev. D}\ }\textbf {\bibinfo {volume} {77}},\ \bibinfo
  {pages} {014505} (\bibinfo {year} {2008})},\ \Eprint
  {http://arxiv.org/abs/0706.3026} {arXiv:0706.3026 [hep-lat]} \BibitemShut
  {NoStop}%
\bibitem [{\citenamefont {Beane}\ \emph {et~al.}(2012)\citenamefont {Beane},
  \citenamefont {Chang}, \citenamefont {Detmold}, \citenamefont {Lin},
  \citenamefont {Luu}, \citenamefont {Orginos}, \citenamefont {Parreno},
  \citenamefont {Savage}, \citenamefont {Torok},\ and\ \citenamefont
  {Walker-Loud}}]{Beane:2011sc}%
  \BibitemOpen
  \bibfield  {author} {\bibinfo {author} {\bibfnamefont {S.R.}\ \bibnamefont
  {Beane}}, \bibinfo {author} {\bibfnamefont {E.}~\bibnamefont {Chang}},
  \bibinfo {author} {\bibfnamefont {W.}~\bibnamefont {Detmold}}, \bibinfo
  {author} {\bibfnamefont {H.W.}\ \bibnamefont {Lin}}, \bibinfo {author}
  {\bibfnamefont {T.C.}\ \bibnamefont {Luu}}, \bibinfo {author} {\bibfnamefont
  {K.}~\bibnamefont {Orginos}}, \bibinfo {author} {\bibfnamefont
  {A.}~\bibnamefont {Parreno}}, \bibinfo {author} {\bibfnamefont {M.J.}\
  \bibnamefont {Savage}}, \bibinfo {author} {\bibfnamefont {A.}~\bibnamefont
  {Torok}}, \ and\ \bibinfo {author} {\bibfnamefont {A.}~\bibnamefont
  {Walker-Loud}} (\bibinfo {collaboration} {NPLQCD}),\ }\bibfield  {title}
  {\enquote {\bibinfo {title} {{The I=2 pipi S-wave Scattering Phase Shift from
  Lattice QCD}},}\ }\href {\doibase 10.1103/PhysRevD.85.034505} {\bibfield
  {journal} {\bibinfo  {journal} {Phys. Rev. D}\ }\textbf {\bibinfo {volume}
  {85}},\ \bibinfo {pages} {034505} (\bibinfo {year} {2012})},\ \Eprint
  {http://arxiv.org/abs/1107.5023} {arXiv:1107.5023 [hep-lat]} \BibitemShut
  {NoStop}%
\bibitem [{\citenamefont {Yagi}\ \emph {et~al.}(2011)\citenamefont {Yagi},
  \citenamefont {Hashimoto}, \citenamefont {Morimatsu},\ and\ \citenamefont
  {Ohtani}}]{Yagi:2011jn}%
  \BibitemOpen
  \bibfield  {author} {\bibinfo {author} {\bibfnamefont {Takuya}\ \bibnamefont
  {Yagi}}, \bibinfo {author} {\bibfnamefont {Shoji}\ \bibnamefont {Hashimoto}},
  \bibinfo {author} {\bibfnamefont {Osamu}\ \bibnamefont {Morimatsu}}, \ and\
  \bibinfo {author} {\bibfnamefont {Munehisa}\ \bibnamefont {Ohtani}},\
  }\bibfield  {title} {\enquote {\bibinfo {title} {{I=2 $\pi$-$\pi$ scattering
  length with dynamical overlap fermion}},}\ }\href@noop {} {\  (\bibinfo
  {year} {2011})},\ \Eprint {http://arxiv.org/abs/1108.2970} {arXiv:1108.2970
  [hep-lat]} \BibitemShut {NoStop}%
\bibitem [{\citenamefont {Fu}(2013)}]{Fu:2013ffa}%
  \BibitemOpen
  \bibfield  {author} {\bibinfo {author} {\bibfnamefont {Ziwen}\ \bibnamefont
  {Fu}},\ }\bibfield  {title} {\enquote {\bibinfo {title} {{Lattice QCD study
  of the s-wave $\pi\pi $ scattering lengths in the I=0 and 2 channels}},}\
  }\href {\doibase 10.1103/PhysRevD.87.074501} {\bibfield  {journal} {\bibinfo
  {journal} {Phys. Rev. D}\ }\textbf {\bibinfo {volume} {87}},\ \bibinfo
  {pages} {074501} (\bibinfo {year} {2013})},\ \Eprint
  {http://arxiv.org/abs/1303.0517} {arXiv:1303.0517 [hep-lat]} \BibitemShut
  {NoStop}%
\bibitem [{\citenamefont {Sasaki}\ \emph {et~al.}(2014)\citenamefont {Sasaki},
  \citenamefont {Ishizuka}, \citenamefont {Oka},\ and\ \citenamefont
  {Yamazaki}}]{Sasaki:2013vxa}%
  \BibitemOpen
  \bibfield  {author} {\bibinfo {author} {\bibfnamefont {Kiyoshi}\ \bibnamefont
  {Sasaki}}, \bibinfo {author} {\bibfnamefont {Naruhito}\ \bibnamefont
  {Ishizuka}}, \bibinfo {author} {\bibfnamefont {Makoto}\ \bibnamefont {Oka}},
  \ and\ \bibinfo {author} {\bibfnamefont {Takeshi}\ \bibnamefont {Yamazaki}}
  (\bibinfo {collaboration} {PACS-CS}),\ }\bibfield  {title} {\enquote
  {\bibinfo {title} {{Scattering lengths for two pseudoscalar meson
  systems}},}\ }\href {\doibase 10.1103/PhysRevD.89.054502} {\bibfield
  {journal} {\bibinfo  {journal} {Phys. Rev. D}\ }\textbf {\bibinfo {volume}
  {89}},\ \bibinfo {pages} {054502} (\bibinfo {year} {2014})},\ \Eprint
  {http://arxiv.org/abs/1311.7226} {arXiv:1311.7226 [hep-lat]} \BibitemShut
  {NoStop}%
\bibitem [{\citenamefont {Bijnens}\ \emph {et~al.}(1997)\citenamefont
  {Bijnens}, \citenamefont {Colangelo}, \citenamefont {Ecker}, \citenamefont
  {Gasser},\ and\ \citenamefont {Sainio}}]{Bijnens:1997vq}%
  \BibitemOpen
  \bibfield  {author} {\bibinfo {author} {\bibfnamefont {J.}~\bibnamefont
  {Bijnens}}, \bibinfo {author} {\bibfnamefont {G.}~\bibnamefont {Colangelo}},
  \bibinfo {author} {\bibfnamefont {G.}~\bibnamefont {Ecker}}, \bibinfo
  {author} {\bibfnamefont {J.}~\bibnamefont {Gasser}}, \ and\ \bibinfo {author}
  {\bibfnamefont {M.E.}\ \bibnamefont {Sainio}},\ }\bibfield  {title} {\enquote
  {\bibinfo {title} {{Pion-pion scattering at low energy}},}\ }\href {\doibase
  10.1016/S0550-3213(97)00621-4} {\bibfield  {journal} {\bibinfo  {journal}
  {Nucl. Phys. B}\ }\textbf {\bibinfo {volume} {508}},\ \bibinfo {pages}
  {263--310} (\bibinfo {year} {1997})},\ \bibinfo {note} {[Erratum: Nucl.Phys.B
  517, 639--639 (1998)]},\ \Eprint {http://arxiv.org/abs/hep-ph/9707291}
  {arXiv:hep-ph/9707291} \BibitemShut {NoStop}%
\bibitem [{\citenamefont {{Jülich Supercomputing Centre}}(2015)}]{juqueen}%
  \BibitemOpen
  \bibfield  {author} {\bibinfo {author} {\bibnamefont {{Jülich Supercomputing
  Centre}}},\ }\bibfield  {title} {\enquote {\bibinfo {title} {{JUQUEEN: IBM
  Blue Gene/Q Supercomputer System at the Jülich Supercomputing Centre}},}\
  }\href {\doibase 10.17815/jlsrf-1-18} {\bibfield  {journal} {\bibinfo
  {journal} {Journal of large-scale research facilities}\ }\textbf {\bibinfo
  {volume} {1}} (\bibinfo {year} {2015}),\ 10.17815/jlsrf-1-18}\BibitemShut
  {NoStop}%
\bibitem [{\citenamefont {{Jülich Supercomputing Centre}}(2018)}]{jureca}%
  \BibitemOpen
  \bibfield  {author} {\bibinfo {author} {\bibnamefont {{Jülich Supercomputing
  Centre}}},\ }\bibfield  {title} {\enquote {\bibinfo {title} {{JURECA: Modular
  supercomputer at Jülich Supercomputing Centre}},}\ }\href {\doibase
  10.17815/jlsrf-4-121-1} {\bibfield  {journal} {\bibinfo  {journal} {Journal
  of large-scale research facilities}\ }\textbf {\bibinfo {volume} {4}}
  (\bibinfo {year} {2018}),\ 10.17815/jlsrf-4-121-1}\BibitemShut {NoStop}%
\bibitem [{\citenamefont {{Jülich Supercomputing Centre}}(2019)}]{juwels}%
  \BibitemOpen
  \bibfield  {author} {\bibinfo {author} {\bibnamefont {{Jülich Supercomputing
  Centre}}},\ }\bibfield  {title} {\enquote {\bibinfo {title} {{JUWELS: Modular
  Tier-0/1 Supercomputer at the Jülich Supercomputing Centre}},}\ }\href
  {\doibase 10.17815/jlsrf-5-171} {\bibfield  {journal} {\bibinfo  {journal}
  {Journal of large-scale research facilities}\ }\textbf {\bibinfo {volume}
  {5}} (\bibinfo {year} {2019}),\ 10.17815/jlsrf-5-171}\BibitemShut {NoStop}%
\bibitem [{\citenamefont {Jansen}\ and\ \citenamefont
  {Urbach}(2009)}]{Jansen:2009xp}%
  \BibitemOpen
  \bibfield  {author} {\bibinfo {author} {\bibfnamefont {K.}~\bibnamefont
  {Jansen}}\ and\ \bibinfo {author} {\bibfnamefont {C.}~\bibnamefont
  {Urbach}},\ }\bibfield  {title} {\enquote {\bibinfo {title} {{tmLQCD: A
  Program suite to simulate Wilson Twisted mass Lattice QCD}},}\ }\href
  {\doibase 10.1016/j.cpc.2009.05.016} {\bibfield  {journal} {\bibinfo
  {journal} {Comput.Phys.Commun.}\ }\textbf {\bibinfo {volume} {180}},\
  \bibinfo {pages} {2717--2738} (\bibinfo {year} {2009})},\ \Eprint
  {http://arxiv.org/abs/0905.3331} {arXiv:0905.3331 [hep-lat]} \BibitemShut
  {NoStop}%
\bibitem [{\citenamefont {Abdel-Rehim}\ \emph {et~al.}(2014)\citenamefont
  {Abdel-Rehim}, \citenamefont {Burger}, \citenamefont {Deuzeman},
  \citenamefont {Jansen}, \citenamefont {Kostrzewa}, \citenamefont {Scorzato},\
  and\ \citenamefont {Urbach}}]{Abdel-Rehim:2013wba}%
  \BibitemOpen
  \bibfield  {author} {\bibinfo {author} {\bibfnamefont {Abdou}\ \bibnamefont
  {Abdel-Rehim}}, \bibinfo {author} {\bibfnamefont {Florian}\ \bibnamefont
  {Burger}}, \bibinfo {author} {\bibfnamefont {Alber}\ \bibnamefont
  {Deuzeman}}, \bibinfo {author} {\bibfnamefont {Karl}\ \bibnamefont {Jansen}},
  \bibinfo {author} {\bibfnamefont {Bartosz}\ \bibnamefont {Kostrzewa}},
  \bibinfo {author} {\bibfnamefont {Luigi}\ \bibnamefont {Scorzato}}, \ and\
  \bibinfo {author} {\bibfnamefont {Carsten}\ \bibnamefont {Urbach}},\
  }\bibfield  {title} {\enquote {\bibinfo {title} {{Recent developments in the
  tmLQCD software suite}},}\ }\href {\doibase 10.22323/1.187.0414} {\bibfield
  {journal} {\bibinfo  {journal} {PoS}\ }\textbf {\bibinfo {volume}
  {LATTICE2013}},\ \bibinfo {pages} {414} (\bibinfo {year} {2014})},\ \Eprint
  {http://arxiv.org/abs/1311.5495} {arXiv:1311.5495 [hep-lat]} \BibitemShut
  {NoStop}%
\bibitem [{\citenamefont {Deuzeman}\ \emph {et~al.}(2013)\citenamefont
  {Deuzeman}, \citenamefont {Jansen}, \citenamefont {Kostrzewa},\ and\
  \citenamefont {Urbach}}]{Deuzeman:2013xaa}%
  \BibitemOpen
  \bibfield  {author} {\bibinfo {author} {\bibfnamefont {A.}~\bibnamefont
  {Deuzeman}}, \bibinfo {author} {\bibfnamefont {K.}~\bibnamefont {Jansen}},
  \bibinfo {author} {\bibfnamefont {B.}~\bibnamefont {Kostrzewa}}, \ and\
  \bibinfo {author} {\bibfnamefont {C.}~\bibnamefont {Urbach}},\ }\bibfield
  {title} {\enquote {\bibinfo {title} {{Experiences with OpenMP in tmLQCD}},}\
  }\href@noop {} {\bibfield  {journal} {\bibinfo  {journal} {PoS}\ }\textbf
  {\bibinfo {volume} {LATTICE2013}},\ \bibinfo {pages} {416} (\bibinfo {year}
  {2013})},\ \Eprint {http://arxiv.org/abs/1311.4521} {arXiv:1311.4521
  [hep-lat]} \BibitemShut {NoStop}%
\bibitem [{\citenamefont {Deuzeman}\ \emph {et~al.}(2012)\citenamefont
  {Deuzeman}, \citenamefont {Reker},\ and\ \citenamefont
  {Urbach}}]{Deuzeman:2011wz}%
  \BibitemOpen
  \bibfield  {author} {\bibinfo {author} {\bibfnamefont {Albert}\ \bibnamefont
  {Deuzeman}}, \bibinfo {author} {\bibfnamefont {Siebren}\ \bibnamefont
  {Reker}}, \ and\ \bibinfo {author} {\bibfnamefont {Carsten}\ \bibnamefont
  {Urbach}} (\bibinfo {collaboration} {ETM}),\ }\bibfield  {title} {\enquote
  {\bibinfo {title} {{Lemon: an MPI parallel I/O library for data encapsulation
  using LIME}},}\ }\href {\doibase 10.1016/j.cpc.2012.01.016} {\bibfield
  {journal} {\bibinfo  {journal} {Comput. Phys. Commun.}\ }\textbf {\bibinfo
  {volume} {183}},\ \bibinfo {pages} {1321--1335} (\bibinfo {year} {2012})},\
  \Eprint {http://arxiv.org/abs/1106.4177} {arXiv:1106.4177 [hep-lat]}
  \BibitemShut {NoStop}%
\bibitem [{\citenamefont {Clark}\ \emph {et~al.}(2010)\citenamefont {Clark},
  \citenamefont {Babich}, \citenamefont {Barros}, \citenamefont {Brower},\ and\
  \citenamefont {Rebbi}}]{Clark:2009wm}%
  \BibitemOpen
  \bibfield  {author} {\bibinfo {author} {\bibfnamefont {M.~A.}\ \bibnamefont
  {Clark}}, \bibinfo {author} {\bibfnamefont {R.}~\bibnamefont {Babich}},
  \bibinfo {author} {\bibfnamefont {K.}~\bibnamefont {Barros}}, \bibinfo
  {author} {\bibfnamefont {R.~C.}\ \bibnamefont {Brower}}, \ and\ \bibinfo
  {author} {\bibfnamefont {C.}~\bibnamefont {Rebbi}},\ }\bibfield  {title}
  {\enquote {\bibinfo {title} {{Solving Lattice QCD systems of equations using
  mixed precision solvers on GPUs}},}\ }\href {\doibase
  10.1016/j.cpc.2010.05.002} {\bibfield  {journal} {\bibinfo  {journal}
  {Comput. Phys. Commun.}\ }\textbf {\bibinfo {volume} {181}},\ \bibinfo
  {pages} {1517--1528} (\bibinfo {year} {2010})},\ \Eprint
  {http://arxiv.org/abs/0911.3191} {arXiv:0911.3191 [hep-lat]} \BibitemShut
  {NoStop}%
\bibitem [{\citenamefont {Babich}\ \emph {et~al.}(2011)\citenamefont {Babich},
  \citenamefont {Clark}, \citenamefont {Joo}, \citenamefont {Shi},
  \citenamefont {Brower},\ and\ \citenamefont {Gottlieb}}]{Babich:2011np}%
  \BibitemOpen
  \bibfield  {author} {\bibinfo {author} {\bibfnamefont {R.}~\bibnamefont
  {Babich}}, \bibinfo {author} {\bibfnamefont {M.~A.}\ \bibnamefont {Clark}},
  \bibinfo {author} {\bibfnamefont {B.}~\bibnamefont {Joo}}, \bibinfo {author}
  {\bibfnamefont {G.}~\bibnamefont {Shi}}, \bibinfo {author} {\bibfnamefont
  {R.~C.}\ \bibnamefont {Brower}}, \ and\ \bibinfo {author} {\bibfnamefont
  {S.}~\bibnamefont {Gottlieb}},\ }\bibfield  {title} {\enquote {\bibinfo
  {title} {{Scaling Lattice QCD beyond 100 GPUs}},}\ }in\ \href {\doibase
  10.1145/2063384.2063478} {\emph {\bibinfo {booktitle} {{SC11 International
  Conference for High Performance Computing, Networking, Storage and Analysis
  Seattle, Washington, November 12-18, 2011}}}}\ (\bibinfo {year} {2011})\
  \Eprint {http://arxiv.org/abs/1109.2935} {arXiv:1109.2935 [hep-lat]}
  \BibitemShut {NoStop}%
\bibitem [{\citenamefont {Clark}\ \emph {et~al.}(2016)\citenamefont {Clark},
  \citenamefont {Joó}, \citenamefont {Strelchenko}, \citenamefont {Cheng},
  \citenamefont {Gambhir},\ and\ \citenamefont {Brower}}]{Clark:2016rdz}%
  \BibitemOpen
  \bibfield  {author} {\bibinfo {author} {\bibfnamefont {M.~A.}\ \bibnamefont
  {Clark}}, \bibinfo {author} {\bibfnamefont {Bálint}\ \bibnamefont {Joó}},
  \bibinfo {author} {\bibfnamefont {Alexei}\ \bibnamefont {Strelchenko}},
  \bibinfo {author} {\bibfnamefont {Michael}\ \bibnamefont {Cheng}}, \bibinfo
  {author} {\bibfnamefont {Arjun}\ \bibnamefont {Gambhir}}, \ and\ \bibinfo
  {author} {\bibfnamefont {Richard}\ \bibnamefont {Brower}},\ }\bibfield
  {title} {\enquote {\bibinfo {title} {{Accelerating Lattice QCD Multigrid on
  GPUs Using Fine-Grained Parallelization}},}\ }\href@noop {} {\  (\bibinfo
  {year} {2016})},\ \Eprint {http://arxiv.org/abs/1612.07873} {arXiv:1612.07873
  [hep-lat]} \BibitemShut {NoStop}%
\bibitem [{\citenamefont {{R Core Team}}(2019)}]{R:2019}%
  \BibitemOpen
  \bibfield  {author} {\bibinfo {author} {\bibnamefont {{R Core Team}}},\
  }\href {https://www.R-project.org/} {\emph {\bibinfo {title} {R: A Language
  and Environment for Statistical Computing}}},\ \bibinfo {organization} {R
  Foundation for Statistical Computing},\ \bibinfo {address} {Vienna, Austria}
  (\bibinfo {year} {2019})\BibitemShut {NoStop}%
\bibitem [{\citenamefont {Kostrzewa}\ \emph {et~al.}(2020)\citenamefont
  {Kostrzewa}, \citenamefont {Ostmeyer}, \citenamefont {Ueding},\ and\
  \citenamefont {Urbach}}]{hadron:2020}%
  \BibitemOpen
  \bibfield  {author} {\bibinfo {author} {\bibfnamefont {Bartosz}\ \bibnamefont
  {Kostrzewa}}, \bibinfo {author} {\bibfnamefont {Johann}\ \bibnamefont
  {Ostmeyer}}, \bibinfo {author} {\bibfnamefont {Martin}\ \bibnamefont
  {Ueding}}, \ and\ \bibinfo {author} {\bibfnamefont {Carsten}\ \bibnamefont
  {Urbach}},\ }\href {https://github.com/HISKP-LQCD/hadron} {\enquote {\bibinfo
  {title} {hadron: package to extract hadronic quantities},}\ }\bibinfo
  {howpublished} {https://github.com/HISKP-LQCD/hadron} (\bibinfo {year}
  {2020}),\ \bibinfo {note} {{R} package version 3.0.1}\BibitemShut {NoStop}%
\bibitem [{\citenamefont {Ueding}({\natexlab{b}})}]{paramvalf}%
  \BibitemOpen
  \bibfield  {author} {\bibinfo {author} {\bibfnamefont {Martin}\ \bibnamefont
  {Ueding}},\ }\href@noop {} {\emph {\bibinfo {title} {paramvalf: Parameter
  Value Analysis Framework}}} ({\natexlab{b}}),\ \bibinfo {note} {r package
  version 2.7.0}\BibitemShut {NoStop}%
\bibitem [{\citenamefont {Wolff}(2004)}]{Wolff:2003sm}%
  \BibitemOpen
  \bibfield  {author} {\bibinfo {author} {\bibfnamefont {Ulli}\ \bibnamefont
  {Wolff}} (\bibinfo {collaboration} {ALPHA}),\ }\bibfield  {title} {\enquote
  {\bibinfo {title} {{Monte Carlo errors with less errors}},}\ }\href {\doibase
  10.1016/S0010-4655(03)00467-3; 10.1016/j.cpc.2006.12.001} {\bibfield
  {journal} {\bibinfo  {journal} {Comput. Phys. Commun.}\ }\textbf {\bibinfo
  {volume} {156}},\ \bibinfo {pages} {143–153} (\bibinfo {year} {2004})},\
  \bibinfo {note} {[Erratum: Comput. Phys. Commun.176,383(2007)]},\ \Eprint
  {http://arxiv.org/abs/hep-lat/0306017} {arXiv:hep-lat/0306017 [hep-lat]}
  \BibitemShut {NoStop}%
\end{thebibliography}%
